\makeatletter
\def\cl@chapter{}
\makeatother

\documentclass[twocolumn,english]{svjour3}        

\usepackage[]{algorithm2e}
\usepackage{nicefrac}
\usepackage[pdftex]{graphicx}
\usepackage{pgfplots} \usepackage{float} \usepackage{adjustbox}
\usepackage[font=footnotesize]{subcaption}

\makeatletter
\if@twocolumn
    \renewcommand\normalsize{\@setfontsize\normalsize{9.5pt}{11pt}\abovedisplayskip=3 mm plus6pt minus 4pt
	\belowdisplayskip=3 mm plus6pt minus 4pt
	\abovedisplayshortskip=0.0 mm plus6pt
	\belowdisplayshortskip=2 mm plus4pt minus 4pt
	\let\@listi\@listI}

\renewcommand\small{\@setfontsize\small{8.5pt}{9pt}\abovedisplayskip 8.5\p@ \@plus3\p@ \@minus4\p@
	\abovedisplayshortskip \z@ \@plus2\p@
	\belowdisplayshortskip 4\p@ \@plus2\p@ \@minus2\p@
	\def\@listi{\leftmargin\leftmargini
		\parsep 0\p@ \@plus1\p@ \@minus\p@
		\topsep 4\p@ \@plus2\p@ \@minus4\p@
		\itemsep0\p@}\belowdisplayskip \abovedisplayskip}
\else
  \if@smallext
    \renewcommand\normalsize{\@setfontsize\normalsize\@xpt\@xiipt
      \abovedisplayskip=3 mm plus6pt minus 4pt
      \belowdisplayskip=3 mm plus6pt minus 4pt
      \abovedisplayshortskip=0.0 mm plus6pt
      \belowdisplayshortskip=2 mm plus4pt minus 4pt
      \let\@listi\@listI}

    \renewcommand\small{\@setfontsize\small\@viiipt{9.5pt}\abovedisplayskip 8.5\p@ \@plus3\p@ \@minus4\p@
      \abovedisplayshortskip \z@ \@plus2\p@
      \belowdisplayshortskip 4\p@ \@plus2\p@ \@minus2\p@
      \def\@listi{\leftmargin\leftmargini
        \parsep 0\p@ \@plus1\p@ \@minus\p@
        \topsep 4\p@ \@plus2\p@ \@minus4\p@
        \itemsep0\p@}\belowdisplayskip \abovedisplayskip}
  \else
    \renewcommand\normalsize{\@setfontsize\normalsize{9.5pt}{11.5pt}\abovedisplayskip=3 mm plus6pt minus 4pt
      \belowdisplayskip=3 mm plus6pt minus 4pt
      \abovedisplayshortskip=0.0 mm plus6pt
      \belowdisplayshortskip=2 mm plus4pt minus 4pt
      \let\@listi\@listI}

    \renewcommand\small{\@setfontsize\small\@viiipt{9.25pt}\abovedisplayskip 8.5\p@ \@plus3\p@ \@minus4\p@
      \abovedisplayshortskip \z@ \@plus2\p@
      \belowdisplayshortskip 4\p@ \@plus2\p@ \@minus2\p@
      \def\@listi{\leftmargin\leftmargini
        \parsep 0\p@ \@plus1\p@ \@minus\p@
        \topsep 4\p@ \@plus2\p@ \@minus4\p@
        \itemsep0\p@}\belowdisplayskip \abovedisplayskip}
  \fi
\fi
\let\footnotesize\small
\makeatother

\usepackage{cmap}

\RequirePackage[shrink=40,
  final,expansion=alltext,protrusion=alltext-nott]{microtype}\DisableLigatures{encoding = T1, family = tt* }

\usepackage[ngerman,main=english]{babel}
\addto\extrasenglish{\languageshorthands{ngerman}\useshorthands{"}}

\usepackage{ifluatex}
\ifluatex
  \usepackage{fontspec}
  \usepackage[english]{selnolig}
\fi

\ifluatex
 \else
\usepackage[rm={oldstyle=false,proportional=true},sf={oldstyle=false,proportional=true},tt={oldstyle=false,proportional=true,variable=false},qt=false]{cfr-lm}
\fi

\ifluatex
\else
\usepackage[T1]{fontenc}
  \usepackage[utf8]{inputenc} \fi

\usepackage{upquote}

\usepackage{booktabs}

\usepackage{paralist}

\usepackage{csquotes}

\usepackage{url}
\makeatletter
\g@addto@macro{\UrlBreaks}{\UrlOrds}
\makeatother

\usepackage{xcolor}

\usepackage{pdfcomment}

\newcommand*{\doi}[1]{\href{https://doi.org/\detokenize{#1}}{DOI: \detokenize{#1}}}

\usepackage{stfloats}
\fnbelowfloat

\usepackage[all]{hypcap}

\usepackage[acronym,indexonlyfirst,nomain]{glossaries}

\usepackage[capitalise,nameinlink]{cleveref}
\crefname{figure}{Fig.}{Fig.}
\crefname{section}{Sect.}{Sect.}
\Crefname{section}{Section}{Sections}
\crefname{listing}{\lstlistingname}{\lstlistingname}
\Crefname{listing}{Listing}{Listings}

\newcommand{\Vlabel}[1]{\label[line]{#1}\hypertarget{#1}{}}
\newcommand{\lref}[1]{\hyperlink{#1}{\FancyVerbLineautorefname~\ref*{#1}}}

\usepackage{xspace}
\newcommand{\ie}{i.\,e.,\ }
\newcommand{\eg}{e.\,g.,\xspace}
\newcommand{\egc}{e.\,g.,\xspace}

\journalname{Journal}

\usepackage[math]{blindtext}
\usepackage{mwe}

\begin{document}

\title{Instance-based Learning with Prototype Reduction for Real-Time Proportional Myocontrol}
\subtitle{A Randomized User Study Demonstrating Accuracy-preserving Data Reduction for Prosthetic Embedded Systems}

\author{Tim Sziburis \and Markus Nowak \and Davide Brunelli}

\institute{
  T.~Sziburis \and M.~Nowak\at German Aerospace Center (DLR), Robotics and Mechatronics Center (RMC), Münchener Str. 20, 82234 We\ss ling, Germany
  \and
  T.~Sziburis\at High Precision Alignment Technologies Section, CERN, 1211 Geneva 23, Switzerland
  \and
  D.~Brunelli\at Department of Industrial Engineering, DII, University of Trento, Via Sommarive, 9, 38123 Trento, Italy
}

\date{}

\maketitle

\begin{abstract}
	This work presents the design, implementation and validation of learning techniques based on the kNN scheme for gesture detection in prosthetic control. To cope with high computational demands in instance-based prediction, methods of dataset reduction are evaluated considering real-time determinism to allow for the reliable integration into battery-powered portable devices.
	
	The influence of parameterization and varying proportionality schemes is analyzed, utilizing an eight-channel-sEMG armband. Besides offline cross-validation accuracy, success rates in real-time pilot experiments (online target achievement tests) are determined.
	
	Based on the assessment of specific dataset reduction techniques' adequacy for embedded control applications regarding accuracy and timing behaviour, Decision Surface Mapping (DSM) proves itself promising when applying kNN on the reduced set.
	
	A randomized, double-blind user study was conducted to evaluate the respective methods (kNN and kNN with DSM-reduction) against Ridge Regression (RR) and RR with Random Fourier Features (RR-RFF). The kNN-based methods performed significantly better ($p < 0.0005$) than the regression techniques. Between DSM-kNN and kNN, there was no statistically significant difference (significance level $0.05$). This is remarkable in consideration of only one sample per class in the reduced set, thus yielding a reduction rate of over 99\% while preserving success rate. The same behaviour could be confirmed in an extended user study.
	
	With $k=1$, which turned out to be an excellent choice, the runtime complexity of both kNN (in every prediction step) as well as DSM-kNN (in the training phase) becomes linear concerning the number of original samples, favouring dependable wearable prosthesis applications.
\end{abstract}

\section{Introduction and Motivation}

The development of prosthetics has been continuously improving since the 20th century. After being solely a cosmetic replacement for amputees, prostheses evolved to body-driven functional devices and, especially beginning in the 1940s, to powered myoelectric systems \cite[p.~32]{Vujaklija}.

From the existing myographic methods to collect data from muscular activity, this paper focuses on surface electromyography (sEMG), which comprises the capturing, processing and analysis of electromyographic signals, i. e. the changes over time in electric potential originating from skeletal muscles (cf. \cite{nih} and \cite{mw}), measured by electrodes on the skin surface.

Prosthetic control describes the general concept behind the process from capturing signal data (sensor side) via processing and analyzing it to forwarding the data interpretation to a prosthetic device (actuator side) -- with potential feedback loops. The (closed-loop) motor-control of the prosthetic actuators themselves, i.~e. the single joints of the prosthetic device by e.~g. direct force control, impedance or admittance control will not be considered in the scope of this work.

There has been a variety of control schemes presented for EMG-based prosthetic devices in the area of open-loop myocontrol \cite[p.~252]{Geethanjali}. As a consequence of their diverse fundamental characteristics, they differ in the achieved granularity, precision and stability of movements (dexterity).

Pattern-recognition-based myoelectric control schemes utilize machine learning methods to correctly detect gestures by means of classification or regression. Specific features can be calculated from the raw or filtered signal in time-, frequency- and time-frequency domain \cite[p.~250]{Geethanjali}. The basic principle of action (intent/intention) detection is to predict a specific action from the sensed biological signal.

This work examines and extends the k-nearest neighbour learning scheme (kNN). The respective methods are based upon detecting specific gestures defined beforehand, leading in the application phase to their real-time recognition. For not explicitly learned, intermediate gestures proportionality scaling techniques are introduced.

Commonly applied pattern recognition methods in myocontrol often show disadvantages in terms of generalizability, intuitive control and robustness regarding ``electrodes shift, varying force levels'' \cite{7591042} (e.~g. overshooting) and others. To cope with these limitations, an extended kNN learning scheme seemed promising due to its simplicity, incrementality and good results in exemplary tests.

Referring to the results of kNN in the context of EMG-based prosthetics in section \ref{state_kNN}, it can be seen that in most cases kNN delivers high performance in terms of accuracy and success rate respectively. \cite{Kuzborskij} states that ``excellent performance can be achieved if sufficient training data is available''.

kNN is shown to be relatively insensitive to noise \cite{5767304}, electrode displacement \cite{6346923} and sampling frequency variation \cite{8122765} which speaks in favour of robustness.

Its low complexity is a main advantage of kNN specifically in the context of implementation for embedded systems. As for all instance-based machine learning techniques, it does not require an explicit model training. However, it has to be coped with high computational demands during the prediction. Thus, this paper introduces mechanisms for dataset reduction, combines them with kNN, and finally analyzes a selected method to be called DSM-kNN. 

The applicability to embedded, wearable systems is specifically important since ``users of modern prosthetics are now given access to applications that can run on external devices capable of fine tuning and setting up gestures or gesture patterns [\ldots] [to] allow[] high-level customization'' \cite{Vujaklija} and comes with non-functional constraints regarding energy consumption, portability, timeliness, safety and dependability.

Precise sensor placement on specific muscles is usually considered as essential for achieving high detection performance from sEMG signals \cite{Raurale}. This placed-sensor approach is not suitable for wearable devices, but could be compensated by (time-)frequency-domain feature analysis, although it ``cannot be realised in real-time using the simple embedded processors housed in EMG wearables'', as mentioned by \cite{Raurale}. They present an operational comparison of applicable features, projection techniques and classifiers. On this basis, they introduce a time-domain algorithm ``suitable for deployment on embedded processors for real-time inference in a portable, battery-operated device'' \cite{Raurale} by reducing clock cycle and therefore power consumption without impairing accuracy. However, their promising approach neither introduced proportionality schemes to classification nor conducted online user studies, and can be seen as a complement.

Although kNN has been applied in several EMG studies, there was no in-depth examination of the strategy's parameters nor has it been particularly combined with data reduction techniques as in our work.

\section{Related Work}

A variety of machine learning strategies has been followed over the years in the context of myocontrol, cf. \cite[p.~251]{Geethanjali}.
This includes:
\begin{inparaitem}
	\item neural networks in different compositions \cite{Guler}, \cite{Sueaseenak}, \cite{Hudgins}, \cite{Jiang}, \cite{Gini}, \cite{Arvetti}, \cite{Englehart1999}, \cite{7437907} \cite{5396091}, \cite{5767304}, \cite{Antfolk2011ACB}, \cite{Shin}, \cite{Geethanjali2011}, \cite{Geethanjali2015}, \cite{Purushothaman2016}, \cite{Robinson}, including such based on adaptive resonance theory \cite{Du},
	\item support vector machines and variants \cite{Kartsch}, \cite{Guler}, \cite{Kakoty}, \cite{Rekhi}, \cite{Zhizeng}, \cite{Zhang2013}, \cite{Shin}, \cite{8122765}, \cite{Robinson},
	\item decision trees \cite{Geethanjali2015},
	\item (naïve) Bayesian classification \cite{Kang}, \cite{Lei}, \cite{Kim2008}, \cite{Shin},
	\item fuzzy logic approaches \cite{Micera2000OnAI}, \cite{Ajiboye},
	\item Gaussian mixture models \cite{Huang2004OptimizedGM}, \cite{1519588}, \cite{Jeong2013},
	\item logistic regression \cite{Geethanjali2015},
	\item logistic model trees \cite{Geethanjali2015},
	\item classification via independent component analysis (ICA) \cite{Sueaseenak}, canonical discriminant analysis \cite{Nagata},
	\item linear discriminant analysis (LDA) \cite{Englehart1999}, \cite{Englehart2003}\cite{Antfolk2011ACB}, \cite{Geethanjali2011}, \cite{Jeong2013}, \cite{Zhang2013}, \cite{KIM2011740}, \cite{6487520}, \cite{7591042}, \cite{Della}, \cite{8122765},
	\item quadratic discriminant analysis (QDA) \cite{KIM2011740}, \cite{6487520}, \cite{Jeong2013}, \cite{Shin},
	\item random forest \cite{Robinson},
	\item extreme learning machines (ELM) \cite{Shin},
	\item hidden Markov models \cite{Chan},
	\item evolvable hardware (EHW): Embedded Cartesian Genetic Programming, Functional Unit Row \cite{4584252}, as well as
	\item kNN, see section \ref{state_kNN},
\end{inparaitem}

The following features and transformations have proven well in the context of pattern-recognition-based myoelectric control (cf. \cite[p.~250-251]{Geethanjali}):
\begin{inparaitem}
	\item linear envelope \cite{Zhang1991ClusteringAA}, \cite[p.~271]{Winter}, \cite{Paek2013}, \cite{BARZILAY2011678},
	\item zero crossings and variance \cite{Saridis1982},
	\item integral absolute value, variance, zero crossing \cite{Saridis1984},
	\item mean absolute value \cite{Antfolk2011ACB}, its slope, wave form length, number of waveform slope sign changes, number of waveform zero crossings (Hudgins set of features) \cite{Hudgins},
	\item frequency spectrum via Fourier transform \cite{Farry1993}, \cite{Guler}, \cite{Sueaseenak}, random Fourier features \cite{Gijs}, \cite{Gijsberts}, as well as local frequency and phase content via short-time Fourier transform \cite{Hannaford}, \cite{Englehart1999,Englehart2003}, \cite{Du},
	\item autoregressive coefficients \cite{Kang}, \cite{Chang}, \cite{Kirlangic},
	\item cepstral coefficients \cite{Kang}, \cite{Chang},
	\item wavelet decomposition coefficients \cite{Englehart1999,Englehart2003}, \cite{Jiang}, \cite{Maitrot}, \cite{Gini}, \cite{Arvetti}, \cite{Kakoty}, \cite{Rekhi} and their Eigenvalues \cite{Zhizeng},
	\item wavelet packet feature sets \cite{Englehart1999,Englehart2003}, motor unit action potentials (MUAPs) via wavelet packet transform and fuzzy C-means clustering \cite{Ren},
	\item signal energy (overall, within Hamming windows, within trapezoidal windows) as temporal features and spectral magnitude as well as spectral moments from short-time Thompson transform \cite{Du},
	\item moving approximate entropy \cite{Ahmad}, and
	\item contraction factors from fractal modeling \cite{Kirlangic}, fractal dimensions \cite{Hu,Arjunan}.
\end{inparaitem}

A review of classification techniques for forearm prostheses is given in \cite[p.~725]{Peerdeman}, along with information about features, performed experiments, selected subjects and achieved results. A review of the multitude of features (table 1), an evaluation thereof on EMG data (tables 2 and 4) with significance analysis (table 3) was conducted in \cite[p.~4834--4838]{PHINYOMARK20134832}.

The following subsections particularly summarize the utilization of the kNN learning scheme in the same context, as well as applicable data reduction techniques.

\subsection{Nearest Neighbour Techniques} \label{state_kNN}
kNN was firstly proposed for the parameter $k$ (numbers of neighbours to consider) set to $1$ as ``nearest neighbor decision rule'' in 1967 \cite{CoverHart}.

The basic principle of kNN consists of comparing new arriving data (instances) with all instances that were captured as reference data in an initial step and considering a subset of them (number of reference instances $k$) for a prediction decision. Although, this initial step does not comprise the generation of a generalized model (training of a model), it is usually called \textit{training} (also in this paper).

The comparison of instances refers to the comparison of distances between a new instance and the training instances by using a specified distance measure.

After calculating the distance, kNN selects a number $k$ of nearest instances to the new instance. If their labels are immediately averaged (which would have to be specified, usually arithmetic mean), this leads to a prediction in the form of a regression method (kNN regression). If instead of averaging a majority vote is applied on the $k$ nearest instances, the label with the most votes is yielded as categorical label prediction (kNN classification). In this sense, the focus of this paper is on kNN classification which will be extended by a proportionality scaling scheme (see \cref{metho_consid}).

In place of directly majority voting after a set of neighbouring instances has been selected (uniform weights), distance-based weighting factors can be calculated for all instances of the selection. For each class, these weightings are summed up, so that the prediction is determined as the class with the maximum sum. This distance weighting is typically introduced to avoid that a majority of class labels from instances which are farther away (but still within the neighbourhood) influences the prediction at the expense of instances which are closer but in minority.

kNN has been applied for EMG-based pattern recognition a variety of times. In 1983, a nearest neighbour classifier ($k=1$) was already chosen in the context of prosthetics \cite{Dening}. Additionally, a variant of prototype reduction (see section \ref{state_red}) was introduced. There was no decrease of performance when reducing the number of samples per gesture from 100 to 2--6 (for power grasp, flexion, extension, pronation), and to 40--46 (for rest and supination), respectively.

Since then, nearest-neighbour-based methods have been evaluated in various EMG-based gesture detection studies rather unsystematically, mainly for comparison to other classifiers, both for able-bodied subjects and amputees (\cref{tab:related}). For example, kNN showed to perform as good as multi-layer neural networks. In this context, it was mentioned that the ``kNN classifier may be considered to be a better choice for classification of continuous EMG signals to actuate the prosthetic drive'' \cite[p.~5]{Purushothaman2016}.

Usually, kNN also exposed similarly good detection accuracy as QDA, SVMs, Gaussian as well as Bayesian methods, and performed comparably or better than LDA (see \ref{tab:related}). In some cases, kNN was shown to perform statistically significantly better than LDA\cite{KIM2011740,Jeong2013} and even QDA\cite{Jeong2013}.
Further work pointed out that ``there was no significant [difference] between weak-load algorithms (NB, KNN, QDA, and ELM) and heavy-load algorithms (SVM and MLP) after applying the dimension reduction'' \cite{Shin}. An experiment with $k=9$ showed that the ``kNN classifier [was] better at classifying the EMG signal with PCA transformed statistical data compared to other classifiers in accuracy, sensitivity and specificity'' \cite{Geethanjali2015}, namely logistic regression, decision and logistic model trees as well as a neural network classifier.

Besides these comparisons of different classifiers without external factors, a study of noise influence on kNN exposed a high stability of detection accuracy even for reduced signal-to-noise ratios if $k$ is chosen properly. For $k=15$, the accuracy decreased from 100\% to 83\% for an increase of simulated noise from 25 to 5\,dB SNR. With this, it showed a higher robustness than a neural network classifier. 

The mid-term performance for discovering the influence of electrode shift on kNN showed a basically constant average performance from one day to another \cite{7860925}. This confirmed a former analysis of performance over time \cite{4584252} and is important for prosthetic devices since repositioning regularly introduces an electrode displacement which otherwise would require immediate retraining.

Finally, kNN has shown to be comparably insensitive to the reduction of recording sample rate. In an experiment, kNN achieved higher accuracies than SVM and LDA at all sampling rates, and the performance reduction for frequency reduction was not as steep as for other classifiers \cite{8122765}. Instead, for a change from 1000\,Hz to 200\,Hz, the accuracy reduced only minimally from 99.9\% to 99\%. At 20\,Hz, it still provided 78\% (vs. 71\%/56\% for SVM/LDA). This behaviour of the kNN method is highly advantageous for embedded systems scenarios, since lower sampling rates lead to lower CPU clock frequencies and therefore reduced powering requirements, which in the end support a low-cost approach and increase portability by requiring smaller dimensions for the prosthetic controller.

\begin{table*}[]
	\centering
	\caption{kNN-based gesture detection in EMG control applications, sorted by year, in studies compared with other classifiers and their accuracies, n/s means not specified}
	\begin{tabular}{@{}l|cc|cc|c|c|p{5.3cm}|c@{}}
		\toprule
		Ref.     & \multicolumn{2}{c|}{kNN Configuration} & \multicolumn{2}{c|}{Subjects} & Actions & Accuracy [\%] & Compared classifiers & Real-time \\
		& k    & Distance  & \#   & Health              &         &               & (accuracy [\%], w.r.t. kNN)          & control  \\
		\midrule
		\cite{Dening}            & 1    & -         & 1    & able-b. & 6     & 72            & -                                                    & no \\
		\cite{4584252}           & 5    & n/s       & 1    & n/s  & 8     & 95.5          & SVM (similar), EHW (91-95.1), DT (91)                & no \\
		\cite{Kim2008}           & 5    & n/s       & 30   & able-b. & 4     & 94            & Bayes (92)                                           & yes \\
		\cite{5396091}           & 31   & -         & n/s  & able-b. & 6     & 59-100        & NN (similar)                                         & no \\
		\cite{5767304}           & 7-17 & Euclidean & n/s  & n/s  & n/s   & 83-100 (k=15) & NN (50-83)                                           & no \\
		\cite{Antfolk2011ACB}    & 16   & Euclidean & 10   & able-b. & 13    & 81            & LDA, NN (similar)                                    & yes \\
		\cite{5704586}           & 8    & Euclidean & 5+5  & both & 7     & 89/79 (amp.)  & -                                                    & yes \\
		\cite{KIM2011740}        & 1-10 & Euclidean & 30   & able-b. & 5     & 85 (k=5)      & QDA (82), LDA (81, signif. diff.)                            & no \\
		\cite{KHUSHABA201210731} & 8    & Euclidean & 8    & able-b. & 10    & 90            & SVM (similar)                                        & yes \\
		\cite{6487520}           & n/s  & Euclidean & 5    & able-b. & 15    & 88-100        & LDA: 51-100, QDA: 82-100                             & no \\
		\cite{Jeong2013}         & 1    & Euclidean & 28   & able-b. & 5     & 95            & Gaussian mixture (similar)\newline Signif. diff.: LDA (92), QDA (94) & no \\
		\cite{Shin}              & 5    & n/s       & 20/8 & able-b. & 10/15 & 87/89         & NB, QDA, SVM, NN, ELM (similar)                      & no \\
		\cite{7437907}           & 1-7  & various   & 4    & able-b. & 9     & 80-86 (k=6)   & NN (93)                                              & no \\
		\cite{Geethanjali2015}   & 9    & Euclidean & 10   & able-b. & 6     & 88            & LR (91), LMT (91), DT (84), NN (90)                  & no \\
		\cite{Purushothaman2016} & n/s  & n/s       & 10   & able-b. & 6     & 78-89         & NN (70-88)                                           & yes \\
		\cite{7591042}           & 1    & n/s       & 6    & amput. & 11    & 74            & LDA (similar)                                        & no \\
		\cite{Robinson}          & 7    & Manhattan & 11   & able-b. & 17    & 85            & NN (35), RF (90), SVM (73)                           & no \\
		\cite{8122765}           & n/s  & n/s       & 5    & able-b. & 8     & 60-100        & SVM (56-100), LDA (40-98)                            & no \\
		\bottomrule
	\end{tabular}
	\label{tab:related}
\end{table*}

\subsection{Training Dataset Reduction Algorithms} \label{state_red}
An important drawback of instance-based learning schemes is the necessity of comparing new arriving instances whose labels are meant to be predicted to all already stored ones (``training'' data). In order to do so, all instances have to be iterated which leads to potentially -- depending on the amount of data -- high computational effort in the prediction phase.

Typically, two main approaches to improve the performance of nearest neighbour classifiers are pointed out \cite{Bajr}. The first is the utilization of efficient, optimized data structures (``ball-tree data structures, hashing'' \cite{Kusner}, ``kd-tree'' \cite{Bajr}). The second approach (thinning) can be seen both in a horizontal (feature-space) as well as in a vertical dimension (sample-space). Aside from that, there are techniques using an approximation of the kNN classification rule, for example Large Margin Nearest Neighbour \cite{Kusner}.

In terms of horizontal thinning, the concept of feature selection has been applied in the context of pattern-recognition-based prosthetic control for large feature set dimensions, for instance biologically inspired methods such as genetic algorithms and particle swarm optimization \cite[p.~251]{Geethanjali}. Horizontal thinning can be generalized (to horizontal data reduction) when feature projection, positioning \cite{Kusner} and discretization \cite{discr} techniques are also considered. These schemes come along with dimensionality reduction algorithms. Examples are principle component analysis (PCA) \cite{Guler} and adaptions thereof \cite{Nagata} as well as variants of linear discriminant analysis (LDA) \cite{7748960}.

However, the examinations made in this work cope with vertical data reduction techniques. The general idea is to reduce the computational effort of prediction steps in instance-based learning by decreasing the number of instances within the training set. This process is usually referred to as instance reduction or prototype reduction \cite{PS,PG}. In principle, \textit{prototype} stands synonymously for data instance or sample. Nevertheless, it already indicates that it refers to specific instances which represent a larger amount of instances to a certain extent.

Prototype reduction methods can be divided into prototype selection (vertical thinning) on the one hand \cite{PS} and prototype generation on the other hand \cite{PG}. While the former selects a subset of instances from the existing ones, the latter creates new instances based on the existing ones to represent the whole dataset.

\section{Requirements and Concept}

The experimental studies and the developments which they are based on are driven by the requirements of \cref{sec:req} and composed of different parts:

\begin{itemize}
	\item First, a pilot dataset of several (full-intensity) gesture exertions is captured from the authors in order to conduct an offline cross-validation analysis of kNN parameters on gesture classification without real-time application.
	\item Second, the obtained kNN parameter configuration is applied in a real-time scenario, in which new (full-intensity) gesture data is gathered. Additionally, an approach of proportionality scaling is introduced here. With that, real-time gesture detection performance is measured in an online target achievement test with just one subject. The success rate in this pilot study is utilized to analyze the influence of proportionality scaling parameters while testing three levels of exertion intensity (but just training on full intensity).
	\item Third, the two determined parameter configurations (kNN and proportionality scaling) are tested in a real-time user study with 12 subjects (and in an extended user study with 4 subjects). Again, target achievement tests are conducted, including three levels of gesture exertion intensity for detection (but just full-intensity for training). No further parameter optimization takes place in this step. Moreover, a data reduction technique is introduced and applied to each subject's data. The success-rate performance of the non-reduced and the reduced data approach are compared.
\end{itemize}

\subsection{Requirements} \label{sec:req}
The requirements listed in \cref{tab:req1} are to be met by the learning strategies developed in this work. While R1--R4 represent general prerequisites, R5 constitutes an additional constraint for embedded system implementations.

R1 and R2 are considered as the minimum standard for myocontrol, while R3 targets the transfer from offline to online scenarios. R4 is motivated by the benefit of home recalibration for prosthetic users \cite{Nowak2023,Kuiken2016}.

For R5, specific sub-requirements have been defined. The general motivation of providing an algorithm suited for embedded systems and still delivering high performance, is the tendency of developing wearable systems that are usable stand-alone without the necessity of connecting standard computers.

\begin{table}[h]
	\centering
	\caption{Requirements for the learning method, providing embedded applicability}
	\begin{tabular}{@{}lp{\columnwidth-1cm}@{}}
		\toprule
		No. & Description                                                                                                                                                   \\ \midrule
		R1  & High accuracy in classification of actions                                                                                                                    \\
		R2  & Providing proportional control                                                                                                                                \\
		R3  & Besides the static accuracy, the users should be satisfied by the method when using it in real scenarios (success rate, robustness, stability, reaction time) \\
		R4  & Incrementality, i. e. providing the possibility for extending the reference data by new instances                                                             \\
		R5 & Applicability to control on embedded systems (non-functional requirements such as portability, battery consumption):                \\
		R5.1 & Coping with real-time demands in prediction                                                       \\
		R5.2 & Low and deterministic memory requirements                                                         \\
		R5.3 & Low and deterministic training (reduction) time \\
		R5.4 & Simplicity of the algorithm for transparency and dependability \\ \bottomrule
	\end{tabular}
	\label{tab:req1}
\end{table}

\subsection{Sensor Hardware}
A product widely used in research -- also in this work -- is the \textit{Myo} wireless armband, produced from 2013--2018 by the Canadian company Thalmic Labs Inc. which is characterized by (cf. \cite{Visconti}):

\begin{itemize}
	\item eight EMG electrodes with ST 78589 operational amplifier per electrode, with
	\item maximum sampling frequency of 200\,Hz,
	\item 9-axes IMU with 3-axis gyroscope, 3-axis accelerometer, 3-axis magnetometer (InvenSense MPU-9150),
	\item Freescale Kinetis ARM Cortex M4 120\,Mhz MK22FN1M microcontroller,
	\item communication via BLE with Nordic nRF51822 to HM-11 BLE dongle,
	\item vibration motor and LEDs for signalling, and
	\item two lithium batteries (3.7\,V, 260\,mAh), USB-charged.
\end{itemize}
No IMU information is utilized in the context of this work.

\subsection{Signal Processing and Nearest-Neighbour-Based Methods}
In general, a kNN-based classification approach will be given priority over kNN regression, as the latter exposed a high extent of instability in preliminary experiments.

To keep the computational demands as low as possible for an embedded prosthetic control system, we aimed at utilizing time-domain features due to their lower complexity. Specifically, the linear envelope of the signal will be used as input feature. It can be shown that the majority of the discriminatory effect in the widely used Hudgins EMG feature set stems from the mean absolute value \cite{Scheme2013}. In this sense, the reduced demands for obtaining the amplitude data by calculating the absolute value are combined with a window length of 1 to not induce further calculations. As in similar publications \cite{Nowak2023}, this is followed by low-pass filtering with a cut-off frequency of 1\,Hz by a second-order Butterworth filter, as ``at least 90\% of the power in the power spectral density estimates were found to be below 1 Hz'' \cite{Paek2013} in the rectified signal.

The gestures chosen to evaluate the performance are selected among rest state (\textit{rs}), power grasp (\textit{pw}), pointing (\textit{pn}), wrist flexion (\textit{fl}), wrist extension (\textit{ex}), wrist pronation (\textit{pr}), and wrist supination (\textit{su}).

In order to evaluate the static performance of the algorithm and specifically to match requirement R1, the cross-validation accuracy on a variety of \textit{Myo} armband EMG datasets captured by the authors will be examined. For this purpose, these datasets comprise four repetitions per gesture. One repetition contains the filtered eight-channel-EMG data when exerting one specific gesture for two seconds at the maximum sample rate of the Myo armband, namely 200 Hz. Multiple repetitions are necessary as the gathered samples within one repetition cannot be considered as independent and identically distributed. The stochastic dependence is abolished across multiple repetitions since there are interruptions in time, specifically because of training other gestures in between, before capturing the next repetition. With that, a block-wise (group-wise) cross-validation is possible so that samples within one block (group/repetition) are not validated against samples within the same block. In this way, a preventive measure against overfitting is established. In particular, a leave-one-group-out cross-validation will be applied, i. e. selecting one block as testing set while the others form the training set, for all possible combinations. In the end, the arithmetic mean of the single accuracies (i.\,e. correct classifications relative to all classifications) is used to characterize the accuracy of the whole dataset. 

The parameters which can be altered in kNN for static cross-validation are the number of neighbours to be considered ($k$), the distance metric for comparing sample differences, and the weighting of selected samples' data values. It is known that kNN's ``performance is critically dependent on the selection of $k$ and a suitable distance measure'' \cite[p.~3]{Kuzborskij} so that these will be subject to a specific analysis.

A problem with kNN classification is the fundamental characteristic that no intermediate states can be predicted. Therefore, kNN classification will be extended by proportionality scaling schemes to provide proportional control.

The following concepts are applicable for kNN classification based on majority-voting regarding the occurrences of individual class labels.

It is assumed that the intensity of an exerted action/gesture is proportional to the amplitude of the EMG signal's linear envelope \cite{s18082553} (averaged for all channels). By analyzing this magnitude, a proportionality scaling can be applied as soon as a gesture has been detected \cite{Hudgins,Scheme}. To obtain a correct gesture classification from samples of a specific gesture at lower intensity levels, the samples are normalized (assuming that the signal shape is similar when comparing signals of the same gesture at different intensity levels).

Furthermore, a threshold for the rest action, i. e. the state where no gesture is exerted, will be introduced (rest magnitude thresholding). The motivation for this is that in the case samples are closer to the rest state than to the specific, real gesture it would be classified as rest, until the transition point in the signal amplitude is reached. The rest state usually resides at around zero signal amplitude, unless distinct postures are considered where this might differ due to the limb position effect.

The concept of the rest magnitude thresholding consists of measuring the average rest activity and basing a threshold of signal amplitude on this value, possibly altered by further parameters. 
If this threshold amplitude is exceeded when executing the prediction on a new sample, the classification takes place and a class label of the available ones except rest is assigned. Otherwise, the new sample is considered as rest.

Requirement R3 will be evaluated by means of target achievement tests. First, the presented concepts will be evaluated in pilot experiments without being statistically representative. The tendencies obtained are used as a baseline for a user study with multiple subjects following afterwards. For both versions of experiments, several gestures will be tested on different signal intensity levels (for instance exerting just one third of a full wrist flexion), after training solely took place on full intensity level. The single gesture has to be reached and held for a certain period of time without deviating too much within some error range in order to consider the task as successful. For this purpose, the subject will see a visual stimulus in the form of a hand model to be followed, as well as another hand model visualizing the current gesture prediction (as in \cref{fig:setup}). The results will be compared to those obtained from state-of-the-art ridge regression methods.

The final user study will be conducted in a double-blind manner in order to provide comparability of the algorithms. Therefore, the selection of gestures and intensity levels during one experiment will be randomized. For the purpose of not favouring a single method over another (if there should be a time-dependency of success), the occurrences of methods and levels will be equally distributed across the available time slots.

Requirement R4 is met inherently by the standard kNN approach since in every prediction step each instance of the training set is compared with the sample to be predicted for obtaining the particular distance. This means, if there are new samples to be stored in the training set they are directly taken into account during prediction, thus leading to incrementality. 

\subsection{Assessment of Embedded Applicability} \label{conc_emb}
The applicability on embedded systems is specified in requirement R5 with its sub-components R5.1--R5.4.

To meet requirement R5.1 it is necessary to reduce the kNN computation effort in the prediction phase. As an instance-based learning technique, kNN suffers from the computational disadvantage mentioned in section \ref{state_red}. Specifically, for each new instance the prediction step comprises the calculation of the distance from the new instance to the $n$ stored ones (runtime complexity of $\mathcal{O}(n)$) and the sorting of these distances to obtain an ascending order of nearest neighbours. Depending on the sorting algorithm, the overall time complexity can reach $\mathcal{O}(n\log n)$ (being the proven lowest possible bound for comparison-based data sorting). Nevertheless, if the number of neighbours to consider is set to $k=1$, no sorting is necessary anymore. Thus, with a minimum search being sufficient instead, the complexity reduces to $\mathcal{O}(n)$.

Possibilities to reduce the computational effort in terms of the number of training samples $n$ to specifically achieve requirement R5.1 have been introduced in section \ref{state_red}. In particular, the concept of prototype reduction is chosen. 
As presented in \cite{szibBIOSIGNALS}, an assessment of the variety of these algorithms has to be made in order to lower the number of instances in the training set for kNN. To meet requirement R5.2, it is necessary that the particular algorithm to be chosen provides a possibility to specify the number of prototypes in the final set or accordingly the reduction rate beforehand. When it comes to prototype selection algorithms reviewed in \cite{PS}, only Random Mutation Hill Climbing (RMHC, \cite{RMHC}) inherently possesses this characteristic as it is the only method with fixed reduction. Nevertheless, RMHC is a wrapper method which means that in each step the decision if to select a prototype or not, a complete kNN evaluation for all instances has to take place. For this reason, long computational times during the reduction process have to be expected. In \cite[p.~425--427]{PS} it is shown that this assumption holds in real use cases for both small and medium-sized datasets. Exemplary tests on EMG datasets confirmed that behaviour so that RMHC was excluded from consideration.

Besides the fixed reduction prototype selection algorithms, there might be also mixed reduction methods which provide the property of determinism with respect to the number of samples contained in the final training set. However, the algorithms of that category described in \cite{PS} are all wrapper methods, too. Due to the respective high execution times as mentioned before, these methods are not considered within the scope of this work.

In terms of prototype generation there is a variety of fixed reduction algorithms. They can be summarized in the following way:
\begin{itemize}
	\item \textit{Positioning adjustment, condensation approaches:} Learning Vector Quantization (LVQ)-based methods \cite{LVQ,DSM},
	\item \textit{Positioning adjustment, hybrid approach:} Particle Swarm Optimization (PSO \cite{PSO}),
	\item \textit{Centroid-based condensation approaches:} Bootstrap Technique for Nearest Neighbor (BTS3 \cite{BTS}), Adaptive Condensing Algorithm Based on Mixtures of Gaussians (\hbox{MGauss} \cite{MGauss}), and
	\item \textit{Space-splitting:} Chen Algorithm \cite{Chen}.
\end{itemize}

While the Chen and BTS3 algorithms are not incremental in the sense of requirement R3, in PSO, MGauss and the LVQ-based methods each step in the reduction process only depends on the former step (where a certain model or prototype configuration is obtained) but not on the instances themselves from the initialization of the whole process. Usually this leads to the characteristic that the reduction process does not depend on the order of decisions, i. e. the order of instances being considered.

The LVQ-based algorithm LVQTC (LVQ with Training Counter, \cite{LVQTC}) turned out to not provide determinism with regard to the final set's size and was therefore not taken into account for further evaluation.

Again, there are mixed algorithms which may also provide the final set size determinism like the fixed ones are supposed to. Some of them are in turn wrapper methods (Evolutionary Nearest Prototype Classifier ENPC \cite{ENPC}, Adaptive Michigan Particle Swarm Optimization AMPSO \cite{AMPSO}) and hence not considered with respect to the previously mentioned reason.

Filter and semi-wrapper methods which might be applicable in principle, are:
\begin{inparaitem}
	\item Gradient Descent and Deterministic Annealing (MSE \cite{MSE}),
	\item Hybrid LVQ3 (HYB \cite{HYB}),
	\item Integrated Concept Prototype Learner (ICPL2 \cite{ICPL}),
	\item LVQ with Pruning (LVQPRU \cite{LVQPRU}), and
	\item Prototype Selection Clonal Selection Algorithm (PSCSA \cite{PSCSA}, artificial immune system model).
\end{inparaitem}

The reason why the first three of these algorithms were not chosen for the evaluation in the end are their non-determinism with respect to the final set size. The remaining algorithms are to be compared. Since they vary with regard to the time needed for the reduction process, this is examined in experiments that are based on datasets of captured rectified and filtered EMG signals (linear envelope). Besides the amount of reduction (requirement R5.2) and the runtime behaviour (R5.3), the achieved accuracies when using the reduced sets in block-wise cross-validation will be assessed. The choice for specific algorithms will be further guided by requirement R5.4, i.~e. taking into account the implementation complexity of the algorithms.

\section{Methods}
In terms of the methodical realization of the algorithm, several characteristics will be pointed out in the following, regarding both the kNN scheme and data reduction techniques.

\subsection{Methodological Considerations for the kNN Approach} \label{metho_consid}
\begin{figure}[h]
	\centering
	\includegraphics[width=.5\linewidth]{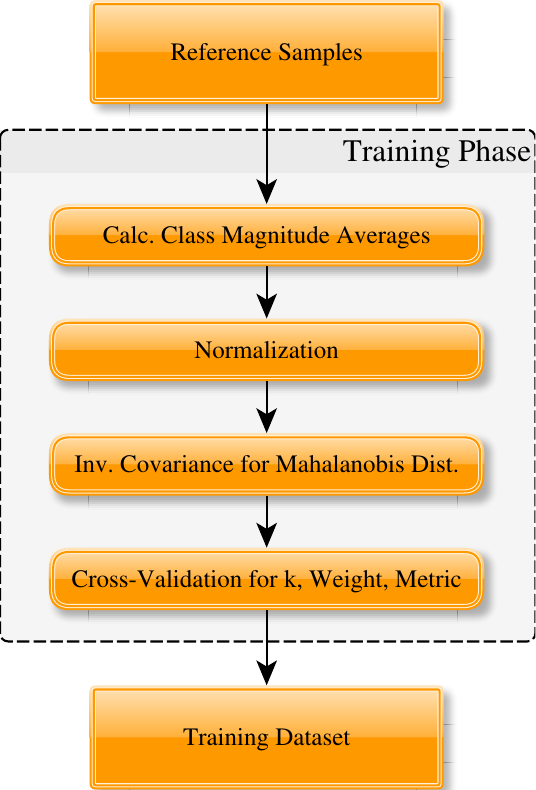}
	\caption{General structure of training phase.}
	\label{fig:struct_train}
\end{figure}

The kNN training process is structured as follows (see also \cref{fig:struct_train}):
\begin{inparaenum}
	\item capturing training data,
	\item calculating class magnitude averages for proportionality scaling, and rest magnitude threshold (if enabled),
	\item executing normalization of this data (if enabled),
	\item calculating the inverse covariance matrix of the data if the Mahalanobis distance is activated,
	\item executing block-wise cross-validation for obtaining the optimal $k$, weighting and metric in terms of accuracy.
\end{inparaenum}

The prediction process comprises:
\begin{inparaenum}
	\item applying rest magnitude thresholding (if enabled),
	\item executing normalization of sample (if enabled),
	\item calculating $k$ nearest neighbours of the sample whose label is supposed to be predicted and their distances,
	\item applying distance weighting on the k selected neighbouring samples (if enabled),
	\item executing direct averaging of neighbour samples in kNN regression, or 
	\item calculating the proportionality scaling factor by analysis of the signal amplitude, before
	\item executing kNN classification by majority voting on the (potentially weighted) samples, i. e. the class with the highest weight sum will be selected for predicting the full gesture, which
	\item will be scaled by applying magnitude proportionality scaling (if enabled).
\end{inparaenum}

\begin{figure}[h]
	\centering
	\includegraphics[width=0.9\linewidth]{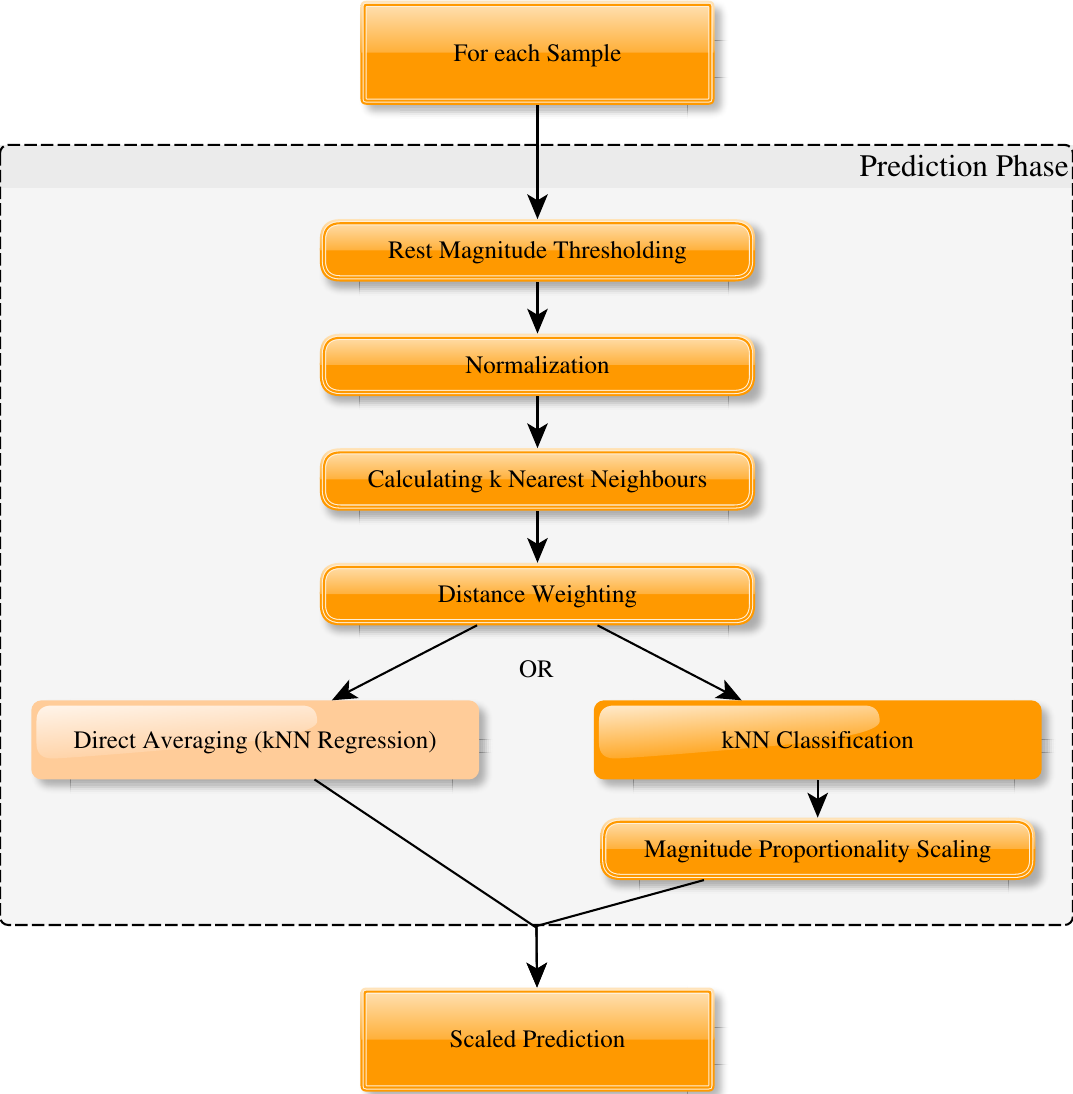}
	\caption{General structure of prediction phase (regression path not chosen).}
	\label{fig:struct_pred}
\end{figure}

These steps are also pointed out in \cref{fig:struct_pred}. Additionally, different windowing schemes could be applied, cf. \cite{szib}.

\subsubsection{Nearest Neighbour Parameter Configurations}
The number of next neighbouring samples ($k$) to consider in a prediction step is varied from $k=1$ (just nearest neighbour) to higher numbers. For each case the particular block-wise cross-validation accuracy is computed, if enabled. Due to the characteristics of this validation scheme, the maximum $k$ cannot exceed the total number of samples minus the size of one repetition block. 

The examined distance measures are the Minkowski-norm based metrics Manhattan ($p=1$), Euclidean ($p=2$) and Chebyshev ($p\rightarrow\infty$), as well the Mahalanobis distance. 
For distance weighting, inversely distance-dependent factors are calculated for each sample and summed up for each class within the selected set of neighbouring samples. Hereby, a weighted majority vote is obtained as classification.

kNN inherently involves $k$ minimum searches to obtain the $k$ nearest neighbours. This is implemented by means of sorting the distances in a descending order and picking the $k$ first entries. For this purpose an appropriate sorting function is called. However, if $k$ is selected to be 1, the sorting procedure can be replaced by a search for the minimum distance within the set.

\subsubsection{Proportionality Scaling and Rest Thresholding}\label{propsca}
As presented in \cite{szibICNR}, the approach for rest magnitude thresholding is realized in a way that the magnitudes of the rest samples gathered during training are averaged and taken as a baseline for rest activity ($t_0$). The threshold of signal amplitude which has to be exceeded for not classifying a gesture as rest anymore is based on the obtained average: $t=g\cdot t_0$ (amplified by gain $g$). Although this enables to reduce unintended actuations, a higher thresholding level $t$ results in a lower proportionality resolution by presuming higher activation forces. Another possibility for calculating a threshold could be to consider other functions applied on the rest activity instead of the mean, such as the median or the maximum (although the latter would require a specific consideration of outliers).

For the non-rest gestures, an approach of proportionality scaling is utilized \cite{szibICNR}. This is implemented in a linear manner, i.~e. intermediate gestures are assumed to be linearly scaled between rest activity and the average training magnitude of the particular gesture set as function maximum. Again, instead of the mean of the individual gestures' magnitudes, other functions might be used.

As mentioned, there is the need for a trade-off between the level of proportionality resolution and suppressing unintended activations. Therefore, a divisor $v$ to scale the proportionality function offset $m_0=\nicefrac{t}{v}$ is moreover introduced. This does not scale the rest threshold $t$ itself.

\begin{figure}
	\centering
	\begin{tikzpicture}[scale=0.6]
	\begin{axis}[xmin=0, xmax=5,
	ymin=0, ymax=1,
	label style={font=\Large},
	tick label style={font=\large},
	xlabel={Magnitude $m$},
	ylabel={Scale $s$},
	xtick={0,0.5,1,1.3,5}, xticklabels={$0$,$t_0$,$m_0$,$t$,$m_{max}$}, 
yticklabels={,$0$,,,,,$1$}
	]
	\addplot[blue, mark=none, domain=0:5] {1/5*x}; \addplot[red, mark=none, domain=0:6] {1/5*x - 1/5*1.3}; \addplot[green, mark=none, domain=0:5] {1/(5-1)*x-1/(5-1)*1}; \end{axis}
	\end{tikzpicture}
	\caption{Example of proportionality scaling: $t=g\cdot t_0$ marks rest threshold, $m_0=\nicefrac{t}{d} $ adjusted offset (for proportionality function, not for thresholding itself), $m_{max}$ gesture magnitude maximum. Blue curve is theoretical ideal behaviour, red exceeds $m_{max}$, green adjusts function slope (implemented).}
	\label{fig:propmag}
\end{figure}
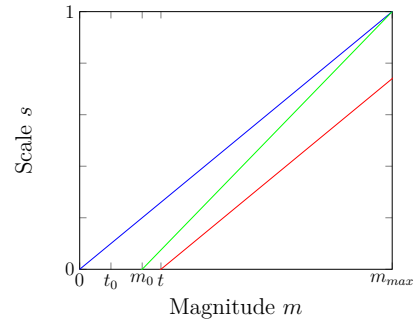

These relations between measured magnitude $m$ and applied scaling factor $s$ are depicted in \cref{fig:propmag}: The blue function describes the theoretical linear proportionality scale, i.~e. the scaling of the predicted gesture starts with $0$ at 0 magnitude, assuming there is no baseline rest activity at all that could lead to wrong classifications. With introducing the rest threshold $t$ as an offset for the scaling function, too, the average activity of the full gesture $m_{max}$ would be required to be exceeded in prediction to reach the maximum scaling. This could be avoided by also adapting the scaling function maximum for $s=1$. Since this would lead to a reduced magnitude resolution, the maximum is pertained and the slope of the function is modified (green curve) as follows:

\begin{equation*}
	s(m) = \frac{1}{m_{max}-m_0}\cdot (m-m_0).
\end{equation*}

An alternative approach could be to use piecewise linear functions or modelling non-linear relationships.

\subsection{Dataset Reduction Algorithms}
For the evaluation of prototype reduction algorithms, the open-source (GPLv3) software tool KEEL (\textit{Knowledge Extraction based on Evolutionary Learning} \cite[p.~1239]{Triguero}) was chosen and extended, in which the particular prototype reduction algorithms from \cite{PS} and \cite{PG} have been implemented. 

Special focus of this work will be put on reduction methods based on learning vector quantization (LVQ). They are composed of the following basic steps:
\begin{enumerate}
	\item initialization by choosing random samples, or
	\item selecting the classes' centres of masses as initial prototypes and potentially adding more samples randomly (as long as the number of prototypes to be chosen is not exceeded, distribute the selection equally over all classes while choosing randomly within each class),
	\item repeating the correction process for a specified number of iterations: for each sample decide if it has a rewarding and/or a penalizing effect on particular prototypes and employ this effect.
\end{enumerate}

The idea of the standard LVQ-based approaches, originally proposed in the context of self-organizing maps in \cite{LVQ} (with prototypes being called codebook vectors) in three different variants (LVQ1, LVQ2, LVQ3), is to represent the probability distribution behind the dataset. An exception is the Decision Surface Mapping (DSM) strategy which instead aims at appropriately modelling the class borders (decision boundaries/surfaces) \cite[p.~335]{Trends}.

As \cite{glvq} points out, standard ``LVQ corresponds to what is usually known as SCL [Simple Competitive Learning] in the neural network literature''. \cite{lvqnn} defines it as ``a single-layer neural network in which the outer layer is made of distance units, referred to as prototypes''.

Two specific variants implemented in the scope of this work are the aforementioned methods DSM and LVQ3. Their correction steps are realized as in \cref{alg:lvq}.

\begin{algorithm}[h]
	\SetAlgoLined
	\KwIn{original samples}
	\KwResult{reduced prototype set}
	initialization: prototypes as centres of masses; if there are more: randomly chosen\;
	\ForEach{iteration i} {
		\ForEach{sample x in the original training set} {
			\If{condA}{
				procA()\;
			}
			\If{condB}{
				procB()\;
			}
		}
	}
	\caption{LVQ-based reduction methods, specific substitutions given in \cref{tab:lvq}}
	\label{alg:lvq}
\end{algorithm}

For DSM and LVQ3 the specific conditions as well as the prototype adjustment actions are defined in \cref{tab:lvq}, which refer to the rewarding and penalization terms in \cref{alg:rew,alg:pen} (the learning rate parameter is set to a fixed number of 0.01).

\begin{table}[h]
	\caption{Specific properties of LVQ3 and DSM}
	\resizebox{\columnwidth}{!}{\begin{tabular}{@{}cllll@{}}
			\toprule
			& \multicolumn{1}{c}{\textit{condA}}                                                                               & \multicolumn{1}{c}{\textit{procA}}                                                                                                                                                                                                       & \multicolumn{1}{c}{\textit{condB}}                                                                                           & \multicolumn{1}{c}{\textit{procB}}                                                \\ \midrule
			\textit{LVQ3} & \begin{tabular}[c]{@{}l@{}}$x$ is inside window\\ of the 2 nearest\\ prototypes $(p_0, p_1)$\end{tabular} & \begin{tabular}[c]{@{}l@{}}If $x$ has same label\\ as $p_0$ but not as $p_1$:\\ $reward(p_0, x)$ and\\ $penalize(p_1, x)$.\\ If $x$ has same label\\ as $p_1$ but not as $p_0$:\\ $reward(p_1, x)$ and\\ $penalize(p_0, x)$.\end{tabular} & \begin{tabular}[c]{@{}l@{}}$(p_0, p_1)$ have\\ same label as $x$\end{tabular}                             & \begin{tabular}[c]{@{}c@{}}$reward(p_0, x)$\\ and\\ $reward(p_1, x)$\end{tabular} \\
			\textit{DSM}  & \begin{tabular}[c]{@{}l@{}}label of $x$ and nearest\\ prototype $p_0$ differ\end{tabular}                 & $penalize(p_0, x)$                                                                                                                                                                                                              & \begin{tabular}[c]{@{}l@{}}\textit{condA} and nearest\\ prototype with\\ same class label\\ $p_s$ as x exists\end{tabular} & $reward(p_s, x)$                                                            \\ \bottomrule
		\end{tabular}}
	\label{tab:lvq}
\end{table}

\begin{algorithm}
	\SetAlgoLined
	\KwIn{prototype p, sample x}
	\KwOut{adjusted prototype $p_s$}
	$ p_s = p + \alpha(x-p) $
	\caption{$reward()$ for LVQ3 and DSM}
	\label{alg:rew}
\end{algorithm}
\begin{algorithm}
	\SetAlgoLined
	\KwIn{prototype p, sample x}
	\KwOut{adjusted prototype $p_s$}
	$ p_s = p - \alpha(x-p) $
	\caption{$penalize()$ for LVQ3 and DSM}
	\label{alg:pen}
\end{algorithm}

Based on these algorithmic descriptions, a specific runtime complexity analysis of DSM is conducted in \cref{runtimecompl}.

\section{Evaluation and Results}
This section presents the experimental outcomes to evaluate the developed strategies. These results were obtained from conducting the following experiments:
\begin{enumerate}
	\item offline tests with datasets from one subject,
	\item online tests with real-time data from one subject (pilot experiments),
	\item online tests with 12 subjects (basic user study), and
	\item online tests with 4 subjects (extended user study).
\end{enumerate}

While the offline tests were primarily evaluated by cross-validation accuracy, the main criterion for the online experiments was the success rate (see also requirements \cref{tab:req1}).

Some experiments include specific gestures in one case but do not include these in another. This applies to the pointing gesture to consider and analyze the assumption that it is not as well separable from rest, power grasp, wrist flexion and wrist extension, as these four are from each other. Furthermore, it applies to wrist pronation and supination (again, with and without the pointing gesture) which were chosen to extend the system by a rotational dimension in order to observe the development of performance with an increasing number of degrees of freedom.

\subsection{Offline Cross-Validation Accuracy} \label{res_meas}
The offline experiments described in this section are based on several series of EMG data captured from the authors. 
They provide a rational measure of the applicability by means of cross-validation accuracy. In the datasets, one training repetition consisted of 400 samples per gesture (two seconds capturing with 200 Hz sample rate). In each set, each gesture was recorded in several repetitions. For each configuration of considered gestures, several sets of data were recorded, see \cref{tab:offline} for the resulting number of samples.

\begin{table}[h]
	\centering
	\caption{Datasets used for the offline tests with overall number of sample vectors (2\,s capturing at 200\,Hz)}
	\begin{tabular}{lccc}
		\toprule
		Classes & Sets & Repetitions & Samples  \\ \midrule
		rs, pw, fl, ex              & 3 & 4 & 19200  \\
		rs, pw, fl, ex, pn         & 3 & 4 & 24000     \\
		rs, pw, fl, ex, pr, su     & 3 & 4 & 28800  \\
		rs, pw, fl, ex, pr, su, pn  & 3 & 4 & 33600  \\
		\bottomrule
	\end{tabular}
	\label{tab:offline}
\end{table}

The main parameter of kNN, the number of neighbours to consider ($k$), is varied throughout all cross-validation accuracy experiments. In order to guarantee comparability of the results concerning specific numbers of $k$ across datasets of different sizes, $k$ is not employed as an absolute number of samples. Instead, $k$ is compared in the sense of a relative value $k_{rel}$, i.~e. as the proportion of $k$ relative to the maximum number of samples in the set ($n$): $ k_{rel} = \nicefrac{k}{n}$.

Since the cross-validation is applied block-wise, the maximum $k$ cannot exceed the total number of samples in the set minus the number of samples in a block.

\subsubsection{Influence of Distance Weighting}
For the evaluation of cross-validation accuracy when changing the distance weighting factor, the different datasets showed the same qualitative behaviour. The distance from the current to the particular other samples is denoted by $d$.

\begin{figure}[h]
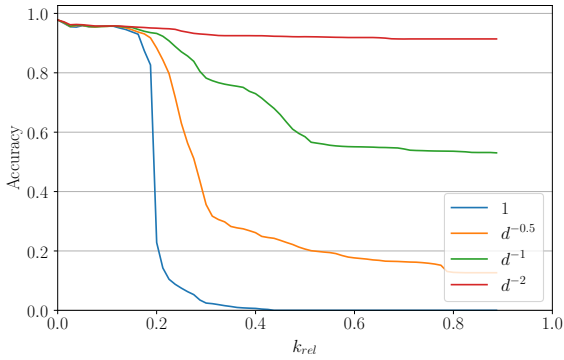
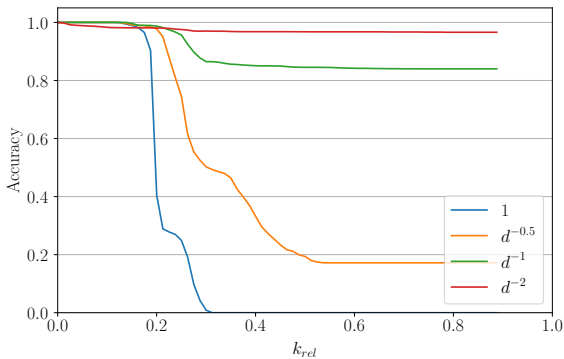

	\centering
	\begin{subfigure}{\linewidth}
		\begin{adjustbox}{max width=\linewidth}
			\begingroup \makeatletter 
\makeatother \endgroup  		\end{adjustbox}
		\caption{Euclidean distance}\label{fig:cv:dw2}
	\end{subfigure}\\
\begin{subfigure}{\linewidth}
		\begin{adjustbox}{max width=\columnwidth}
			\begingroup \makeatletter %
\makeatother \endgroup  		\end{adjustbox}
		\caption{Chebyshev distance}\label{fig:cv:dw4}
	\end{subfigure}
	\caption{Distance weighting influence on cross-validation accuracy for sets of five gestures (rs, pw, pn, fl, ex), $k$ relative to number of samples in the set}\label{fig:cv:dw_c}\end{figure}

Independent of the weighting factor used, it could be observed that high numbers of \textit{k} usually decreased the cross-validation accuracy. Considering a dataset of four gestures (rs, pw, fl, ex), the accuracy stays at about 99\% until $k_{rel}$ is at around 15\% in \cref{fig:cv:dw1} (using Euclidean distance) for all weighting factors. This threshold value of $k_{rel}$ is even higher in \cref{fig:cv:dw3}, namely at about 30\% (using Chebyshev distance). Furthermore, in this case the threshold only applies for weightings of $1$ or $\nicefrac{1}{\sqrt{d}}$. For $\nicefrac{1}{d}$ and $\nicefrac{1}{d^2}$ the accuracy stays above 99\%. In all cases the highest accuracy can be noticed with a weighting of $\nicefrac{1}{d^2}$, followed by $\nicefrac{1}{d}$, $\nicefrac{1}{\sqrt{d}}$ and $1$. The effect of decreasing accuracy is the most apparent in the case no weighting is applied (decreasing until 0 at about 40--50\% relative $k$). All accuracies stabilize at some point.

When also including the pointing gesture into the comparison, the behaviour is principally similar. Starting at around 98\% accuracy in \cref{fig:cv:dw2} and even 100\% in \cref{fig:cv:dw4} respectively for all weightings at $k=1$, it drops to 0 for higher $k$s when using no weighting. Again, the decrease at weighting $1$ is the highest, followed by $\nicefrac{1}{\sqrt{d}}$, $\nicefrac{1}{d}$ and finally $\nicefrac{1}{d^2}$. The above mentioned threshold level of decrease lies at about $k_{rel}=20\%$.

It can be stated that as soon as a low number of $k$ is meant to be used ($k=1$ seems suitable in all cases), the weighting scheme does not matter. This means in this case that for the sake of computation resources even no weighting could be applied. Nevertheless, if higher numbers of k should be necessary, 
a higher exponent in the weighting factor's divisor should be introduced. $\nicefrac{1}{d^2}$ seems to be a good choice for that, without increasing the computation effort notably.

This observation is also confirmed in further tests: \Cref{fig:cv:dm3} shows this for the Chebyshev distance while additionally including pronation and supination (obtaining a threshold of about $k_{rel}=15\%$); and \cref{fig:cv:dm4} for the Manhattan distance with pronation and supination included without pointing (threshold value some percents higher). In the latter case, the even better performance of a distance weighting of $\nicefrac{1}{d^3}$ is additionally depicted, although the difference only appears after reaching a relative $k$ of 25\% and is neglectable due to its small value (0.25\%).

\subsubsection{Influence of Distance Metric}
The variation of the distance metric showed almost no effect in the case of the four gestures (rs, pw, fl, ex), see \cref{fig:cv:dm1} in the appendix (using a weighting factor of $\nicefrac{1}{d^2}$). An exception is the Mahalanobis distance which only provided about 96\% of accuracy at low numbers of $k$ while the other metrics achieved 100\%. Furthermore in the case of the Mahalanobis distance the accuracy dropped fast when increasing $k$ until it stabilized at around 69\% for $k_{rel} > 50\%$. The accuracy when using the other metrics stayed constant at about 100\% (Euclidean drops slightly to 99\%). 

\begin{figure}[h]
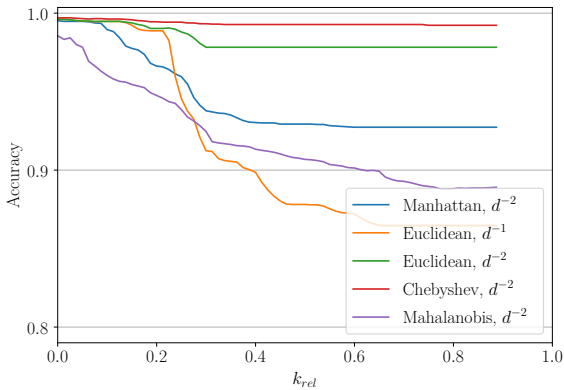

	\centering
	\begin{adjustbox}{max width=\columnwidth}
		\begingroup \makeatletter 
\makeatother \endgroup  \end{adjustbox}
	\caption{Distance metric influence on cross-validation accuracy, gestures (rs, pw, pn, fl, ex), $k$ relative to number of samples in the set}\label{fig:cv:dm2}\end{figure}

The observed behaviour in the case of data which included the pointing gesture exposed the following differences (\cref{fig:cv:dm2}): While the accuracy when applying the Mahalanobis distance showed the same tendency (starting from at about 98\% going down to 89\%), it also dropped for the other distance metric cases when increasing $k_{rel}$ over 10\%. This was mostly noticeable when looking at the Manhattan metric as the accuracy started at about 99\% for low numbers of $k$ and decreased until 93\%. For the other metrics, it went down from almost 100\% to 99\% (Chebyshev) and 98\% (Euclidean) respectively.

Including the wrist rotation gestures without pointing (weighting factor of $\nicefrac{1}{d^2}$, appendix \cref{fig:cv:dm4}) showed qualitatively the same behaviour as in the already described case where pointing was not included. This means, the Mahalanobis distance started at lower accuracy values than the others (98\% instead of 100\%) and dropped until it stabilized at 84\% (the Chebyshev norm dropped to 99.5\%, the Euclidean norm to 99.6\% and the Manhattan norm to 99.7\%).

When additional including the pointing gesture again, the effect was comparable, although pointing influenced the Minkowski-norm-based distances slightly more. For the Mahalanobis distance the accuracy dropped from 97\% until it reached a stabilization level of about 88\%. The Minkowski-based norms started at 100\% accuracy for low numbers of $k$ and decreased at a relative $k$ of about 15\% until they reached an accuracy of 96\% (Chebyshev), 98.4\% (Euclidean), and 99.3\% (Manhattan) respectively.

\Cref{fig:cv:dm3} shows also that the behaviour is the same when applying a distance weighting of $\nicefrac{1}{d}$ instead of $\nicefrac{1}{d^2}$ for the cases of Euclidean and Chebyshev norm, although the drop in accuracy is higher.

The evaluation of the distance metrics showed that differences are not evident in all cases. It can be summarized that the Mahalanobis distance is not recommended to be used for the present data. Due to the necessary calculation of the covariance matrix and its inverse it is also of disadvantage with respect to computational resources.

The Minkowski-distance-based metrics differ regarding the chosen order of norm, especially for high numbers of $k$. In some cases the accuracy gets better the higher the order of norm gets (Chebyshev ($p\rightarrow\infty$) is best, followed by Euclidean ($p=2$) and Manhattan ($p=1$) in the end). However, when pronation and supination are included, the effect is reversed (both with and without pointing). In fact, this reversed effect is lower than original effect. Nevertheless, the Euclidean norm seems to be a good trade-off to compensate both effects.

\subsubsection{Summary} \label{res_meas_sum}
All in all, it could be observed that for the cross-validation accuracy in the case of low numbers of $k$ (relative $k$ until about 5--10\%), neither the weighting factor nor the distance metric are of essential importance as long as a Minkowski-based distance norm is applied. However,  \cref{fig:cv:mw4} shows an exceptional case where there was a clear accuracy difference between the Euclidean (99\%) and the Chebyshev distance (91\%) even at low numbers of $k$. In the sense of computational demands, for the lower range of $k_{rel}$ a distance weighting of $1$ (i.~e. no further arithmetic operations) is recommended. If also considering $k_{rel}$ higher than 5--10\%, a weighting factor of $\nicefrac{1}{d^2}$ might be the best choice, together with the Euclidean norm. These recommendations hold for all tested sets of gestures. Further evaluations of other datasets which confirm this observation are depicted in the appendix in \cref{fig:cv:mw} with respect to the Euclidean and the Chebyshev distance as well as several weighting factors.

With that, the Euclidean distance and a weighting of $\nicefrac{1}{d^2}$ can be seen as a general recommendation in terms of accuracy for a broad range of $k$. However, with regard to requirement R5.1, also the Euclidean distance might not be preferred since its calculation (8 subtractions, 8 multiplications, 7 additions in each prediction step due to 8 EMG channels) is more computationally expensive than both the Manhattan distance calculation (8 subtractions, 8 absolute value calculations, 7 additions) and the one of the Chebyshev distance (8 subtractions, 8 absolute value calculations, 7 comparisons for maximum search) which do not require multiplication operations. The individual requirements must be balanced with respect to the specific use case.

\subsection{Real-time Pilot Experiments}\label{pil}
The pilot study experiments described in this section were only evaluated on one subject. Although the results obtained from these target achievement tests are therefore not representative, they may give insights on how different means and adaptions in the used algorithms can affect the achieved online success rates in gesture recognition with kNN (with $k$ set to 1 and 10 respectively, equally distributed, results averaged), especially when it comes to intermediate intensity levels of gestures. Following the results from \cref{res_meas} for a broad range of $k$, for kNN the Euclidean norm was chosen as distance metric with a weighting of $\nicefrac{1}{d^2}$.

For each pilot experiment, the user first trained the system by capturing data from the exertion of the full-intensity gestures. Each gesture had to be held for two seconds -- as in the offline training, resulting in 400 training samples per gesture and repetition. This time, three training repetitions were gathered, i.~e. 1200 8-value sample vectors per gesture.

In the prediction phase of each pilot experiment, all gestures (apart from rest) were not only tested on full-intensity exertion, but on three different intensity levels ($\nicefrac{1}{3}$, $\nicefrac{2}{3}$, full gesture). For this proportional control, proportionality scaling as described in \cref{propsca} was implemented. To consider a trial as success, the user had to mimic a virtual stimulus, while the real-time continuous prediction was shown in a hand model, and provide spatial matching within a time margin of several seconds. Each combination of gesture and exertion level was tested twice. In this way, the number of prediction samples was several magnitudes higher than the number of training samples, so that issues of overfitting can be further excluded.

As a measure of comparison, the accuracy of ridge regression with Random Fourier Features (RR-RFF) as state-of-the-art gesture recognition method was also evaluated in each test run.

\begin{table}[h]
	\centering
	\caption{Captured training data for online pilot experiments and online user studies with overall number of sample vectors resulting from numbers of participants, repetitions and  2\,s capturing at 200\,Hz}
	\begin{tabular}{llccc}
		\toprule
		& Classes & Part. & Rep. & Samples  \\ \midrule
		\multicolumn{5}{c}{\textit{Pilot Experiments}}       \\ \midrule
		& rs, pw, fl, ex          & 1  & 5 & 8000     \\
		& rs, pw, fl, ex, pn            & 1  & 5 & 10000      \\
		& rs, pw, fl, ex, pr, su        & 1  & 5 & 12000      \\
		& rs, pw, fl, ex, pr, su, pn    & 1  & 5 & 14000      \\
		\midrule
		\multicolumn{5}{c}{\textit{User Studies}}        \\
		\midrule
		\textit{Basic} & rs, pw, pn, fl, ex       & 12 & 3 & 72000  \\
		\textit{Ext.} & rs, pw, pn, fl, ex, pr, su & 4  & 3 & 33600  \\
		\bottomrule
	\end{tabular}
	\label{tab:online}
\end{table}

\subsubsection{Rest Class Thresholding: Rest Magnitude Threshold}
\label{pil_restmagthresh}
The rest magnitude threshold was introduced to cope with the problem of separating intermediate gestures from the rest class in the proposed proportional control. In order to evaluate the influence on the user success rate, multiple tests were conducted with the gesture sets (rs, pw, fl, ex) and (rs, pw, pn, fl, ex). 

\begin{figure}[h]
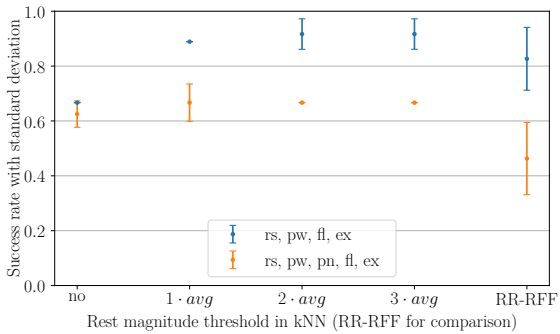

	\centering
	\begin{adjustbox}{max width=\columnwidth}
		\begingroup \makeatletter 
\makeatother \endgroup  	\end{adjustbox}
	\caption{Influence of rest magnitude thresholding, pilot experiments}
	\label{fig:cv:p1}
\end{figure}

\Cref{fig:cv:p1} shows that the standard approach without any rest thresholding yielded averaged success rates of 65\% on average for both types of dataset. While the success rates in the variant with pointing could not be considerably increased (only by 4\%), it was beneficial for the variant without pointing. 92\% success rate could be achieved for two times the mean rest signal magnitude ($g=2$) as well as three times mean rest magnitude ($g=3$) as threshold. Furthermore, it is noticeable that even without thresholding kNN performed better than RR-RFF when including pointing (63\% vs. 46\%). When not including point, kNN without thresholding performed worse than RR-RFF (67\% vs. 83\%). But with thresholding in the latter case, kNN's success rate could exceed RR-RFF's (92\% vs. 83\%).

\begin{figure}[h]
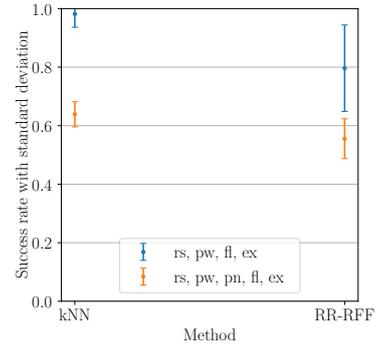

	\centering
	\begin{adjustbox}{max width=0.6\columnwidth}
		\begingroup \makeatletter %
\makeatother \endgroup  	\end{adjustbox}
	\caption{Influence of selected gestures, introducing rest magnitude threshold ($g=2.5$), pilot experiments}
	\label{fig:cv:p2}
\end{figure}

Since there was no difference recognizable between the success rates of $g=2$ and $g=3$, $g=2.5$ was chosen as a compromise for further experiments. The expected performance of this choice could be confirmed in \cref{fig:cv:p2}. 98\% success rate could be achieved for kNN without including pointing (in comparison to 80\% for RR-RFF) and 64\% when including pointing (56\% for RR-RFF).

It has to be noted that the results of RR-RFF yielded larger standard deviations than in all kNN cases. This could signify that kNN performs more stable and robust with less nondeterminism in the algorithm's behaviour.

\subsubsection{Proportionality Offset Scaling: Scale Offset Divisor}
As described, besides the rest magnitude threshold, a proportionality offset was introduced. This offset is divided by the scale offset divisor $v$ with the purpose of adjusting the proportionality scaling for intermediate gestures. The target achievement tests described in the following refer to a variety of runs with datasets comprising (rs, pw, fl, ex, pr, su) and (rs, pw, pn, fl, ex, pr, su), respectively. Besides kNN with different scale offset divisors, the performance of RR-RFF and standard ridge regression (RR) was also captured. For the evaluation, a rest magnitude threshold with $g=2.5$ was chosen, as motivated in \cref{pil_restmagthresh}.

\begin{figure}[h]
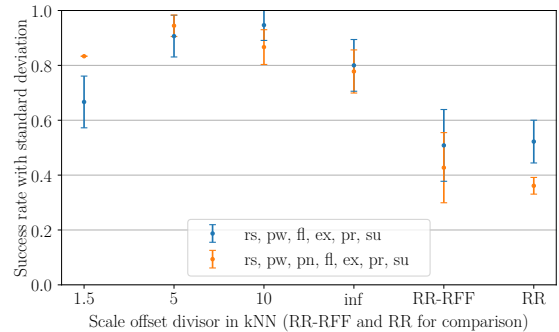

	\centering
	\begin{adjustbox}{max width=\columnwidth}
		\begingroup \makeatletter 
\makeatother \endgroup  	\end{adjustbox}
	\caption{Influence of scale offset divisor (with rest thresholding, $g=2.5$), pilot experiments} \label{fig:cv:p5}
\end{figure}

\Cref{fig:cv:p5} depicts the particular results. It is observable that the increase of $v$ could initially improve the average success rate for the used datasets. After reaching a maximum around 5--10, the success rates started to decrease again, probably because of low-intensity levels of gestures getting less reachable due to misclassification with rest. Nevertheless, the approach without any offset (corresponding to an infinite scale offset divisor) still performed clearly better than RR-RFF and standard RR.

Higher averaged success rates were achievable for all scale offset divisors in kNN than for RR and RR-RFF. The best averaged performance when the pointing gesture was included could be achieved for $v=5$ (94\% vs. 43\% for RR-RFF); and for $v=10$ (95\% vs. 51\% for RR-RFF) when pointing was not included.

With higher scale offset divisors, low-intensity level gestures $(\nicefrac{1}{3})$ get less reachable. This property has been assessed as more severely influencing the motivation of subjects than a reduced magnitude value range, since jumps between rest condition and low-intensity gestures appeared rather difficult than reduced sensitivity perceived as ``missing damping''.

Because of this, 5 was favoured over 10, although their performance appeared to be comparable (with 5 providing a slightly better performance when averaging over all dataset configurations, i. e. 93\% vs. 91\% with a comparable standard deviation).

\subsection{Evaluation of Prototype Reduction Algorithms}
\label{res_red}
In order to evaluate the performance of the chosen prototype reduction algorithms (see \cref{conc_emb}), the datasets captured for offline tests (\cref{tab:offline}) were transferred to KEEL and utilized as baseline.

These were pilot results to test the algorithms' accuracy and processing times with reduced datasets.

The reduction was executed on each cross-validation fold of the dataset individually, followed by the actual validation. As the considered algorithms comprise kNN-calculations inside, specifically for obtaining the validation accuracy, its parameters had to be defined. Following the recommendations in \cref{res_meas_sum}, a $k$ of 1, using $\nicefrac{1}{d^2}$ weighting, and the Euclidean distance as metric were configured. 

\begin{figure}[h]
	\centering
	\includegraphics[width=0.8\linewidth]{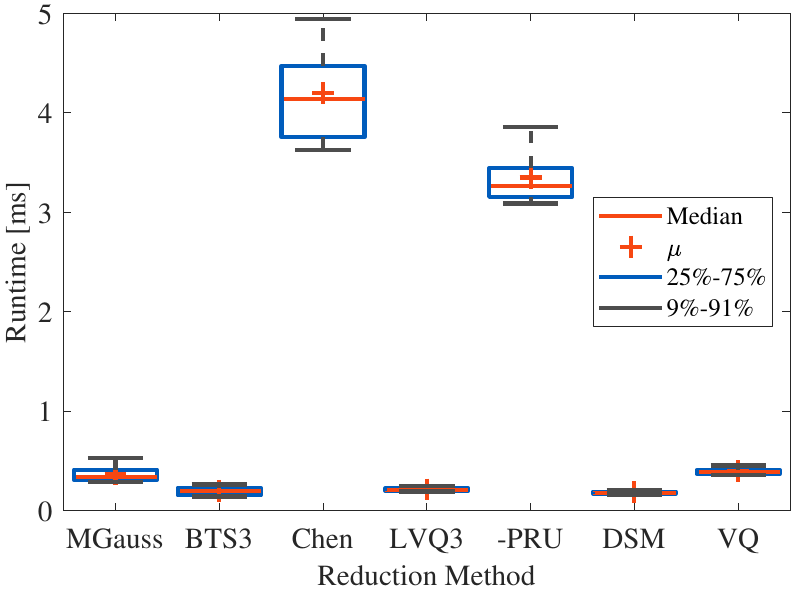}
	\caption{Average reduction times for seven gestures (rs, pw, pn, fl, ex, pr, su).}\label{fig:keel:k4}\end{figure}

The detailed examinations and results for varying datasets were presented in previous work \cite{szibBIOSIGNALS}. From this data, it could be seen that BTS3 and VQ were the lowest performing algorithms in terms of cross-validation accuracy so that these algorithms were disregarded. It also described the exclusion of PSCSA due to slow timing characteristics. Further conclusions drawn in that paper regarding timing can be representatively seen in \cref{fig:keel:k4}, where the time needed for reduction to 20 prototypes is depicted. This reduction time adds up with the cross-validation time to constitute the easily measurable overall runtime. Since the validation is the same process for each fold, the validation time can be disregarded so that the runtime qualitatively describes the algorithms' reduction times for comparison.

With respect to reduction time, MGauss, Chen and LVQPRU exposed a broad variance, leading to the presumption of reduced time determinism. Furthermore, these showed the highest means and medians of runtime, so that MGauss, LVQPRU and Chen were disregarded, too.

With that, LVQ3 and DSM (also based on LVQ) were the techniques to be chosen for a real-time implementation. With a low runtime of about 0.2 ms in most cases and a low time variance \cite{szibBIOSIGNALS} they turned out to be suitable for real-time scenarios, thus fulfilling requirement R5.1. For the present study, particularly DSM was selected to be examined in any further steps and proved itself as appropriate.

In order to deeper analyze DSM's suitability for embedded systems, an assessment of the runtime complexity will be made in the subsequent section.

\subsection{Runtime Complexity of DSM}
\label{runtimecompl}
To analyze the DSM prototype generation algorithm with regard to its runtime complexity, two phases can be distinguished, namely initialization and actual reduction. The phases will be analyzed on their worst-case runtime.

The following conventions are made: $ N \equiv $ number of samples in the original training set,
$ M \equiv $ number of prototypes in the final reduced set,
$ C \equiv $ number of gestures/classes, and
$ I \equiv $ number of iterations.

The results of this analysis are shown in \cref{a_DSM}. All operations are considered per EMG channel. The initialization process is designed in a way that there is at least one prototype per class by using the class centres as initial prototypes which become adjusted later on by penalizing or rewarding them in the reduction phase.

Summarizing \cref{a_DSM}, this yields the following running time complexity in initialization:
\begin{equation*}
	\mathcal{O}(C\cdot N + (M-C)\cdot N) = \mathcal{O}(M\cdot N)
\end{equation*}
and in reduction:
\begin{equation*}
	\mathcal{O}\left( I\cdot N\cdot \left(M + M\cdot \log{}M\right) \right) = \mathcal{O}(I\cdot N\cdot M\cdot \log{}M).
\end{equation*}

When assuming the number of classes to be constant with e. g. $C=7$ for (rs, pw, pn, fl, ex, pr, su) and also thinking of the number of iterations as a constant, e. g. $I=20$, an overall running-time complexity of $ \mathcal{O}(N\cdot M\cdot \log{}M) $ can be derived.

It has to be noted that the time complexity in reduction can principally be reduced from $\mathcal{O}(N\cdot M\cdot \log{}M)$ to $\mathcal{O}(N\cdot M)$ since no complete sorting of the distances between the currently selected sample and the single prototypes is necessary. Instead, a minimum search for the closest sample (1NN approach) and another minimum search for the closest sample with an identical class label would be sufficient -- thus leading to two times iterating the full prototype set at most (comparing the distances in the first case and comparing both distance and class label in the second case).

Generally speaking, if $k=1$ is used in kNN, the runtime complexity can be reduced to linear instead of logarithmic-linear (quasilinear).

Due to the fact that the number of prototypes $M$ is selected small and configured as a constant for the purpose of final prototype set size determinism (e. g. $M=20$), it might also be disregarded with regard to runtime, leading to an overall complexity of $\mathcal{O}(N)$ in the best case.

Interestingly, this would mean that DSM has the same runtime complexity in reduction (which is only performed once) as standard kNN in each single prediction step (or even better if a higher number of $k$ is used in kNN which would require sorting). Depending on the number of prototypes, the computational effort in a prediction step of DSM-reduced kNN is neglectable, in particular if $k=1$ is set inside prediction.

\subsection{Real-time User Studies with Multiple Subjects}
\begin{figure}[h]
	\centering
	\includegraphics[width=0.5\linewidth]{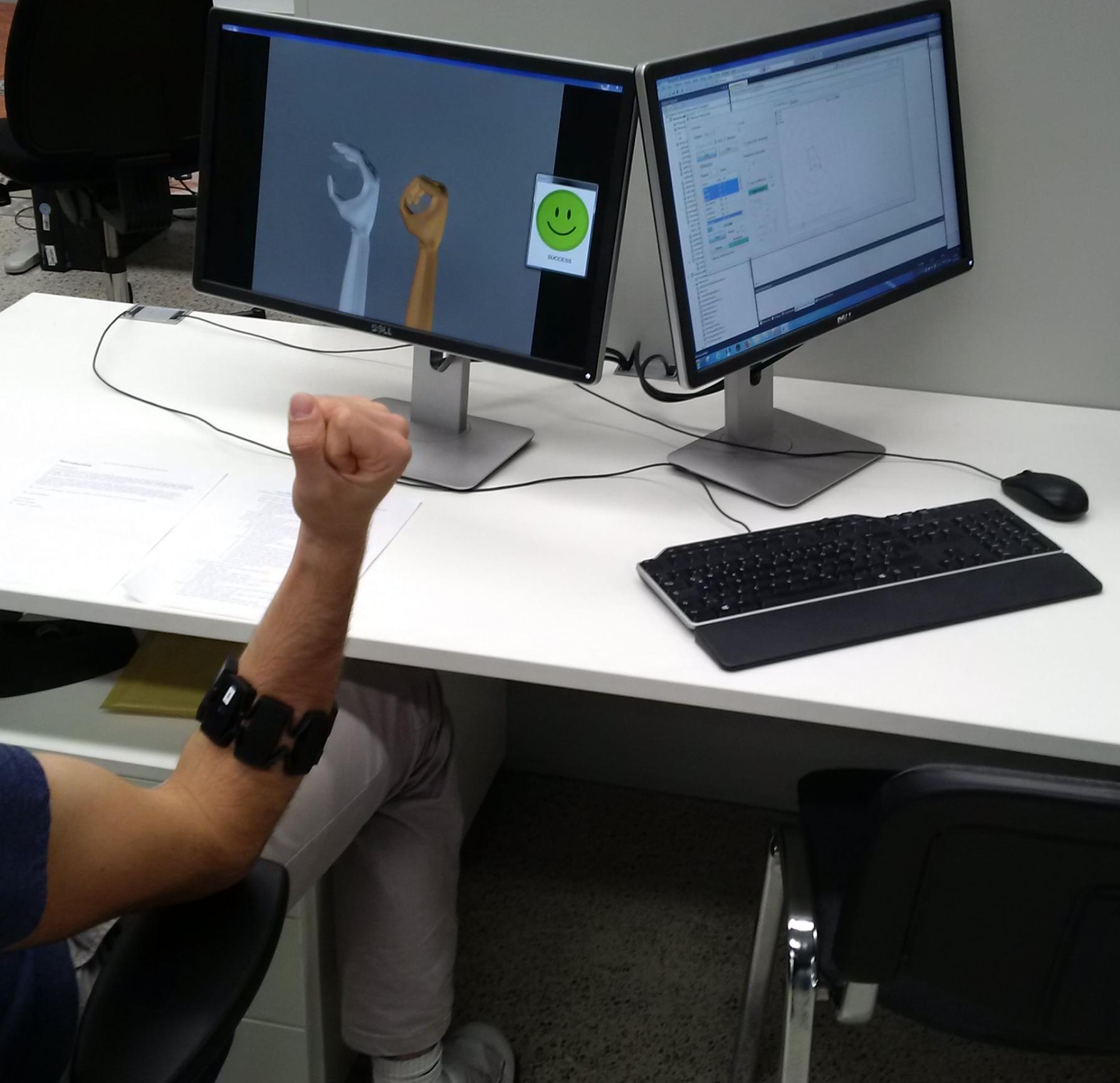}
	\caption{Experimental set-up of the user study, the subject is positioned to only view the left screen while the experimenter can control the experiment on the right screen, e.~g. for pausing if needed (without seeing the currently used method)}
	\label{fig:setup}
\end{figure}

In order to analyze if requirement R4 can be fulfilled by the proposed algorithms, online user studies with multiple subjects were conducted for the evaluation of suitability in practical scenarios. The set-up of the experiment is shown in \cref{fig:setup}. In the basic user study, it was chosen to compare the following four methods:
\begin{itemize}
	\item kNN parameterized according to the configuration obtained in the pilot experiments,
	\item kNN after training dataset reduction by means of DSM,
	\item ridge regression with Random Fourier Features (RR-RFF), and
	\item standard ridge regression (RF).
\end{itemize}

In the extended user study, RR was not examined due to a higher number of analyzed gestures.

Following the general recommendations from \cref{res_meas,pil}, the configuration of the standard kNN algorithm was set to $k=1$, the Euclidean distance metric, a weighting of $\nicefrac{1}{d^2}$, a rest magnitude threshold of $g=2.5$, and proportionality scaling with $v=5$.

For DSM-kNN, the same parameters were used within the prediction phase. For the reduction phase, DSM was configured to generate 7 prototypes in the final set with 40 iterations enabled. The results obtained will be explained in the following.

All statistical tests conducted in the following refer to a significance level of $\alpha=0.05$.

\subsubsection{Basic User Study (Five Classes)}
The subjects provided informed consent and statistical information as follows:
\begin{itemize}
	\item age range from 21--34 (mean 25, median 24),
	\item 3 female, 9 male,
	\item 1 left-handed, 11 right-handed,
	\item 4 already participated in many EMG experiments, 3 in a few, 5 without any EMG experience.
\end{itemize}

The experimental procedure for the real-time user study followed the same structure as the pilot experiments. The participants put on the Myo armband on their dominant forearm. Afterwards, for the training phase, they followed the visual stimulus (as in \cref{fig:setup}) by performing a repetitive series of hand and wrist movements (classes rs, pw, pn, fl, ex) one after another in three repetitions for two seconds each (all with full-intensity exertion). At maximum sampling rate of 200\,Hz, this results in 6000 training sample vectors (8 channels) per person ($= 5\cdot 3\cdot 2s\cdot 200\frac{1}{s}$).

In the prediction and test phase, they were asked to follow the stimulus again, in a total of 96 tasks (randomized but equally distributed among the subjects: 4 gestures (the rest class was not tested), 3 exertion levels, 4 methods, 2 repetitions) with breaks after a quarter, the half and three quarters of tasks. In this phase they furthermore saw the prediction of the currently exerted gesture in a second hand model. The goal was to match the stimulus and the predicted gesture within some spatial margin and time frame. Success was signalled by a green visualization. Otherwise, a yellow visualization was shown as visual feedback.

\begin{figure}[h]
	\centering
	\begin{tikzpicture}[x=1pt,y=1pt, scale=0.8]
\definecolor{fillColor}{RGB}{255,255,255}
\path[use as bounding box,fill=fillColor,fill opacity=0.00] (0,0) rectangle (216.81,216.81);
\begin{scope}
\path[clip] (  0.00,  0.00) rectangle (216.81,216.81);

\path[] (  0.00,  0.00) rectangle (216.81,216.81);
\end{scope}
\begin{scope}
\path[clip] ( 31.44, 28.09) rectangle (211.31,211.31);

\path[] ( 31.44, 28.09) rectangle (211.31,211.31);
\definecolor{drawColor}{RGB}{255,255,255}

\path[draw=drawColor,line width= 0.3pt,line join=round] ( 31.44, 51.56) --
	(211.31, 51.56);

\path[draw=drawColor,line width= 0.3pt,line join=round] ( 31.44, 81.84) --
	(211.31, 81.84);

\path[draw=drawColor,line width= 0.3pt,line join=round] ( 31.44,112.13) --
	(211.31,112.13);

\path[draw=drawColor,line width= 0.3pt,line join=round] ( 31.44,142.41) --
	(211.31,142.41);

\path[draw=drawColor,line width= 0.3pt,line join=round] ( 31.44,172.70) --
	(211.31,172.70);

\path[draw=drawColor,line width= 0.3pt,line join=round] ( 31.44,202.98) --
	(211.31,202.98);
\definecolor{drawColor}{RGB}{190,190,190}

\path[draw=drawColor,line width= 0.6pt,line join=round] ( 31.44, 36.42) --
	(211.31, 36.42);

\path[draw=drawColor,line width= 0.6pt,line join=round] ( 31.44, 66.70) --
	(211.31, 66.70);

\path[draw=drawColor,line width= 0.6pt,line join=round] ( 31.44, 96.99) --
	(211.31, 96.99);

\path[draw=drawColor,line width= 0.6pt,line join=round] ( 31.44,127.27) --
	(211.31,127.27);

\path[draw=drawColor,line width= 0.6pt,line join=round] ( 31.44,157.56) --
	(211.31,157.56);

\path[draw=drawColor,line width= 0.6pt,line join=round] ( 31.44,187.84) --
	(211.31,187.84);
\definecolor{drawColor}{RGB}{0,125,197}

\path[draw=drawColor,line width= 0.6pt,line join=round] ( 52.85,187.84) --
	( 61.42,187.84);

\path[draw=drawColor,line width= 0.6pt,line join=round] ( 57.13,187.84) --
	( 57.13,112.13);

\path[draw=drawColor,line width= 0.6pt,line join=round] ( 52.85,112.13) --
	( 61.42,112.13);

\path[draw=drawColor,line width= 0.6pt,line join=round] ( 95.68,181.53) --
	(104.24,181.53);

\path[draw=drawColor,line width= 0.6pt,line join=round] ( 99.96,181.53) --
	( 99.96,118.44);

\path[draw=drawColor,line width= 0.6pt,line join=round] ( 95.68,118.44) --
	(104.24,118.44);

\path[draw=drawColor,line width= 0.6pt,line join=round] (138.50,156.29) --
	(147.07,156.29);

\path[draw=drawColor,line width= 0.6pt,line join=round] (142.79,156.29) --
	(142.79, 36.42);

\path[draw=drawColor,line width= 0.6pt,line join=round] (138.50, 36.42) --
	(147.07, 36.42);

\path[draw=drawColor,line width= 0.6pt,line join=round] (181.33,137.37) --
	(189.90,137.37);

\path[draw=drawColor,line width= 0.6pt,line join=round] (185.61,137.37) --
	(185.61, 49.04);

\path[draw=drawColor,line width= 0.6pt,line join=round] (181.33, 49.04) --
	(189.90, 49.04);

\path[draw=drawColor,line width= 0.6pt,line join=round] ( 57.13,164.18) -- ( 57.13,187.84);

\path[draw=drawColor,line width= 0.6pt,line join=round] ( 57.13,131.06) -- ( 57.13,112.13);
\definecolor{fillColor}{RGB}{148,188,228}

\path[draw=drawColor,line width= 0.6pt,line join=round,line cap=round,fill=fillColor] ( 41.07,164.18) --
	( 41.07,131.06) --
	( 73.19,131.06) --
	( 73.19,164.18) --
	( 41.07,164.18) --
	cycle;

\path[draw=drawColor,line width= 1.1pt,line join=round] ( 41.07,137.37) -- ( 73.19,137.37);

\path[draw=drawColor,line width= 0.6pt,line join=round] ( 99.96,159.45) -- ( 99.96,181.53);

\path[draw=drawColor,line width= 0.6pt,line join=round] ( 99.96,131.06) -- ( 99.96,118.44);

\path[draw=drawColor,line width= 0.6pt,line join=round,line cap=round,fill=fillColor] ( 83.90,159.45) --
	( 83.90,131.06) --
	(116.02,131.06) --
	(116.02,159.45) --
	( 83.90,159.45) --
	cycle;

\path[draw=drawColor,line width= 1.1pt,line join=round] ( 83.90,143.68) -- (116.02,143.68);

\path[draw=drawColor,line width= 0.6pt,line join=round] (142.79,112.13) -- (142.79,156.29);

\path[draw=drawColor,line width= 0.6pt,line join=round] (142.79, 71.12) -- (142.79, 36.42);

\path[draw=drawColor,line width= 0.6pt,line join=round,line cap=round,fill=fillColor] (126.73,112.13) --
	(126.73, 71.12) --
	(158.85, 71.12) --
	(158.85,112.13) --
	(126.73,112.13) --
	cycle;

\path[draw=drawColor,line width= 1.1pt,line join=round] (126.73, 90.05) -- (158.85, 90.05);

\path[draw=drawColor,line width= 0.6pt,line join=round] (185.61, 94.78) -- (185.61,137.37);

\path[draw=drawColor,line width= 0.6pt,line join=round] (185.61, 61.66) -- (185.61, 49.04);

\path[draw=drawColor,line width= 0.6pt,line join=round,line cap=round,fill=fillColor] (169.55, 94.78) --
	(169.55, 61.66) --
	(201.67, 61.66) --
	(201.67, 94.78) --
	(169.55, 94.78) --
	cycle;

\path[draw=drawColor,line width= 1.1pt,line join=round] (169.55, 74.27) -- (201.67, 74.27);

\path[draw=drawColor,line width= 0.4pt,line join=round,line cap=round] ( 54.36,144.73) -- ( 59.91,144.73);

\path[draw=drawColor,line width= 0.4pt,line join=round,line cap=round] ( 57.13,141.95) -- ( 57.13,147.50);

\path[draw=drawColor,line width= 0.4pt,line join=round,line cap=round] ( 97.18,147.36) -- (102.73,147.36);

\path[draw=drawColor,line width= 0.4pt,line join=round,line cap=round] ( 99.96,144.58) -- ( 99.96,150.13);

\path[draw=drawColor,line width= 0.4pt,line join=round,line cap=round] (140.01, 92.15) -- (145.56, 92.15);

\path[draw=drawColor,line width= 0.4pt,line join=round,line cap=round] (142.79, 89.37) -- (142.79, 94.92);

\path[draw=drawColor,line width= 0.4pt,line join=round,line cap=round] (182.84, 81.63) -- (188.39, 81.63);

\path[draw=drawColor,line width= 0.4pt,line join=round,line cap=round] (185.61, 78.86) -- (185.61, 84.41);
\definecolor{drawColor}{RGB}{0,0,0}

\path[draw=drawColor,line width= 0.6pt,line join=round,line cap=round] ( 57.13,192.38) -- ( 57.13,195.41);

\path[draw=drawColor,line width= 0.6pt,line join=round,line cap=round] ( 57.13,195.41) -- ( 99.96,195.41);

\path[draw=drawColor,line width= 0.6pt,line join=round,line cap=round] ( 99.96,195.41) -- ( 99.96,192.38);

\path[draw=drawColor,line width= 0.6pt,line join=round,line cap=round] (142.79,192.38) -- (142.79,195.41);

\path[draw=drawColor,line width= 0.6pt,line join=round,line cap=round] (142.79,195.41) -- (185.61,195.41);

\path[draw=drawColor,line width= 0.6pt,line join=round,line cap=round] (185.61,195.41) -- (185.61,192.38);

\node[text=drawColor,anchor=base,inner sep=0pt, outer sep=0pt, scale=  0.85] at (121.37,204.81) {$p<.0005$};

\path[draw=drawColor,line width= 0.6pt,line join=round,line cap=round] ( 78.55,199.95) -- ( 78.55,202.98);

\path[draw=drawColor,line width= 0.6pt,line join=round,line cap=round] ( 78.55,202.98) -- (164.20,202.98);

\path[draw=drawColor,line width= 0.6pt,line join=round,line cap=round] (164.20,202.98) -- (164.20,199.95);
\end{scope}
\begin{scope}
\path[clip] (  0.00,  0.00) rectangle (216.81,216.81);
\definecolor{drawColor}{RGB}{0,0,0}

\node[text=drawColor,anchor=base east,inner sep=0pt, outer sep=0pt, scale=  0.80] at ( 26.49, 33.66) {0.0};

\node[text=drawColor,anchor=base east,inner sep=0pt, outer sep=0pt, scale=  0.80] at ( 26.49, 63.95) {0.2};

\node[text=drawColor,anchor=base east,inner sep=0pt, outer sep=0pt, scale=  0.80] at ( 26.49, 94.23) {0.4};

\node[text=drawColor,anchor=base east,inner sep=0pt, outer sep=0pt, scale=  0.80] at ( 26.49,124.52) {0.6};

\node[text=drawColor,anchor=base east,inner sep=0pt, outer sep=0pt, scale=  0.80] at ( 26.49,154.80) {0.8};

\node[text=drawColor,anchor=base east,inner sep=0pt, outer sep=0pt, scale=  0.80] at ( 26.49,185.08) {1.0};
\end{scope}
\begin{scope}
\path[clip] (  0.00,  0.00) rectangle (216.81,216.81);
\definecolor{drawColor}{gray}{0.20}

\path[draw=drawColor,line width= 0.6pt,line join=round] ( 28.69, 36.42) --
	( 31.44, 36.42);

\path[draw=drawColor,line width= 0.6pt,line join=round] ( 28.69, 66.70) --
	( 31.44, 66.70);

\path[draw=drawColor,line width= 0.6pt,line join=round] ( 28.69, 96.99) --
	( 31.44, 96.99);

\path[draw=drawColor,line width= 0.6pt,line join=round] ( 28.69,127.27) --
	( 31.44,127.27);

\path[draw=drawColor,line width= 0.6pt,line join=round] ( 28.69,157.56) --
	( 31.44,157.56);

\path[draw=drawColor,line width= 0.6pt,line join=round] ( 28.69,187.84) --
	( 31.44,187.84);
\end{scope}
\begin{scope}
\path[clip] (  0.00,  0.00) rectangle (216.81,216.81);
\definecolor{drawColor}{gray}{0.20}

\path[draw=drawColor,line width= 0.6pt,line join=round] ( 57.13, 25.34) --
	( 57.13, 28.09);

\path[draw=drawColor,line width= 0.6pt,line join=round] ( 99.96, 25.34) --
	( 99.96, 28.09);

\path[draw=drawColor,line width= 0.6pt,line join=round] (142.79, 25.34) --
	(142.79, 28.09);

\path[draw=drawColor,line width= 0.6pt,line join=round] (185.61, 25.34) --
	(185.61, 28.09);
\end{scope}
\begin{scope}
\path[clip] (  0.00,  0.00) rectangle (216.81,216.81);
\definecolor{drawColor}{RGB}{0,0,0}

\node[text=drawColor,anchor=base,inner sep=0pt, outer sep=0pt, scale=  0.80] at ( 57.13, 17.63) {kNN};

\node[text=drawColor,anchor=base,inner sep=0pt, outer sep=0pt, scale=  0.80] at ( 99.96, 17.63) {DSM-kNN};

\node[text=drawColor,anchor=base,inner sep=0pt, outer sep=0pt, scale=  0.80] at (142.79, 17.63) {RR-RFF};

\node[text=drawColor,anchor=base,inner sep=0pt, outer sep=0pt, scale=  0.80] at (185.61, 17.63) {RR};
\end{scope}
\begin{scope}
\path[clip] (  0.00,  0.00) rectangle (216.81,216.81);
\definecolor{drawColor}{RGB}{0,0,0}

\node[text=drawColor,anchor=base,inner sep=0pt, outer sep=0pt, scale=  0.90] at (121.37,  7.32) {Method};
\end{scope}
\begin{scope}
\path[clip] (  0.00,  0.00) rectangle (216.81,216.81);
\definecolor{drawColor}{RGB}{0,0,0}

\node[text=drawColor,rotate= 90.00,anchor=base,inner sep=0pt, outer sep=0pt, scale=  0.90] at ( 11.70,119.70) {Success Rate};\end{scope}

\end{tikzpicture}
 	\caption{Basic user study: success rates (over subjects) and significance from ANOVA ($\alpha=0.05$), showing significant difference in success rates for kNN-based compared to RR-based methods}
	\label{fig:userstudy1_overall}
\end{figure}
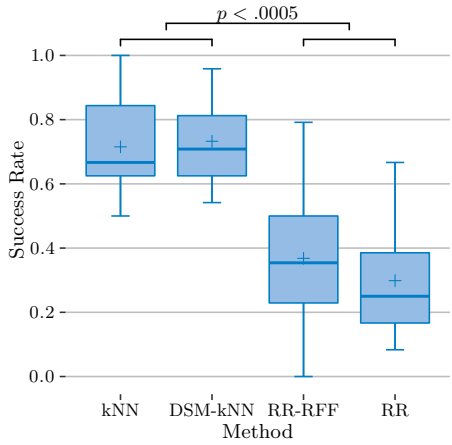

The summarized performance of each examined method for the 12 subjects is depicted in \cref{fig:userstudy1_overall}, after first averaging the per-level- and -gesture-performances for each subject-method combination. This yields the variance and median of the success rates in a subject-based manner. Overall, it is observable that the success rates achieved with kNN-based methods exceeded the ones from RR-based methods. DSM-reduced kNN performed as good as standard kNN (success rates of 73\% and 71\% mean, 71\% and 67\% median respectively), while RR-RFF and RR showed success rates at a lower level (37\% and 30\% mean, 25\% and 25\% median). An ANOVA pointed out significance between the groups of kNN-based methods and the group of RR-based methods ($p<0.0005$), while there is no significant difference within each of the groups.

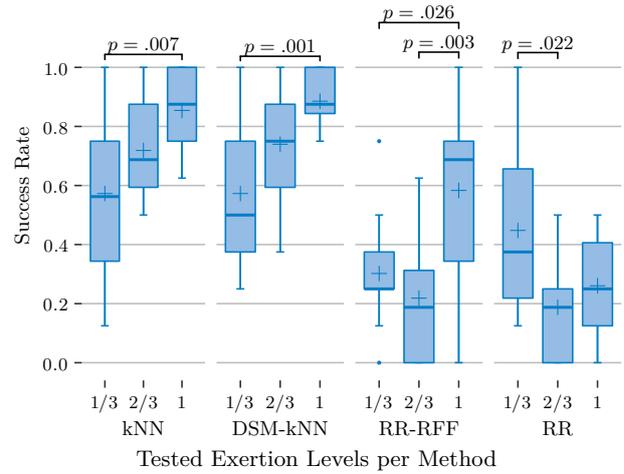
\begin{figure}[h]
	\begin{tikzpicture}[x=1pt,y=1pt]
\definecolor{fillColor}{RGB}{255,255,255}
\path[use as bounding box,fill=fillColor,fill opacity=0.00] (0,0) rectangle ( 83.11,180.67);
\begin{scope}
\path[clip] (  0.00,  0.00) rectangle ( 83.11,180.67);

\path[] ( -0.00,  0.00) rectangle ( 83.11,180.68);
\end{scope}
\begin{scope}
\path[clip] ( 31.44, 28.09) rectangle ( 77.61,175.17);

\path[] ( 31.44, 28.09) rectangle ( 77.61,175.18);
\definecolor{drawColor}{RGB}{255,255,255}

\path[draw=drawColor,line width= 0.3pt,line join=round] ( 31.44, 45.92) --
	( 77.61, 45.92);

\path[draw=drawColor,line width= 0.3pt,line join=round] ( 31.44, 68.20) --
	( 77.61, 68.20);

\path[draw=drawColor,line width= 0.3pt,line join=round] ( 31.44, 90.49) --
	( 77.61, 90.49);

\path[draw=drawColor,line width= 0.3pt,line join=round] ( 31.44,112.78) --
	( 77.61,112.78);

\path[draw=drawColor,line width= 0.3pt,line join=round] ( 31.44,135.06) --
	( 77.61,135.06);

\path[draw=drawColor,line width= 0.3pt,line join=round] ( 31.44,157.35) --
	( 77.61,157.35);

\path[draw=drawColor,line width= 0.3pt,line join=round] ( 31.44,168.49) --
	( 77.61,168.49);
\definecolor{drawColor}{RGB}{190,190,190}

\path[draw=drawColor,line width= 0.6pt,line join=round] ( 31.44, 34.78) --
	( 77.61, 34.78);

\path[draw=drawColor,line width= 0.6pt,line join=round] ( 31.44, 57.06) --
	( 77.61, 57.06);

\path[draw=drawColor,line width= 0.6pt,line join=round] ( 31.44, 79.35) --
	( 77.61, 79.35);

\path[draw=drawColor,line width= 0.6pt,line join=round] ( 31.44,101.63) --
	( 77.61,101.63);

\path[draw=drawColor,line width= 0.6pt,line join=round] ( 31.44,123.92) --
	( 77.61,123.92);

\path[draw=drawColor,line width= 0.6pt,line join=round] ( 31.44,146.20) --
	( 77.61,146.20);
\definecolor{drawColor}{RGB}{0,125,197}

\path[draw=drawColor,line width= 0.6pt,line join=round] ( 38.65,146.20) --
	( 41.54,146.20);

\path[draw=drawColor,line width= 0.6pt,line join=round] ( 40.09,146.20) --
	( 40.09, 48.70);

\path[draw=drawColor,line width= 0.6pt,line join=round] ( 38.65, 48.70) --
	( 41.54, 48.70);

\path[draw=drawColor,line width= 0.6pt,line join=round] ( 53.08,146.20) --
	( 55.97,146.20);

\path[draw=drawColor,line width= 0.6pt,line join=round] ( 54.52,146.20) --
	( 54.52, 90.49);

\path[draw=drawColor,line width= 0.6pt,line join=round] ( 53.08, 90.49) --
	( 55.97, 90.49);

\path[draw=drawColor,line width= 0.6pt,line join=round] ( 67.51,146.20) --
	( 70.40,146.20);

\path[draw=drawColor,line width= 0.6pt,line join=round] ( 68.95,146.20) --
	( 68.95,104.42);

\path[draw=drawColor,line width= 0.6pt,line join=round] ( 67.51,104.42) --
	( 70.40,104.42);

\path[draw=drawColor,line width= 0.6pt,line join=round] ( 40.09,118.35) -- ( 40.09,146.20);

\path[draw=drawColor,line width= 0.6pt,line join=round] ( 40.09, 73.08) -- ( 40.09, 48.70);
\definecolor{fillColor}{RGB}{148,188,228}

\path[draw=drawColor,line width= 0.6pt,line join=round,line cap=round,fill=fillColor] ( 34.68,118.35) --
	( 34.68, 73.08) --
	( 45.51, 73.08) --
	( 45.51,118.35) --
	( 34.68,118.35) --
	cycle;

\path[draw=drawColor,line width= 1.1pt,line join=round] ( 34.68, 97.45) -- ( 45.51, 97.45);

\path[draw=drawColor,line width= 0.6pt,line join=round] ( 54.52,132.28) -- ( 54.52,146.20);

\path[draw=drawColor,line width= 0.6pt,line join=round] ( 54.52,100.94) -- ( 54.52, 90.49);

\path[draw=drawColor,line width= 0.6pt,line join=round,line cap=round,fill=fillColor] ( 49.11,132.28) --
	( 49.11,100.94) --
	( 59.93,100.94) --
	( 59.93,132.28) --
	( 49.11,132.28) --
	cycle;

\path[draw=drawColor,line width= 1.1pt,line join=round] ( 49.11,111.38) -- ( 59.93,111.38);

\path[draw=drawColor,line width= 0.6pt,line join=round] ( 68.95,146.20) -- ( 68.95,146.20);

\path[draw=drawColor,line width= 0.6pt,line join=round] ( 68.95,118.35) -- ( 68.95,104.42);

\path[draw=drawColor,line width= 0.6pt,line join=round,line cap=round,fill=fillColor] ( 63.54,146.20) --
	( 63.54,118.35) --
	( 74.36,118.35) --
	( 74.36,146.20) --
	( 63.54,146.20) --
	cycle;

\path[draw=drawColor,line width= 1.1pt,line join=round] ( 63.54,132.28) -- ( 74.36,132.28);

\path[draw=drawColor,line width= 0.4pt,line join=round,line cap=round] ( 37.32, 98.61) -- ( 42.87, 98.61);

\path[draw=drawColor,line width= 0.4pt,line join=round,line cap=round] ( 40.09, 95.84) -- ( 40.09,101.39);

\path[draw=drawColor,line width= 0.4pt,line join=round,line cap=round] ( 51.75,114.86) -- ( 57.30,114.86);

\path[draw=drawColor,line width= 0.4pt,line join=round,line cap=round] ( 54.52,112.09) -- ( 54.52,117.64);

\path[draw=drawColor,line width= 0.4pt,line join=round,line cap=round] ( 66.18,129.95) -- ( 71.73,129.95);

\path[draw=drawColor,line width= 0.4pt,line join=round,line cap=round] ( 68.95,127.18) -- ( 68.95,132.73);
\definecolor{drawColor}{RGB}{0,0,0}

\node[text=drawColor,anchor=base,inner sep=0pt, outer sep=0pt, scale=  0.85] at ( 54.52,152.55) {$p=.007$};

\path[draw=drawColor,line width= 0.6pt,line join=round,line cap=round] ( 40.09,149.13) -- ( 40.09,151.08);

\path[draw=drawColor,line width= 0.6pt,line join=round,line cap=round] ( 40.09,151.08) -- ( 68.95,151.08);

\path[draw=drawColor,line width= 0.6pt,line join=round,line cap=round] ( 68.95,151.08) -- ( 68.95,149.13);
\end{scope}
\begin{scope}
\path[clip] (  0.00,  0.00) rectangle ( 83.11,180.67);
\definecolor{drawColor}{RGB}{0,0,0}

\node[text=drawColor,anchor=base east,inner sep=0pt, outer sep=0pt, scale=  0.80] at ( 26.49, 32.02) {0.0};

\node[text=drawColor,anchor=base east,inner sep=0pt, outer sep=0pt, scale=  0.80] at ( 26.49, 54.31) {0.2};

\node[text=drawColor,anchor=base east,inner sep=0pt, outer sep=0pt, scale=  0.80] at ( 26.49, 76.59) {0.4};

\node[text=drawColor,anchor=base east,inner sep=0pt, outer sep=0pt, scale=  0.80] at ( 26.49, 98.88) {0.6};

\node[text=drawColor,anchor=base east,inner sep=0pt, outer sep=0pt, scale=  0.80] at ( 26.49,121.16) {0.8};

\node[text=drawColor,anchor=base east,inner sep=0pt, outer sep=0pt, scale=  0.80] at ( 26.49,143.45) {1.0};
\end{scope}
\begin{scope}
\path[clip] (  0.00,  0.00) rectangle ( 83.11,180.67);
\definecolor{drawColor}{gray}{0.20}

\path[draw=drawColor,line width= 0.6pt,line join=round] ( 28.69, 34.78) --
	( 31.44, 34.78);

\path[draw=drawColor,line width= 0.6pt,line join=round] ( 28.69, 57.06) --
	( 31.44, 57.06);

\path[draw=drawColor,line width= 0.6pt,line join=round] ( 28.69, 79.35) --
	( 31.44, 79.35);

\path[draw=drawColor,line width= 0.6pt,line join=round] ( 28.69,101.63) --
	( 31.44,101.63);

\path[draw=drawColor,line width= 0.6pt,line join=round] ( 28.69,123.92) --
	( 31.44,123.92);

\path[draw=drawColor,line width= 0.6pt,line join=round] ( 28.69,146.20) --
	( 31.44,146.20);
\end{scope}
\begin{scope}
\path[clip] (  0.00,  0.00) rectangle ( 83.11,180.67);
\definecolor{drawColor}{gray}{0.20}

\path[draw=drawColor,line width= 0.6pt,line join=round] ( 40.09, 25.34) --
	( 40.09, 28.09);

\path[draw=drawColor,line width= 0.6pt,line join=round] ( 54.52, 25.34) --
	( 54.52, 28.09);

\path[draw=drawColor,line width= 0.6pt,line join=round] ( 68.95, 25.34) --
	( 68.95, 28.09);
\end{scope}
\begin{scope}
\path[clip] (  0.00,  0.00) rectangle ( 83.11,180.67);
\definecolor{drawColor}{RGB}{0,0,0}

\node[text=drawColor,anchor=base,inner sep=0pt, outer sep=0pt, scale=  0.80] at ( 40.09, 17.63) {1/3};

\node[text=drawColor,anchor=base,inner sep=0pt, outer sep=0pt, scale=  0.80] at ( 54.52, 17.63) {2/3};

\node[text=drawColor,anchor=base,inner sep=0pt, outer sep=0pt, scale=  0.80] at ( 68.95, 17.63) {1};
\end{scope}
\begin{scope}
\path[clip] (  0.00,  0.00) rectangle ( 83.11,180.67);
\definecolor{drawColor}{RGB}{0,0,0}

\node[text=drawColor,anchor=base,inner sep=0pt, outer sep=0pt, scale=  0.90] at ( 54.52,  7.32) {kNN};
\end{scope}
\begin{scope}
\path[clip] (  0.00,  0.00) rectangle ( 83.11,180.67);
\definecolor{drawColor}{RGB}{0,0,0}

\node[text=drawColor,rotate= 90.00,anchor=base,inner sep=0pt, outer sep=0pt, scale=  0.90] at ( 11.70,101.63) {Success Rate};
\end{scope}
\end{tikzpicture}
 \hspace{-0.5cm}
	\begin{tikzpicture}[x=1pt,y=1pt]
\definecolor{fillColor}{RGB}{255,255,255}
\path[use as bounding box,fill=fillColor,fill opacity=0.00] (0,0) rectangle ( 61.43,180.67);
\begin{scope}
\path[clip] (  0.00,  0.00) rectangle ( 61.43,180.67);

\path[] (  0.00,  0.00) rectangle ( 61.43,180.68);
\end{scope}
\begin{scope}
\path[clip] (  8.25, 28.09) rectangle ( 55.93,175.17);

\path[] (  8.25, 28.09) rectangle ( 55.93,175.18);
\definecolor{drawColor}{RGB}{255,255,255}

\path[draw=drawColor,line width= 0.3pt,line join=round] (  8.25, 45.92) --
	( 55.93, 45.92);

\path[draw=drawColor,line width= 0.3pt,line join=round] (  8.25, 68.20) --
	( 55.93, 68.20);

\path[draw=drawColor,line width= 0.3pt,line join=round] (  8.25, 90.49) --
	( 55.93, 90.49);

\path[draw=drawColor,line width= 0.3pt,line join=round] (  8.25,112.78) --
	( 55.93,112.78);

\path[draw=drawColor,line width= 0.3pt,line join=round] (  8.25,135.06) --
	( 55.93,135.06);

\path[draw=drawColor,line width= 0.3pt,line join=round] (  8.25,157.35) --
	( 55.93,157.35);

\path[draw=drawColor,line width= 0.3pt,line join=round] (  8.25,168.49) --
	( 55.93,168.49);
\definecolor{drawColor}{RGB}{190,190,190}

\path[draw=drawColor,line width= 0.6pt,line join=round] (  8.25, 34.78) --
	( 55.93, 34.78);

\path[draw=drawColor,line width= 0.6pt,line join=round] (  8.25, 57.06) --
	( 55.93, 57.06);

\path[draw=drawColor,line width= 0.6pt,line join=round] (  8.25, 79.35) --
	( 55.93, 79.35);

\path[draw=drawColor,line width= 0.6pt,line join=round] (  8.25,101.63) --
	( 55.93,101.63);

\path[draw=drawColor,line width= 0.6pt,line join=round] (  8.25,123.92) --
	( 55.93,123.92);

\path[draw=drawColor,line width= 0.6pt,line join=round] (  8.25,146.20) --
	( 55.93,146.20);
\definecolor{drawColor}{RGB}{0,125,197}

\path[draw=drawColor,line width= 0.6pt,line join=round] ( 15.70,146.20) --
	( 18.68,146.20);

\path[draw=drawColor,line width= 0.6pt,line join=round] ( 17.19,146.20) --
	( 17.19, 62.63);

\path[draw=drawColor,line width= 0.6pt,line join=round] ( 15.70, 62.63) --
	( 18.68, 62.63);

\path[draw=drawColor,line width= 0.6pt,line join=round] ( 30.60,146.20) --
	( 33.58,146.20);

\path[draw=drawColor,line width= 0.6pt,line join=round] ( 32.09,146.20) --
	( 32.09, 76.56);

\path[draw=drawColor,line width= 0.6pt,line join=round] ( 30.60, 76.56) --
	( 33.58, 76.56);

\path[draw=drawColor,line width= 0.6pt,line join=round] ( 45.50,146.20) --
	( 48.48,146.20);

\path[draw=drawColor,line width= 0.6pt,line join=round] ( 46.99,146.20) --
	( 46.99,118.35);

\path[draw=drawColor,line width= 0.6pt,line join=round] ( 45.50,118.35) --
	( 48.48,118.35);

\path[draw=drawColor,line width= 0.6pt,line join=round] ( 17.19,118.35) -- ( 17.19,146.20);

\path[draw=drawColor,line width= 0.6pt,line join=round] ( 17.19, 76.56) -- ( 17.19, 62.63);
\definecolor{fillColor}{RGB}{148,188,228}

\path[draw=drawColor,line width= 0.6pt,line join=round,line cap=round,fill=fillColor] ( 11.60,118.35) --
	( 11.60, 76.56) --
	( 22.78, 76.56) --
	( 22.78,118.35) --
	( 11.60,118.35) --
	cycle;

\path[draw=drawColor,line width= 1.1pt,line join=round] ( 11.60, 90.49) -- ( 22.78, 90.49);

\path[draw=drawColor,line width= 0.6pt,line join=round] ( 32.09,132.28) -- ( 32.09,146.20);

\path[draw=drawColor,line width= 0.6pt,line join=round] ( 32.09,100.94) -- ( 32.09, 76.56);

\path[draw=drawColor,line width= 0.6pt,line join=round,line cap=round,fill=fillColor] ( 26.50,132.28) --
	( 26.50,100.94) --
	( 37.68,100.94) --
	( 37.68,132.28) --
	( 26.50,132.28) --
	cycle;

\path[draw=drawColor,line width= 1.1pt,line join=round] ( 26.50,118.35) -- ( 37.68,118.35);

\path[draw=drawColor,line width= 0.6pt,line join=round] ( 46.99,146.20) -- ( 46.99,146.20);

\path[draw=drawColor,line width= 0.6pt,line join=round] ( 46.99,128.79) -- ( 46.99,118.35);

\path[draw=drawColor,line width= 0.6pt,line join=round,line cap=round,fill=fillColor] ( 41.40,146.20) --
	( 41.40,128.79) --
	( 52.58,128.79) --
	( 52.58,146.20) --
	( 41.40,146.20) --
	cycle;

\path[draw=drawColor,line width= 1.1pt,line join=round] ( 41.40,132.28) -- ( 52.58,132.28);

\path[draw=drawColor,line width= 0.4pt,line join=round,line cap=round] ( 14.42, 98.61) -- ( 19.96, 98.61);

\path[draw=drawColor,line width= 0.4pt,line join=round,line cap=round] ( 17.19, 95.84) -- ( 17.19,101.39);

\path[draw=drawColor,line width= 0.4pt,line join=round,line cap=round] ( 29.31,117.19) -- ( 34.86,117.19);

\path[draw=drawColor,line width= 0.4pt,line join=round,line cap=round] ( 32.09,114.41) -- ( 32.09,119.96);

\path[draw=drawColor,line width= 0.4pt,line join=round,line cap=round] ( 44.21,133.44) -- ( 49.76,133.44);

\path[draw=drawColor,line width= 0.4pt,line join=round,line cap=round] ( 46.99,130.66) -- ( 46.99,136.21);
\definecolor{drawColor}{RGB}{0,0,0}

\node[text=drawColor,anchor=base,inner sep=0pt, outer sep=0pt, scale=  0.85] at ( 32.09,151.85) {$p=.001$};

\path[draw=drawColor,line width= 0.6pt,line join=round,line cap=round] ( 17.19,148.71) -- ( 17.19,150.38);

\path[draw=drawColor,line width= 0.6pt,line join=round,line cap=round] ( 17.19,150.38) -- ( 46.99,150.38);

\path[draw=drawColor,line width= 0.6pt,line join=round,line cap=round] ( 46.99,150.38) -- ( 46.99,148.71);
\end{scope}
\begin{scope}
\path[clip] (  0.00,  0.00) rectangle ( 61.43,180.67);
\definecolor{drawColor}{gray}{0.20}

\path[draw=drawColor,line width= 0.6pt,line join=round] ( 17.19, 25.34) --
	( 17.19, 28.09);

\path[draw=drawColor,line width= 0.6pt,line join=round] ( 32.09, 25.34) --
	( 32.09, 28.09);

\path[draw=drawColor,line width= 0.6pt,line join=round] ( 46.99, 25.34) --
	( 46.99, 28.09);
\end{scope}
\begin{scope}
\path[clip] (  0.00,  0.00) rectangle ( 61.43,180.67);
\definecolor{drawColor}{RGB}{0,0,0}

\node[text=drawColor,anchor=base,inner sep=0pt, outer sep=0pt, scale=  0.80] at ( 17.19, 17.63) {1/3};

\node[text=drawColor,anchor=base,inner sep=0pt, outer sep=0pt, scale=  0.80] at ( 32.09, 17.63) {2/3};

\node[text=drawColor,anchor=base,inner sep=0pt, outer sep=0pt, scale=  0.80] at ( 46.99, 17.63) {1};
\end{scope}
\begin{scope}
\path[clip] (  0.00,  0.00) rectangle ( 61.43,180.67);
\definecolor{drawColor}{RGB}{0,0,0}

\node[text=drawColor,anchor=base,inner sep=0pt, outer sep=0pt, scale=  0.90] at ( 32.09,  7.32) {DSM-kNN};
\end{scope}
\end{tikzpicture}
 \hspace{-0.5cm}
	\begin{tikzpicture}[x=1pt,y=1pt]
\definecolor{fillColor}{RGB}{255,255,255}
\path[use as bounding box,fill=fillColor,fill opacity=0.00] (0,0) rectangle ( 61.43,180.67);
\begin{scope}
\path[clip] (  0.00,  0.00) rectangle ( 61.43,180.67);

\path[] (  0.00,  0.00) rectangle ( 61.43,180.68);
\end{scope}
\begin{scope}
\path[clip] (  8.25, 28.09) rectangle ( 55.93,175.17);

\path[] (  8.25, 28.09) rectangle ( 55.93,175.18);
\definecolor{drawColor}{RGB}{255,255,255}

\path[draw=drawColor,line width= 0.3pt,line join=round] (  8.25, 45.92) --
	( 55.93, 45.92);

\path[draw=drawColor,line width= 0.3pt,line join=round] (  8.25, 68.20) --
	( 55.93, 68.20);

\path[draw=drawColor,line width= 0.3pt,line join=round] (  8.25, 90.49) --
	( 55.93, 90.49);

\path[draw=drawColor,line width= 0.3pt,line join=round] (  8.25,112.78) --
	( 55.93,112.78);

\path[draw=drawColor,line width= 0.3pt,line join=round] (  8.25,135.06) --
	( 55.93,135.06);

\path[draw=drawColor,line width= 0.3pt,line join=round] (  8.25,157.35) --
	( 55.93,157.35);

\path[draw=drawColor,line width= 0.3pt,line join=round] (  8.25,168.49) --
	( 55.93,168.49);
\definecolor{drawColor}{RGB}{190,190,190}

\path[draw=drawColor,line width= 0.6pt,line join=round] (  8.25, 34.78) --
	( 55.93, 34.78);

\path[draw=drawColor,line width= 0.6pt,line join=round] (  8.25, 57.06) --
	( 55.93, 57.06);

\path[draw=drawColor,line width= 0.6pt,line join=round] (  8.25, 79.35) --
	( 55.93, 79.35);

\path[draw=drawColor,line width= 0.6pt,line join=round] (  8.25,101.63) --
	( 55.93,101.63);

\path[draw=drawColor,line width= 0.6pt,line join=round] (  8.25,123.92) --
	( 55.93,123.92);

\path[draw=drawColor,line width= 0.6pt,line join=round] (  8.25,146.20) --
	( 55.93,146.20);
\definecolor{drawColor}{RGB}{0,125,197}

\path[draw=drawColor,line width= 0.6pt,line join=round] ( 15.70, 90.49) --
	( 18.68, 90.49);

\path[draw=drawColor,line width= 0.6pt,line join=round] ( 17.19, 90.49) --
	( 17.19, 48.70);

\path[draw=drawColor,line width= 0.6pt,line join=round] ( 15.70, 48.70) --
	( 18.68, 48.70);

\path[draw=drawColor,line width= 0.6pt,line join=round] ( 30.60,104.42) --
	( 33.58,104.42);

\path[draw=drawColor,line width= 0.6pt,line join=round] ( 32.09,104.42) --
	( 32.09, 34.78);

\path[draw=drawColor,line width= 0.6pt,line join=round] ( 30.60, 34.78) --
	( 33.58, 34.78);

\path[draw=drawColor,line width= 0.6pt,line join=round] ( 45.50,146.20) --
	( 48.48,146.20);

\path[draw=drawColor,line width= 0.6pt,line join=round] ( 46.99,146.20) --
	( 46.99, 34.78);

\path[draw=drawColor,line width= 0.6pt,line join=round] ( 45.50, 34.78) --
	( 48.48, 34.78);
\definecolor{fillColor}{RGB}{0,125,197}

\path[draw=drawColor,line width= 0.4pt,line join=round,line cap=round,fill=fillColor] ( 17.19, 34.78) circle (  0.57);

\path[draw=drawColor,line width= 0.4pt,line join=round,line cap=round,fill=fillColor] ( 17.19,118.35) circle (  0.57);

\path[draw=drawColor,line width= 0.6pt,line join=round] ( 17.19, 76.56) -- ( 17.19, 90.49);

\path[draw=drawColor,line width= 0.6pt,line join=round] ( 17.19, 62.63) -- ( 17.19, 48.70);
\definecolor{fillColor}{RGB}{148,188,228}

\path[draw=drawColor,line width= 0.6pt,line join=round,line cap=round,fill=fillColor] ( 11.60, 76.56) --
	( 11.60, 62.63) --
	( 22.78, 62.63) --
	( 22.78, 76.56) --
	( 11.60, 76.56) --
	cycle;

\path[draw=drawColor,line width= 1.1pt,line join=round] ( 11.60, 62.63) -- ( 22.78, 62.63);

\path[draw=drawColor,line width= 0.6pt,line join=round] ( 32.09, 69.60) -- ( 32.09,104.42);

\path[draw=drawColor,line width= 0.6pt,line join=round] ( 32.09, 34.78) -- ( 32.09, 34.78);

\path[draw=drawColor,line width= 0.6pt,line join=round,line cap=round,fill=fillColor] ( 26.50, 69.60) --
	( 26.50, 34.78) --
	( 37.68, 34.78) --
	( 37.68, 69.60) --
	( 26.50, 69.60) --
	cycle;

\path[draw=drawColor,line width= 1.1pt,line join=round] ( 26.50, 55.67) -- ( 37.68, 55.67);

\path[draw=drawColor,line width= 0.6pt,line join=round] ( 46.99,118.35) -- ( 46.99,146.20);

\path[draw=drawColor,line width= 0.6pt,line join=round] ( 46.99, 73.08) -- ( 46.99, 34.78);

\path[draw=drawColor,line width= 0.6pt,line join=round,line cap=round,fill=fillColor] ( 41.40,118.35) --
	( 41.40, 73.08) --
	( 52.58, 73.08) --
	( 52.58,118.35) --
	( 41.40,118.35) --
	cycle;

\path[draw=drawColor,line width= 1.1pt,line join=round] ( 41.40,111.38) -- ( 52.58,111.38);

\path[draw=drawColor,line width= 0.4pt,line join=round,line cap=round] ( 14.42, 68.44) -- ( 19.96, 68.44);

\path[draw=drawColor,line width= 0.4pt,line join=round,line cap=round] ( 17.19, 65.66) -- ( 17.19, 71.21);

\path[draw=drawColor,line width= 0.4pt,line join=round,line cap=round] ( 29.31, 59.15) -- ( 34.86, 59.15);

\path[draw=drawColor,line width= 0.4pt,line join=round,line cap=round] ( 32.09, 56.38) -- ( 32.09, 61.93);

\path[draw=drawColor,line width= 0.4pt,line join=round,line cap=round] ( 44.21, 99.78) -- ( 49.76, 99.78);

\path[draw=drawColor,line width= 0.4pt,line join=round,line cap=round] ( 46.99, 97.00) -- ( 46.99,102.55);
\definecolor{drawColor}{RGB}{0,0,0}

\node[text=drawColor,anchor=base,inner sep=0pt, outer sep=0pt, scale=  0.85] at ( 32.09,164.55) {$p=.026$};

\path[draw=drawColor,line width= 0.6pt,line join=round,line cap=round] ( 17.19,160.88) -- ( 17.19,163.09);

\path[draw=drawColor,line width= 0.6pt,line join=round,line cap=round] ( 17.19,163.09) -- ( 46.99,163.09);

\path[draw=drawColor,line width= 0.6pt,line join=round,line cap=round] ( 46.99,163.09) -- ( 46.99,160.88);

\node[text=drawColor,anchor=base,inner sep=0pt, outer sep=0pt, scale=  0.85] at ( 39.54,153.41) {$p=.003$};

\path[draw=drawColor,line width= 0.6pt,line join=round,line cap=round] ( 32.09,149.74) -- ( 32.09,151.95);

\path[draw=drawColor,line width= 0.6pt,line join=round,line cap=round] ( 32.09,151.95) -- ( 46.99,151.95);

\path[draw=drawColor,line width= 0.6pt,line join=round,line cap=round] ( 46.99,151.95) -- ( 46.99,149.74);
\end{scope}
\begin{scope}
\path[clip] (  0.00,  0.00) rectangle ( 61.43,180.67);
\definecolor{drawColor}{gray}{0.20}

\path[draw=drawColor,line width= 0.6pt,line join=round] ( 17.19, 25.34) --
	( 17.19, 28.09);

\path[draw=drawColor,line width= 0.6pt,line join=round] ( 32.09, 25.34) --
	( 32.09, 28.09);

\path[draw=drawColor,line width= 0.6pt,line join=round] ( 46.99, 25.34) --
	( 46.99, 28.09);
\end{scope}
\begin{scope}
\path[clip] (  0.00,  0.00) rectangle ( 61.43,180.67);
\definecolor{drawColor}{RGB}{0,0,0}

\node[text=drawColor,anchor=base,inner sep=0pt, outer sep=0pt, scale=  0.80] at ( 17.19, 17.63) {1/3};

\node[text=drawColor,anchor=base,inner sep=0pt, outer sep=0pt, scale=  0.80] at ( 32.09, 17.63) {2/3};

\node[text=drawColor,anchor=base,inner sep=0pt, outer sep=0pt, scale=  0.80] at ( 46.99, 17.63) {1};
\end{scope}
\begin{scope}
\path[clip] (  0.00,  0.00) rectangle ( 61.43,180.67);
\definecolor{drawColor}{RGB}{0,0,0}

\node[text=drawColor,anchor=base,inner sep=0pt, outer sep=0pt, scale=  0.90] at ( 32.09,  7.32) {RR-RFF};
\end{scope}
\end{tikzpicture}
 \hspace{-0.5cm}
	\begin{tikzpicture}[x=1pt,y=1pt]
\definecolor{fillColor}{RGB}{255,255,255}
\path[use as bounding box,fill=fillColor,fill opacity=0.00] (0,0) rectangle ( 61.43,180.67);
\begin{scope}
\path[clip] (  0.00,  0.00) rectangle ( 61.43,180.67);

\path[] (  0.00,  0.00) rectangle ( 61.43,180.68);
\end{scope}
\begin{scope}
\path[clip] (  8.25, 28.09) rectangle ( 55.93,175.17);

\path[] (  8.25, 28.09) rectangle ( 55.93,175.18);
\definecolor{drawColor}{RGB}{255,255,255}

\path[draw=drawColor,line width= 0.3pt,line join=round] (  8.25, 45.92) --
	( 55.93, 45.92);

\path[draw=drawColor,line width= 0.3pt,line join=round] (  8.25, 68.20) --
	( 55.93, 68.20);

\path[draw=drawColor,line width= 0.3pt,line join=round] (  8.25, 90.49) --
	( 55.93, 90.49);

\path[draw=drawColor,line width= 0.3pt,line join=round] (  8.25,112.78) --
	( 55.93,112.78);

\path[draw=drawColor,line width= 0.3pt,line join=round] (  8.25,135.06) --
	( 55.93,135.06);

\path[draw=drawColor,line width= 0.3pt,line join=round] (  8.25,157.35) --
	( 55.93,157.35);

\path[draw=drawColor,line width= 0.3pt,line join=round] (  8.25,168.49) --
	( 55.93,168.49);
\definecolor{drawColor}{RGB}{190,190,190}

\path[draw=drawColor,line width= 0.6pt,line join=round] (  8.25, 34.78) --
	( 55.93, 34.78);

\path[draw=drawColor,line width= 0.6pt,line join=round] (  8.25, 57.06) --
	( 55.93, 57.06);

\path[draw=drawColor,line width= 0.6pt,line join=round] (  8.25, 79.35) --
	( 55.93, 79.35);

\path[draw=drawColor,line width= 0.6pt,line join=round] (  8.25,101.63) --
	( 55.93,101.63);

\path[draw=drawColor,line width= 0.6pt,line join=round] (  8.25,123.92) --
	( 55.93,123.92);

\path[draw=drawColor,line width= 0.6pt,line join=round] (  8.25,146.20) --
	( 55.93,146.20);
\definecolor{drawColor}{RGB}{0,125,197}

\path[draw=drawColor,line width= 0.6pt,line join=round] ( 15.70,146.20) --
	( 18.68,146.20);

\path[draw=drawColor,line width= 0.6pt,line join=round] ( 17.19,146.20) --
	( 17.19, 48.70);

\path[draw=drawColor,line width= 0.6pt,line join=round] ( 15.70, 48.70) --
	( 18.68, 48.70);

\path[draw=drawColor,line width= 0.6pt,line join=round] ( 30.60, 90.49) --
	( 33.58, 90.49);

\path[draw=drawColor,line width= 0.6pt,line join=round] ( 32.09, 90.49) --
	( 32.09, 34.78);

\path[draw=drawColor,line width= 0.6pt,line join=round] ( 30.60, 34.78) --
	( 33.58, 34.78);

\path[draw=drawColor,line width= 0.6pt,line join=round] ( 45.50, 90.49) --
	( 48.48, 90.49);

\path[draw=drawColor,line width= 0.6pt,line join=round] ( 46.99, 90.49) --
	( 46.99, 34.78);

\path[draw=drawColor,line width= 0.6pt,line join=round] ( 45.50, 34.78) --
	( 48.48, 34.78);

\path[draw=drawColor,line width= 0.6pt,line join=round] ( 17.19,107.90) -- ( 17.19,146.20);

\path[draw=drawColor,line width= 0.6pt,line join=round] ( 17.19, 59.15) -- ( 17.19, 48.70);
\definecolor{fillColor}{RGB}{148,188,228}

\path[draw=drawColor,line width= 0.6pt,line join=round,line cap=round,fill=fillColor] ( 11.60,107.90) --
	( 11.60, 59.15) --
	( 22.78, 59.15) --
	( 22.78,107.90) --
	( 11.60,107.90) --
	cycle;

\path[draw=drawColor,line width= 1.1pt,line join=round] ( 11.60, 76.56) -- ( 22.78, 76.56);

\path[draw=drawColor,line width= 0.6pt,line join=round] ( 32.09, 62.63) -- ( 32.09, 90.49);

\path[draw=drawColor,line width= 0.6pt,line join=round] ( 32.09, 34.78) -- ( 32.09, 34.78);

\path[draw=drawColor,line width= 0.6pt,line join=round,line cap=round,fill=fillColor] ( 26.50, 62.63) --
	( 26.50, 34.78) --
	( 37.68, 34.78) --
	( 37.68, 62.63) --
	( 26.50, 62.63) --
	cycle;

\path[draw=drawColor,line width= 1.1pt,line join=round] ( 26.50, 55.67) -- ( 37.68, 55.67);

\path[draw=drawColor,line width= 0.6pt,line join=round] ( 46.99, 80.04) -- ( 46.99, 90.49);

\path[draw=drawColor,line width= 0.6pt,line join=round] ( 46.99, 48.70) -- ( 46.99, 34.78);

\path[draw=drawColor,line width= 0.6pt,line join=round,line cap=round,fill=fillColor] ( 41.40, 80.04) --
	( 41.40, 48.70) --
	( 52.58, 48.70) --
	( 52.58, 80.04) --
	( 41.40, 80.04) --
	cycle;

\path[draw=drawColor,line width= 1.1pt,line join=round] ( 41.40, 62.63) -- ( 52.58, 62.63);

\path[draw=drawColor,line width= 0.4pt,line join=round,line cap=round] ( 14.42, 84.69) -- ( 19.96, 84.69);

\path[draw=drawColor,line width= 0.4pt,line join=round,line cap=round] ( 17.19, 81.91) -- ( 17.19, 87.46);

\path[draw=drawColor,line width= 0.4pt,line join=round,line cap=round] ( 29.31, 55.67) -- ( 34.86, 55.67);

\path[draw=drawColor,line width= 0.4pt,line join=round,line cap=round] ( 32.09, 52.89) -- ( 32.09, 58.44);

\path[draw=drawColor,line width= 0.4pt,line join=round,line cap=round] ( 44.21, 63.79) -- ( 49.76, 63.79);

\path[draw=drawColor,line width= 0.4pt,line join=round,line cap=round] ( 46.99, 61.02) -- ( 46.99, 66.57);
\definecolor{drawColor}{RGB}{0,0,0}

\node[text=drawColor,anchor=base,inner sep=0pt, outer sep=0pt, scale=  0.85] at ( 24.64,153.25) {$p=.022$};

\path[draw=drawColor,line width= 0.6pt,line join=round,line cap=round] ( 17.19,149.55) -- ( 17.19,151.78);

\path[draw=drawColor,line width= 0.6pt,line join=round,line cap=round] ( 17.19,151.78) -- ( 32.09,151.78);

\path[draw=drawColor,line width= 0.6pt,line join=round,line cap=round] ( 32.09,151.78) -- ( 32.09,149.55);
\end{scope}
\begin{scope}
\path[clip] (  0.00,  0.00) rectangle ( 61.43,180.67);
\definecolor{drawColor}{gray}{0.20}

\path[draw=drawColor,line width= 0.6pt,line join=round] ( 17.19, 25.34) --
	( 17.19, 28.09);

\path[draw=drawColor,line width= 0.6pt,line join=round] ( 32.09, 25.34) --
	( 32.09, 28.09);

\path[draw=drawColor,line width= 0.6pt,line join=round] ( 46.99, 25.34) --
	( 46.99, 28.09);
\end{scope}
\begin{scope}
\path[clip] (  0.00,  0.00) rectangle ( 61.43,180.67);
\definecolor{drawColor}{RGB}{0,0,0}

\node[text=drawColor,anchor=base,inner sep=0pt, outer sep=0pt, scale=  0.80] at ( 17.19, 17.63) {1/3};

\node[text=drawColor,anchor=base,inner sep=0pt, outer sep=0pt, scale=  0.80] at ( 32.09, 17.63) {2/3};

\node[text=drawColor,anchor=base,inner sep=0pt, outer sep=0pt, scale=  0.80] at ( 46.99, 17.63) {1};
\end{scope}
\begin{scope}
\path[clip] (  0.00,  0.00) rectangle ( 61.43,180.67);
\definecolor{drawColor}{RGB}{0,0,0}

\node[text=drawColor,anchor=base,inner sep=0pt, outer sep=0pt, scale=  0.90] at ( 32.09,  7.32) {RR};
\end{scope}
\end{tikzpicture}
 	\vspace{-0.7cm}
	\begin{center}
		\footnotesize Tested Exertion Levels per Method
	\end{center}
	\caption{Basic user study: performance of the examined methods for individual gesture exertion levels and significance from ANOVA ($\alpha=0.05$)}
	\label{fig:userstudy1_level}\end{figure}

\Cref{fig:userstudy1_level} splits the achieved success rates additionally per gesture exertion level, after first averaging the per-action-performances for each level-subject-combination. The subject- and level-based variances and medians are depicted for each of the methods. Again, kNN and DSM-kNN do not show major differences, despite the intensive dataset reduction of DSM-kNN. It is apparent that these methods performed better at higher intensity levels (median of 87.5\% for full intensity). An exertion level of $\nicefrac{2}{3}$ exhibits intermediate performance, while gestures with only $\nicefrac{1}{3}$ of intensity yielded a mean success rate of 57\% for each with high variance. In both methods, the difference between the lowest and the highest level success rate was significant.

When looking into the results from the RR-based methods, it can be observed that there is no level of intensity where those would have outperformed the kNN-based methods in median and mean of the achieved success rate. Interestingly, standard RR yielded a higher number of successes for low-intensity signal amplitudes than when incorporating Random Fourier Features. In contrast, RR-RFF performed better than RR for gestures of full intensity. For gestures of $\nicefrac{2}{3}$ exertion level, both had a similar performance, with a mean of 19\% success rate, the lowest across the intensity levels. This resulted in significance between lowest and intermediate level for RR, as well as between intermediate and highest level for RR-RFF.

At the level of $\nicefrac{1}{3}$, there was no significance between any of the methods ($\alpha=0.05$). At intermediate level, both kNN and DSM-kNN performed significantly better than RR and RR-RFF ($p<10^{-5}$). For the full intensity gestures, both kNN-based methods were significantly better than RR and RR-RFF, while also RR-RFF exposed significantly better performance than RR ($p<0.01$).

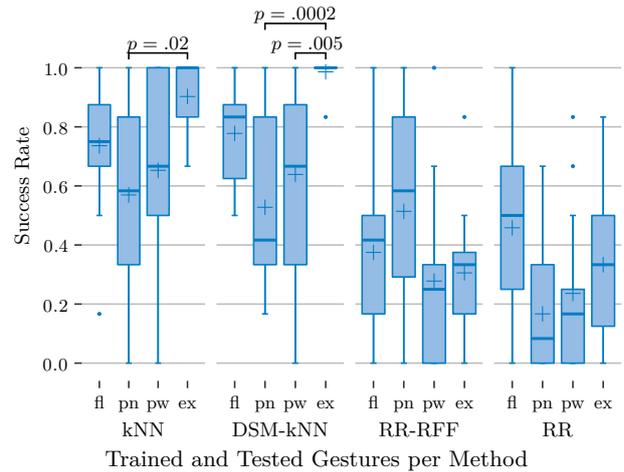
\begin{figure}[h]
	\begin{tikzpicture}[x=1pt,y=1pt]
\definecolor{fillColor}{RGB}{255,255,255}
\path[use as bounding box,fill=fillColor,fill opacity=0.00] (0,0) rectangle ( 83.11,180.67);
\begin{scope}
\path[clip] (  0.00,  0.00) rectangle ( 83.11,180.67);

\path[] ( -0.00,  0.00) rectangle ( 83.11,180.68);
\end{scope}
\begin{scope}
\path[clip] ( 31.44, 28.09) rectangle ( 77.61,175.17);

\path[] ( 31.44, 28.09) rectangle ( 77.61,175.18);
\definecolor{drawColor}{RGB}{255,255,255}

\path[draw=drawColor,line width= 0.3pt,line join=round] ( 31.44, 45.92) --
	( 77.61, 45.92);

\path[draw=drawColor,line width= 0.3pt,line join=round] ( 31.44, 68.20) --
	( 77.61, 68.20);

\path[draw=drawColor,line width= 0.3pt,line join=round] ( 31.44, 90.49) --
	( 77.61, 90.49);

\path[draw=drawColor,line width= 0.3pt,line join=round] ( 31.44,112.78) --
	( 77.61,112.78);

\path[draw=drawColor,line width= 0.3pt,line join=round] ( 31.44,135.06) --
	( 77.61,135.06);

\path[draw=drawColor,line width= 0.3pt,line join=round] ( 31.44,157.35) --
	( 77.61,157.35);

\path[draw=drawColor,line width= 0.3pt,line join=round] ( 31.44,168.49) --
	( 77.61,168.49);
\definecolor{drawColor}{RGB}{190,190,190}

\path[draw=drawColor,line width= 0.6pt,line join=round] ( 31.44, 34.78) --
	( 77.61, 34.78);

\path[draw=drawColor,line width= 0.6pt,line join=round] ( 31.44, 57.06) --
	( 77.61, 57.06);

\path[draw=drawColor,line width= 0.6pt,line join=round] ( 31.44, 79.35) --
	( 77.61, 79.35);

\path[draw=drawColor,line width= 0.6pt,line join=round] ( 31.44,101.63) --
	( 77.61,101.63);

\path[draw=drawColor,line width= 0.6pt,line join=round] ( 31.44,123.92) --
	( 77.61,123.92);

\path[draw=drawColor,line width= 0.6pt,line join=round] ( 31.44,146.20) --
	( 77.61,146.20);
\definecolor{drawColor}{RGB}{0,125,197}

\path[draw=drawColor,line width= 0.6pt,line join=round] ( 36.93,146.20) --
	( 39.13,146.20);

\path[draw=drawColor,line width= 0.6pt,line join=round] ( 38.03,146.20) --
	( 38.03, 90.49);

\path[draw=drawColor,line width= 0.6pt,line join=round] ( 36.93, 90.49) --
	( 39.13, 90.49);

\path[draw=drawColor,line width= 0.6pt,line join=round] ( 47.93,146.20) --
	( 50.13,146.20);

\path[draw=drawColor,line width= 0.6pt,line join=round] ( 49.03,146.20) --
	( 49.03, 34.78);

\path[draw=drawColor,line width= 0.6pt,line join=round] ( 47.93, 34.78) --
	( 50.13, 34.78);

\path[draw=drawColor,line width= 0.6pt,line join=round] ( 58.92,146.20) --
	( 61.12,146.20);

\path[draw=drawColor,line width= 0.6pt,line join=round] ( 60.02,146.20) --
	( 60.02, 34.78);

\path[draw=drawColor,line width= 0.6pt,line join=round] ( 58.92, 34.78) --
	( 61.12, 34.78);

\path[draw=drawColor,line width= 0.6pt,line join=round] ( 69.91,146.20) --
	( 72.11,146.20);

\path[draw=drawColor,line width= 0.6pt,line join=round] ( 71.01,146.20) --
	( 71.01,109.06);

\path[draw=drawColor,line width= 0.6pt,line join=round] ( 69.91,109.06) --
	( 72.11,109.06);
\definecolor{fillColor}{RGB}{0,125,197}

\path[draw=drawColor,line width= 0.4pt,line join=round,line cap=round,fill=fillColor] ( 38.03, 53.35) circle (  0.57);

\path[draw=drawColor,line width= 0.6pt,line join=round] ( 38.03,132.28) -- ( 38.03,146.20);

\path[draw=drawColor,line width= 0.6pt,line join=round] ( 38.03,109.06) -- ( 38.03, 90.49);
\definecolor{fillColor}{RGB}{148,188,228}

\path[draw=drawColor,line width= 0.6pt,line join=round,line cap=round,fill=fillColor] ( 33.91,132.28) --
	( 33.91,109.06) --
	( 42.16,109.06) --
	( 42.16,132.28) --
	( 33.91,132.28) --
	cycle;

\path[draw=drawColor,line width= 1.1pt,line join=round] ( 33.91,118.35) -- ( 42.16,118.35);

\path[draw=drawColor,line width= 0.6pt,line join=round] ( 49.03,127.63) -- ( 49.03,146.20);

\path[draw=drawColor,line width= 0.6pt,line join=round] ( 49.03, 71.92) -- ( 49.03, 34.78);

\path[draw=drawColor,line width= 0.6pt,line join=round,line cap=round,fill=fillColor] ( 44.90,127.63) --
	( 44.90, 71.92) --
	( 53.15, 71.92) --
	( 53.15,127.63) --
	( 44.90,127.63) --
	cycle;

\path[draw=drawColor,line width= 1.1pt,line join=round] ( 44.90, 99.78) -- ( 53.15, 99.78);

\path[draw=drawColor,line width= 0.6pt,line join=round] ( 60.02,146.20) -- ( 60.02,146.20);

\path[draw=drawColor,line width= 0.6pt,line join=round] ( 60.02, 90.49) -- ( 60.02, 34.78);

\path[draw=drawColor,line width= 0.6pt,line join=round,line cap=round,fill=fillColor] ( 55.90,146.20) --
	( 55.90, 90.49) --
	( 64.14, 90.49) --
	( 64.14,146.20) --
	( 55.90,146.20) --
	cycle;

\path[draw=drawColor,line width= 1.1pt,line join=round] ( 55.90,109.06) -- ( 64.14,109.06);

\path[draw=drawColor,line width= 0.6pt,line join=round] ( 71.01,146.20) -- ( 71.01,146.20);

\path[draw=drawColor,line width= 0.6pt,line join=round] ( 71.01,127.63) -- ( 71.01,109.06);

\path[draw=drawColor,line width= 0.6pt,line join=round,line cap=round,fill=fillColor] ( 66.89,146.20) --
	( 66.89,127.63) --
	( 75.14,127.63) --
	( 75.14,146.20) --
	( 66.89,146.20) --
	cycle;

\path[draw=drawColor,line width= 1.1pt,line join=round] ( 66.89,146.20) -- ( 75.14,146.20);

\path[draw=drawColor,line width= 0.4pt,line join=round,line cap=round] ( 35.26,116.80) -- ( 40.81,116.80);

\path[draw=drawColor,line width= 0.4pt,line join=round,line cap=round] ( 38.03,114.02) -- ( 38.03,119.57);

\path[draw=drawColor,line width= 0.4pt,line join=round,line cap=round] ( 46.25, 98.23) -- ( 51.80, 98.23);

\path[draw=drawColor,line width= 0.4pt,line join=round,line cap=round] ( 49.03, 95.45) -- ( 49.03,101.00);

\path[draw=drawColor,line width= 0.4pt,line join=round,line cap=round] ( 57.25,107.51) -- ( 62.80,107.51);

\path[draw=drawColor,line width= 0.4pt,line join=round,line cap=round] ( 60.02,104.74) -- ( 60.02,110.29);

\path[draw=drawColor,line width= 0.4pt,line join=round,line cap=round] ( 68.24,135.37) -- ( 73.79,135.37);

\path[draw=drawColor,line width= 0.4pt,line join=round,line cap=round] ( 71.01,132.60) -- ( 71.01,138.15);
\definecolor{drawColor}{RGB}{0,0,0}

\node[text=drawColor,anchor=base,inner sep=0pt, outer sep=0pt, scale=  0.85] at ( 60.02,153.25) {$p=.02$};

\path[draw=drawColor,line width= 0.6pt,line join=round,line cap=round] ( 71.01,149.55) -- ( 71.01,151.78);

\path[draw=drawColor,line width= 0.6pt,line join=round,line cap=round] ( 71.01,151.78) -- ( 49.03,151.78);

\path[draw=drawColor,line width= 0.6pt,line join=round,line cap=round] ( 49.03,151.78) -- ( 49.03,149.55);
\end{scope}
\begin{scope}
\path[clip] (  0.00,  0.00) rectangle ( 83.11,180.67);
\definecolor{drawColor}{RGB}{0,0,0}

\node[text=drawColor,anchor=base east,inner sep=0pt, outer sep=0pt, scale=  0.80] at ( 26.49, 32.02) {0.0};

\node[text=drawColor,anchor=base east,inner sep=0pt, outer sep=0pt, scale=  0.80] at ( 26.49, 54.31) {0.2};

\node[text=drawColor,anchor=base east,inner sep=0pt, outer sep=0pt, scale=  0.80] at ( 26.49, 76.59) {0.4};

\node[text=drawColor,anchor=base east,inner sep=0pt, outer sep=0pt, scale=  0.80] at ( 26.49, 98.88) {0.6};

\node[text=drawColor,anchor=base east,inner sep=0pt, outer sep=0pt, scale=  0.80] at ( 26.49,121.16) {0.8};

\node[text=drawColor,anchor=base east,inner sep=0pt, outer sep=0pt, scale=  0.80] at ( 26.49,143.45) {1.0};
\end{scope}
\begin{scope}
\path[clip] (  0.00,  0.00) rectangle ( 83.11,180.67);
\definecolor{drawColor}{gray}{0.20}

\path[draw=drawColor,line width= 0.6pt,line join=round] ( 28.69, 34.78) --
	( 31.44, 34.78);

\path[draw=drawColor,line width= 0.6pt,line join=round] ( 28.69, 57.06) --
	( 31.44, 57.06);

\path[draw=drawColor,line width= 0.6pt,line join=round] ( 28.69, 79.35) --
	( 31.44, 79.35);

\path[draw=drawColor,line width= 0.6pt,line join=round] ( 28.69,101.63) --
	( 31.44,101.63);

\path[draw=drawColor,line width= 0.6pt,line join=round] ( 28.69,123.92) --
	( 31.44,123.92);

\path[draw=drawColor,line width= 0.6pt,line join=round] ( 28.69,146.20) --
	( 31.44,146.20);
\end{scope}
\begin{scope}
\path[clip] (  0.00,  0.00) rectangle ( 83.11,180.67);
\definecolor{drawColor}{gray}{0.20}

\path[draw=drawColor,line width= 0.6pt,line join=round] ( 38.03, 25.34) --
	( 38.03, 28.09);

\path[draw=drawColor,line width= 0.6pt,line join=round] ( 49.03, 25.34) --
	( 49.03, 28.09);

\path[draw=drawColor,line width= 0.6pt,line join=round] ( 60.02, 25.34) --
	( 60.02, 28.09);

\path[draw=drawColor,line width= 0.6pt,line join=round] ( 71.01, 25.34) --
	( 71.01, 28.09);
\end{scope}
\begin{scope}
\path[clip] (  0.00,  0.00) rectangle ( 83.11,180.67);
\definecolor{drawColor}{RGB}{0,0,0}

\node[text=drawColor,anchor=base,inner sep=0pt, outer sep=0pt, scale=  0.80] at ( 38.03, 17.63) {fl};

\node[text=drawColor,anchor=base,inner sep=0pt, outer sep=0pt, scale=  0.80] at ( 49.03, 17.63) {pn};

\node[text=drawColor,anchor=base,inner sep=0pt, outer sep=0pt, scale=  0.80] at ( 60.02, 17.63) {pw};

\node[text=drawColor,anchor=base,inner sep=0pt, outer sep=0pt, scale=  0.80] at ( 71.01, 17.63) {ex};
\end{scope}
\begin{scope}
\path[clip] (  0.00,  0.00) rectangle ( 83.11,180.67);
\definecolor{drawColor}{RGB}{0,0,0}

\node[text=drawColor,anchor=base,inner sep=0pt, outer sep=0pt, scale=  0.90] at ( 54.52,  7.32) {kNN};
\end{scope}
\begin{scope}
\path[clip] (  0.00,  0.00) rectangle ( 83.11,180.67);
\definecolor{drawColor}{RGB}{0,0,0}

\node[text=drawColor,rotate= 90.00,anchor=base,inner sep=0pt, outer sep=0pt, scale=  0.90] at ( 11.70,101.63) {Success Rate};
\end{scope}
\end{tikzpicture}
 \hspace{-0.5cm}
	\begin{tikzpicture}[x=1pt,y=1pt]
\definecolor{fillColor}{RGB}{255,255,255}
\path[use as bounding box,fill=fillColor,fill opacity=0.00] (0,0) rectangle ( 61.43,180.67);
\begin{scope}
\path[clip] (  0.00,  0.00) rectangle ( 61.43,180.67);

\path[] (  0.00,  0.00) rectangle ( 61.43,180.68);
\end{scope}
\begin{scope}
\path[clip] (  8.25, 28.09) rectangle ( 55.93,175.17);

\path[] (  8.25, 28.09) rectangle ( 55.93,175.18);
\definecolor{drawColor}{RGB}{255,255,255}

\path[draw=drawColor,line width= 0.3pt,line join=round] (  8.25, 45.92) --
	( 55.93, 45.92);

\path[draw=drawColor,line width= 0.3pt,line join=round] (  8.25, 68.20) --
	( 55.93, 68.20);

\path[draw=drawColor,line width= 0.3pt,line join=round] (  8.25, 90.49) --
	( 55.93, 90.49);

\path[draw=drawColor,line width= 0.3pt,line join=round] (  8.25,112.78) --
	( 55.93,112.78);

\path[draw=drawColor,line width= 0.3pt,line join=round] (  8.25,135.06) --
	( 55.93,135.06);

\path[draw=drawColor,line width= 0.3pt,line join=round] (  8.25,157.35) --
	( 55.93,157.35);

\path[draw=drawColor,line width= 0.3pt,line join=round] (  8.25,168.49) --
	( 55.93,168.49);
\definecolor{drawColor}{RGB}{190,190,190}

\path[draw=drawColor,line width= 0.6pt,line join=round] (  8.25, 34.78) --
	( 55.93, 34.78);

\path[draw=drawColor,line width= 0.6pt,line join=round] (  8.25, 57.06) --
	( 55.93, 57.06);

\path[draw=drawColor,line width= 0.6pt,line join=round] (  8.25, 79.35) --
	( 55.93, 79.35);

\path[draw=drawColor,line width= 0.6pt,line join=round] (  8.25,101.63) --
	( 55.93,101.63);

\path[draw=drawColor,line width= 0.6pt,line join=round] (  8.25,123.92) --
	( 55.93,123.92);

\path[draw=drawColor,line width= 0.6pt,line join=round] (  8.25,146.20) --
	( 55.93,146.20);
\definecolor{drawColor}{RGB}{0,125,197}

\path[draw=drawColor,line width= 0.6pt,line join=round] ( 13.93,146.20) --
	( 16.20,146.20);

\path[draw=drawColor,line width= 0.6pt,line join=round] ( 15.06,146.20) --
	( 15.06, 90.49);

\path[draw=drawColor,line width= 0.6pt,line join=round] ( 13.93, 90.49) --
	( 16.20, 90.49);

\path[draw=drawColor,line width= 0.6pt,line join=round] ( 25.28,146.20) --
	( 27.55,146.20);

\path[draw=drawColor,line width= 0.6pt,line join=round] ( 26.41,146.20) --
	( 26.41, 53.35);

\path[draw=drawColor,line width= 0.6pt,line join=round] ( 25.28, 53.35) --
	( 27.55, 53.35);

\path[draw=drawColor,line width= 0.6pt,line join=round] ( 36.63,146.20) --
	( 38.90,146.20);

\path[draw=drawColor,line width= 0.6pt,line join=round] ( 37.77,146.20) --
	( 37.77, 34.78);

\path[draw=drawColor,line width= 0.6pt,line join=round] ( 36.63, 34.78) --
	( 38.90, 34.78);

\path[draw=drawColor,line width= 0.6pt,line join=round] ( 47.98,146.20) --
	( 50.25,146.20);

\path[draw=drawColor,line width= 0.6pt,line join=round] ( 49.12,146.20) --
	( 49.12,146.20);

\path[draw=drawColor,line width= 0.6pt,line join=round] ( 47.98,146.20) --
	( 50.25,146.20);

\path[draw=drawColor,line width= 0.6pt,line join=round] ( 15.06,132.28) -- ( 15.06,146.20);

\path[draw=drawColor,line width= 0.6pt,line join=round] ( 15.06,104.42) -- ( 15.06, 90.49);
\definecolor{fillColor}{RGB}{148,188,228}

\path[draw=drawColor,line width= 0.6pt,line join=round,line cap=round,fill=fillColor] ( 10.80,132.28) --
	( 10.80,104.42) --
	( 19.32,104.42) --
	( 19.32,132.28) --
	( 10.80,132.28) --
	cycle;

\path[draw=drawColor,line width= 1.1pt,line join=round] ( 10.80,127.63) -- ( 19.32,127.63);

\path[draw=drawColor,line width= 0.6pt,line join=round] ( 26.41,127.63) -- ( 26.41,146.20);

\path[draw=drawColor,line width= 0.6pt,line join=round] ( 26.41, 71.92) -- ( 26.41, 53.35);

\path[draw=drawColor,line width= 0.6pt,line join=round,line cap=round,fill=fillColor] ( 22.16,127.63) --
	( 22.16, 71.92) --
	( 30.67, 71.92) --
	( 30.67,127.63) --
	( 22.16,127.63) --
	cycle;

\path[draw=drawColor,line width= 1.1pt,line join=round] ( 22.16, 81.20) -- ( 30.67, 81.20);

\path[draw=drawColor,line width= 0.6pt,line join=round] ( 37.77,132.28) -- ( 37.77,146.20);

\path[draw=drawColor,line width= 0.6pt,line join=round] ( 37.77, 71.92) -- ( 37.77, 34.78);

\path[draw=drawColor,line width= 0.6pt,line join=round,line cap=round,fill=fillColor] ( 33.51,132.28) --
	( 33.51, 71.92) --
	( 42.02, 71.92) --
	( 42.02,132.28) --
	( 33.51,132.28) --
	cycle;

\path[draw=drawColor,line width= 1.1pt,line join=round] ( 33.51,109.06) -- ( 42.02,109.06);
\definecolor{fillColor}{RGB}{0,125,197}

\path[draw=drawColor,line width= 0.4pt,line join=round,line cap=round,fill=fillColor] ( 49.12,127.63) circle (  0.57);

\path[draw=drawColor,line width= 0.6pt,line join=round] ( 49.12,146.20) -- ( 49.12,146.20);

\path[draw=drawColor,line width= 0.6pt,line join=round] ( 49.12,146.20) -- ( 49.12,146.20);
\definecolor{fillColor}{RGB}{148,188,228}

\path[draw=drawColor,line width= 0.6pt,line join=round,line cap=round,fill=fillColor] ( 44.86,146.20) --
	( 44.86,146.20) --
	( 53.38,146.20) --
	( 53.38,146.20) --
	( 44.86,146.20) --
	cycle;

\path[draw=drawColor,line width= 1.1pt,line join=round] ( 44.86,146.20) -- ( 53.38,146.20);

\path[draw=drawColor,line width= 0.4pt,line join=round,line cap=round] ( 12.29,121.44) -- ( 17.84,121.44);

\path[draw=drawColor,line width= 0.4pt,line join=round,line cap=round] ( 15.06,118.67) -- ( 15.06,124.22);

\path[draw=drawColor,line width= 0.4pt,line join=round,line cap=round] ( 23.64, 93.58) -- ( 29.19, 93.58);

\path[draw=drawColor,line width= 0.4pt,line join=round,line cap=round] ( 26.41, 90.81) -- ( 26.41, 96.36);

\path[draw=drawColor,line width= 0.4pt,line join=round,line cap=round] ( 34.99,105.97) -- ( 40.54,105.97);

\path[draw=drawColor,line width= 0.4pt,line join=round,line cap=round] ( 37.77,103.19) -- ( 37.77,108.74);

\path[draw=drawColor,line width= 0.4pt,line join=round,line cap=round] ( 46.34,144.66) -- ( 51.89,144.66);

\path[draw=drawColor,line width= 0.4pt,line join=round,line cap=round] ( 49.12,141.88) -- ( 49.12,147.43);
\definecolor{drawColor}{RGB}{0,0,0}

\node[text=drawColor,anchor=base,inner sep=0pt, outer sep=0pt, scale=  0.85] at ( 37.77,164.39) {$p=.0002$};

\path[draw=drawColor,line width= 0.6pt,line join=round,line cap=round] ( 49.12,160.69) -- ( 49.12,162.92);

\path[draw=drawColor,line width= 0.6pt,line join=round,line cap=round] ( 49.12,162.92) -- ( 26.41,162.92);

\path[draw=drawColor,line width= 0.6pt,line join=round,line cap=round] ( 26.41,162.92) -- ( 26.41,160.69);

\node[text=drawColor,anchor=base,inner sep=0pt, outer sep=0pt, scale=  0.85] at ( 43.44,153.25) {$p=.005$ ~};

\path[draw=drawColor,line width= 0.6pt,line join=round,line cap=round] ( 49.12,149.55) -- ( 49.12,151.78);

\path[draw=drawColor,line width= 0.6pt,line join=round,line cap=round] ( 49.12,151.78) -- ( 37.77,151.78);

\path[draw=drawColor,line width= 0.6pt,line join=round,line cap=round] ( 37.77,151.78) -- ( 37.77,149.55);
\end{scope}
\begin{scope}
\path[clip] (  0.00,  0.00) rectangle ( 61.43,180.67);
\definecolor{drawColor}{gray}{0.20}

\path[draw=drawColor,line width= 0.6pt,line join=round] ( 15.06, 25.34) --
	( 15.06, 28.09);

\path[draw=drawColor,line width= 0.6pt,line join=round] ( 26.41, 25.34) --
	( 26.41, 28.09);

\path[draw=drawColor,line width= 0.6pt,line join=round] ( 37.77, 25.34) --
	( 37.77, 28.09);

\path[draw=drawColor,line width= 0.6pt,line join=round] ( 49.12, 25.34) --
	( 49.12, 28.09);
\end{scope}
\begin{scope}
\path[clip] (  0.00,  0.00) rectangle ( 61.43,180.67);
\definecolor{drawColor}{RGB}{0,0,0}

\node[text=drawColor,anchor=base,inner sep=0pt, outer sep=0pt, scale=  0.80] at ( 15.06, 17.63) {fl};

\node[text=drawColor,anchor=base,inner sep=0pt, outer sep=0pt, scale=  0.80] at ( 26.41, 17.63) {pn};

\node[text=drawColor,anchor=base,inner sep=0pt, outer sep=0pt, scale=  0.80] at ( 37.77, 17.63) {pw};

\node[text=drawColor,anchor=base,inner sep=0pt, outer sep=0pt, scale=  0.80] at ( 49.12, 17.63) {ex};
\end{scope}
\begin{scope}
\path[clip] (  0.00,  0.00) rectangle ( 61.43,180.67);
\definecolor{drawColor}{RGB}{0,0,0}

\node[text=drawColor,anchor=base,inner sep=0pt, outer sep=0pt, scale=  0.90] at ( 32.09,  7.32) {DSM-kNN};
\end{scope}
\end{tikzpicture}
 \hspace{-0.5cm}
	\begin{tikzpicture}[x=1pt,y=1pt]
\definecolor{fillColor}{RGB}{255,255,255}
\path[use as bounding box,fill=fillColor,fill opacity=0.00] (0,0) rectangle ( 61.43,180.67);
\begin{scope}
\path[clip] (  0.00,  0.00) rectangle ( 61.43,180.67);

\path[] (  0.00,  0.00) rectangle ( 61.43,180.68);
\end{scope}
\begin{scope}
\path[clip] (  8.25, 28.09) rectangle ( 55.93,175.17);

\path[] (  8.25, 28.09) rectangle ( 55.93,175.18);
\definecolor{drawColor}{RGB}{255,255,255}

\path[draw=drawColor,line width= 0.3pt,line join=round] (  8.25, 45.92) --
	( 55.93, 45.92);

\path[draw=drawColor,line width= 0.3pt,line join=round] (  8.25, 68.20) --
	( 55.93, 68.20);

\path[draw=drawColor,line width= 0.3pt,line join=round] (  8.25, 90.49) --
	( 55.93, 90.49);

\path[draw=drawColor,line width= 0.3pt,line join=round] (  8.25,112.78) --
	( 55.93,112.78);

\path[draw=drawColor,line width= 0.3pt,line join=round] (  8.25,135.06) --
	( 55.93,135.06);

\path[draw=drawColor,line width= 0.3pt,line join=round] (  8.25,157.35) --
	( 55.93,157.35);

\path[draw=drawColor,line width= 0.3pt,line join=round] (  8.25,168.49) --
	( 55.93,168.49);
\definecolor{drawColor}{RGB}{190,190,190}

\path[draw=drawColor,line width= 0.6pt,line join=round] (  8.25, 34.78) --
	( 55.93, 34.78);

\path[draw=drawColor,line width= 0.6pt,line join=round] (  8.25, 57.06) --
	( 55.93, 57.06);

\path[draw=drawColor,line width= 0.6pt,line join=round] (  8.25, 79.35) --
	( 55.93, 79.35);

\path[draw=drawColor,line width= 0.6pt,line join=round] (  8.25,101.63) --
	( 55.93,101.63);

\path[draw=drawColor,line width= 0.6pt,line join=round] (  8.25,123.92) --
	( 55.93,123.92);

\path[draw=drawColor,line width= 0.6pt,line join=round] (  8.25,146.20) --
	( 55.93,146.20);
\definecolor{drawColor}{RGB}{0,125,197}

\path[draw=drawColor,line width= 0.6pt,line join=round] ( 13.93,146.20) --
	( 16.20,146.20);

\path[draw=drawColor,line width= 0.6pt,line join=round] ( 15.06,146.20) --
	( 15.06, 34.78);

\path[draw=drawColor,line width= 0.6pt,line join=round] ( 13.93, 34.78) --
	( 16.20, 34.78);

\path[draw=drawColor,line width= 0.6pt,line join=round] ( 25.28,146.20) --
	( 27.55,146.20);

\path[draw=drawColor,line width= 0.6pt,line join=round] ( 26.41,146.20) --
	( 26.41, 34.78);

\path[draw=drawColor,line width= 0.6pt,line join=round] ( 25.28, 34.78) --
	( 27.55, 34.78);

\path[draw=drawColor,line width= 0.6pt,line join=round] ( 36.63,109.06) --
	( 38.90,109.06);

\path[draw=drawColor,line width= 0.6pt,line join=round] ( 37.77,109.06) --
	( 37.77, 34.78);

\path[draw=drawColor,line width= 0.6pt,line join=round] ( 36.63, 34.78) --
	( 38.90, 34.78);

\path[draw=drawColor,line width= 0.6pt,line join=round] ( 47.98, 90.49) --
	( 50.25, 90.49);

\path[draw=drawColor,line width= 0.6pt,line join=round] ( 49.12, 90.49) --
	( 49.12, 34.78);

\path[draw=drawColor,line width= 0.6pt,line join=round] ( 47.98, 34.78) --
	( 50.25, 34.78);

\path[draw=drawColor,line width= 0.6pt,line join=round] ( 15.06, 90.49) -- ( 15.06,146.20);

\path[draw=drawColor,line width= 0.6pt,line join=round] ( 15.06, 53.35) -- ( 15.06, 34.78);
\definecolor{fillColor}{RGB}{148,188,228}

\path[draw=drawColor,line width= 0.6pt,line join=round,line cap=round,fill=fillColor] ( 10.80, 90.49) --
	( 10.80, 53.35) --
	( 19.32, 53.35) --
	( 19.32, 90.49) --
	( 10.80, 90.49) --
	cycle;

\path[draw=drawColor,line width= 1.1pt,line join=round] ( 10.80, 81.20) -- ( 19.32, 81.20);

\path[draw=drawColor,line width= 0.6pt,line join=round] ( 26.41,127.63) -- ( 26.41,146.20);

\path[draw=drawColor,line width= 0.6pt,line join=round] ( 26.41, 67.28) -- ( 26.41, 34.78);

\path[draw=drawColor,line width= 0.6pt,line join=round,line cap=round,fill=fillColor] ( 22.16,127.63) --
	( 22.16, 67.28) --
	( 30.67, 67.28) --
	( 30.67,127.63) --
	( 22.16,127.63) --
	cycle;

\path[draw=drawColor,line width= 1.1pt,line join=round] ( 22.16, 99.78) -- ( 30.67, 99.78);
\definecolor{fillColor}{RGB}{0,125,197}

\path[draw=drawColor,line width= 0.4pt,line join=round,line cap=round,fill=fillColor] ( 37.77,146.20) circle (  0.57);

\path[draw=drawColor,line width= 0.6pt,line join=round] ( 37.77, 71.92) -- ( 37.77,109.06);

\path[draw=drawColor,line width= 0.6pt,line join=round] ( 37.77, 34.78) -- ( 37.77, 34.78);
\definecolor{fillColor}{RGB}{148,188,228}

\path[draw=drawColor,line width= 0.6pt,line join=round,line cap=round,fill=fillColor] ( 33.51, 71.92) --
	( 33.51, 34.78) --
	( 42.02, 34.78) --
	( 42.02, 71.92) --
	( 33.51, 71.92) --
	cycle;

\path[draw=drawColor,line width= 1.1pt,line join=round] ( 33.51, 62.63) -- ( 42.02, 62.63);
\definecolor{fillColor}{RGB}{0,125,197}

\path[draw=drawColor,line width= 0.4pt,line join=round,line cap=round,fill=fillColor] ( 49.12,127.63) circle (  0.57);

\path[draw=drawColor,line width= 0.6pt,line join=round] ( 49.12, 76.56) -- ( 49.12, 90.49);

\path[draw=drawColor,line width= 0.6pt,line join=round] ( 49.12, 53.35) -- ( 49.12, 34.78);
\definecolor{fillColor}{RGB}{148,188,228}

\path[draw=drawColor,line width= 0.6pt,line join=round,line cap=round,fill=fillColor] ( 44.86, 76.56) --
	( 44.86, 53.35) --
	( 53.38, 53.35) --
	( 53.38, 76.56) --
	( 44.86, 76.56) --
	cycle;

\path[draw=drawColor,line width= 1.1pt,line join=round] ( 44.86, 71.92) -- ( 53.38, 71.92);

\path[draw=drawColor,line width= 0.4pt,line join=round,line cap=round] ( 12.29, 76.56) -- ( 17.84, 76.56);

\path[draw=drawColor,line width= 0.4pt,line join=round,line cap=round] ( 15.06, 73.79) -- ( 15.06, 79.34);

\path[draw=drawColor,line width= 0.4pt,line join=round,line cap=round] ( 23.64, 92.04) -- ( 29.19, 92.04);

\path[draw=drawColor,line width= 0.4pt,line join=round,line cap=round] ( 26.41, 89.26) -- ( 26.41, 94.81);

\path[draw=drawColor,line width= 0.4pt,line join=round,line cap=round] ( 34.99, 65.73) -- ( 40.54, 65.73);

\path[draw=drawColor,line width= 0.4pt,line join=round,line cap=round] ( 37.77, 62.95) -- ( 37.77, 68.50);

\path[draw=drawColor,line width= 0.4pt,line join=round,line cap=round] ( 46.34, 68.82) -- ( 51.89, 68.82);

\path[draw=drawColor,line width= 0.4pt,line join=round,line cap=round] ( 49.12, 66.05) -- ( 49.12, 71.60);
\end{scope}
\begin{scope}
\path[clip] (  0.00,  0.00) rectangle ( 61.43,180.67);
\definecolor{drawColor}{gray}{0.20}

\path[draw=drawColor,line width= 0.6pt,line join=round] ( 15.06, 25.34) --
	( 15.06, 28.09);

\path[draw=drawColor,line width= 0.6pt,line join=round] ( 26.41, 25.34) --
	( 26.41, 28.09);

\path[draw=drawColor,line width= 0.6pt,line join=round] ( 37.77, 25.34) --
	( 37.77, 28.09);

\path[draw=drawColor,line width= 0.6pt,line join=round] ( 49.12, 25.34) --
	( 49.12, 28.09);
\end{scope}
\begin{scope}
\path[clip] (  0.00,  0.00) rectangle ( 61.43,180.67);
\definecolor{drawColor}{RGB}{0,0,0}

\node[text=drawColor,anchor=base,inner sep=0pt, outer sep=0pt, scale=  0.80] at ( 15.06, 17.63) {fl};

\node[text=drawColor,anchor=base,inner sep=0pt, outer sep=0pt, scale=  0.80] at ( 26.41, 17.63) {pn};

\node[text=drawColor,anchor=base,inner sep=0pt, outer sep=0pt, scale=  0.80] at ( 37.77, 17.63) {pw};

\node[text=drawColor,anchor=base,inner sep=0pt, outer sep=0pt, scale=  0.80] at ( 49.12, 17.63) {ex};
\end{scope}
\begin{scope}
\path[clip] (  0.00,  0.00) rectangle ( 61.43,180.67);
\definecolor{drawColor}{RGB}{0,0,0}

\node[text=drawColor,anchor=base,inner sep=0pt, outer sep=0pt, scale=  0.90] at ( 32.09,  7.32) {RR-RFF};
\end{scope}
\end{tikzpicture}
 \hspace{-0.5cm}
	\begin{tikzpicture}[x=1pt,y=1pt]
\definecolor{fillColor}{RGB}{255,255,255}
\path[use as bounding box,fill=fillColor,fill opacity=0.00] (0,0) rectangle ( 61.43,180.67);
\begin{scope}
\path[clip] (  0.00,  0.00) rectangle ( 61.43,180.67);

\path[] (  0.00,  0.00) rectangle ( 61.43,180.68);
\end{scope}
\begin{scope}
\path[clip] (  8.25, 28.09) rectangle ( 55.93,175.17);

\path[] (  8.25, 28.09) rectangle ( 55.93,175.18);
\definecolor{drawColor}{RGB}{255,255,255}

\path[draw=drawColor,line width= 0.3pt,line join=round] (  8.25, 45.92) --
	( 55.93, 45.92);

\path[draw=drawColor,line width= 0.3pt,line join=round] (  8.25, 68.20) --
	( 55.93, 68.20);

\path[draw=drawColor,line width= 0.3pt,line join=round] (  8.25, 90.49) --
	( 55.93, 90.49);

\path[draw=drawColor,line width= 0.3pt,line join=round] (  8.25,112.78) --
	( 55.93,112.78);

\path[draw=drawColor,line width= 0.3pt,line join=round] (  8.25,135.06) --
	( 55.93,135.06);

\path[draw=drawColor,line width= 0.3pt,line join=round] (  8.25,157.35) --
	( 55.93,157.35);

\path[draw=drawColor,line width= 0.3pt,line join=round] (  8.25,168.49) --
	( 55.93,168.49);
\definecolor{drawColor}{RGB}{190,190,190}

\path[draw=drawColor,line width= 0.6pt,line join=round] (  8.25, 34.78) --
	( 55.93, 34.78);

\path[draw=drawColor,line width= 0.6pt,line join=round] (  8.25, 57.06) --
	( 55.93, 57.06);

\path[draw=drawColor,line width= 0.6pt,line join=round] (  8.25, 79.35) --
	( 55.93, 79.35);

\path[draw=drawColor,line width= 0.6pt,line join=round] (  8.25,101.63) --
	( 55.93,101.63);

\path[draw=drawColor,line width= 0.6pt,line join=round] (  8.25,123.92) --
	( 55.93,123.92);

\path[draw=drawColor,line width= 0.6pt,line join=round] (  8.25,146.20) --
	( 55.93,146.20);
\definecolor{drawColor}{RGB}{0,125,197}

\path[draw=drawColor,line width= 0.6pt,line join=round] ( 13.93,146.20) --
	( 16.20,146.20);

\path[draw=drawColor,line width= 0.6pt,line join=round] ( 15.06,146.20) --
	( 15.06, 34.78);

\path[draw=drawColor,line width= 0.6pt,line join=round] ( 13.93, 34.78) --
	( 16.20, 34.78);

\path[draw=drawColor,line width= 0.6pt,line join=round] ( 25.28,109.06) --
	( 27.55,109.06);

\path[draw=drawColor,line width= 0.6pt,line join=round] ( 26.41,109.06) --
	( 26.41, 34.78);

\path[draw=drawColor,line width= 0.6pt,line join=round] ( 25.28, 34.78) --
	( 27.55, 34.78);

\path[draw=drawColor,line width= 0.6pt,line join=round] ( 36.63, 90.49) --
	( 38.90, 90.49);

\path[draw=drawColor,line width= 0.6pt,line join=round] ( 37.77, 90.49) --
	( 37.77, 34.78);

\path[draw=drawColor,line width= 0.6pt,line join=round] ( 36.63, 34.78) --
	( 38.90, 34.78);

\path[draw=drawColor,line width= 0.6pt,line join=round] ( 47.98,127.63) --
	( 50.25,127.63);

\path[draw=drawColor,line width= 0.6pt,line join=round] ( 49.12,127.63) --
	( 49.12, 34.78);

\path[draw=drawColor,line width= 0.6pt,line join=round] ( 47.98, 34.78) --
	( 50.25, 34.78);

\path[draw=drawColor,line width= 0.6pt,line join=round] ( 15.06,109.06) -- ( 15.06,146.20);

\path[draw=drawColor,line width= 0.6pt,line join=round] ( 15.06, 62.63) -- ( 15.06, 34.78);
\definecolor{fillColor}{RGB}{148,188,228}

\path[draw=drawColor,line width= 0.6pt,line join=round,line cap=round,fill=fillColor] ( 10.80,109.06) --
	( 10.80, 62.63) --
	( 19.32, 62.63) --
	( 19.32,109.06) --
	( 10.80,109.06) --
	cycle;

\path[draw=drawColor,line width= 1.1pt,line join=round] ( 10.80, 90.49) -- ( 19.32, 90.49);

\path[draw=drawColor,line width= 0.6pt,line join=round] ( 26.41, 71.92) -- ( 26.41,109.06);

\path[draw=drawColor,line width= 0.6pt,line join=round] ( 26.41, 34.78) -- ( 26.41, 34.78);

\path[draw=drawColor,line width= 0.6pt,line join=round,line cap=round,fill=fillColor] ( 22.16, 71.92) --
	( 22.16, 34.78) --
	( 30.67, 34.78) --
	( 30.67, 71.92) --
	( 22.16, 71.92) --
	cycle;

\path[draw=drawColor,line width= 1.1pt,line join=round] ( 22.16, 44.06) -- ( 30.67, 44.06);
\definecolor{fillColor}{RGB}{0,125,197}

\path[draw=drawColor,line width= 0.4pt,line join=round,line cap=round,fill=fillColor] ( 37.77,109.06) circle (  0.57);

\path[draw=drawColor,line width= 0.4pt,line join=round,line cap=round,fill=fillColor] ( 37.77,127.63) circle (  0.57);

\path[draw=drawColor,line width= 0.6pt,line join=round] ( 37.77, 62.63) -- ( 37.77, 90.49);

\path[draw=drawColor,line width= 0.6pt,line join=round] ( 37.77, 34.78) -- ( 37.77, 34.78);
\definecolor{fillColor}{RGB}{148,188,228}

\path[draw=drawColor,line width= 0.6pt,line join=round,line cap=round,fill=fillColor] ( 33.51, 62.63) --
	( 33.51, 34.78) --
	( 42.02, 34.78) --
	( 42.02, 62.63) --
	( 33.51, 62.63) --
	cycle;

\path[draw=drawColor,line width= 1.1pt,line join=round] ( 33.51, 53.35) -- ( 42.02, 53.35);

\path[draw=drawColor,line width= 0.6pt,line join=round] ( 49.12, 90.49) -- ( 49.12,127.63);

\path[draw=drawColor,line width= 0.6pt,line join=round] ( 49.12, 48.70) -- ( 49.12, 34.78);

\path[draw=drawColor,line width= 0.6pt,line join=round,line cap=round,fill=fillColor] ( 44.86, 90.49) --
	( 44.86, 48.70) --
	( 53.38, 48.70) --
	( 53.38, 90.49) --
	( 44.86, 90.49) --
	cycle;

\path[draw=drawColor,line width= 1.1pt,line join=round] ( 44.86, 71.92) -- ( 53.38, 71.92);

\path[draw=drawColor,line width= 0.4pt,line join=round,line cap=round] ( 12.29, 85.85) -- ( 17.84, 85.85);

\path[draw=drawColor,line width= 0.4pt,line join=round,line cap=round] ( 15.06, 83.07) -- ( 15.06, 88.62);

\path[draw=drawColor,line width= 0.4pt,line join=round,line cap=round] ( 23.64, 53.35) -- ( 29.19, 53.35);

\path[draw=drawColor,line width= 0.4pt,line join=round,line cap=round] ( 26.41, 50.57) -- ( 26.41, 56.12);

\path[draw=drawColor,line width= 0.4pt,line join=round,line cap=round] ( 34.99, 61.09) -- ( 40.54, 61.09);

\path[draw=drawColor,line width= 0.4pt,line join=round,line cap=round] ( 37.77, 58.31) -- ( 37.77, 63.86);

\path[draw=drawColor,line width= 0.4pt,line join=round,line cap=round] ( 46.34, 71.92) -- ( 51.89, 71.92);

\path[draw=drawColor,line width= 0.4pt,line join=round,line cap=round] ( 49.12, 69.14) -- ( 49.12, 74.69);
\end{scope}
\begin{scope}
\path[clip] (  0.00,  0.00) rectangle ( 61.43,180.67);
\definecolor{drawColor}{gray}{0.20}

\path[draw=drawColor,line width= 0.6pt,line join=round] ( 15.06, 25.34) --
	( 15.06, 28.09);

\path[draw=drawColor,line width= 0.6pt,line join=round] ( 26.41, 25.34) --
	( 26.41, 28.09);

\path[draw=drawColor,line width= 0.6pt,line join=round] ( 37.77, 25.34) --
	( 37.77, 28.09);

\path[draw=drawColor,line width= 0.6pt,line join=round] ( 49.12, 25.34) --
	( 49.12, 28.09);
\end{scope}
\begin{scope}
\path[clip] (  0.00,  0.00) rectangle ( 61.43,180.67);
\definecolor{drawColor}{RGB}{0,0,0}

\node[text=drawColor,anchor=base,inner sep=0pt, outer sep=0pt, scale=  0.80] at ( 15.06, 17.63) {fl};

\node[text=drawColor,anchor=base,inner sep=0pt, outer sep=0pt, scale=  0.80] at ( 26.41, 17.63) {pn};

\node[text=drawColor,anchor=base,inner sep=0pt, outer sep=0pt, scale=  0.80] at ( 37.77, 17.63) {pw};

\node[text=drawColor,anchor=base,inner sep=0pt, outer sep=0pt, scale=  0.80] at ( 49.12, 17.63) {ex};
\end{scope}
\begin{scope}
\path[clip] (  0.00,  0.00) rectangle ( 61.43,180.67);
\definecolor{drawColor}{RGB}{0,0,0}

\node[text=drawColor,anchor=base,inner sep=0pt, outer sep=0pt, scale=  0.90] at ( 32.09,  7.32) {RR};
\end{scope}
\end{tikzpicture}
 	\vspace{-0.7cm}
	\begin{center}
		\footnotesize Trained and Tested Gestures per Method
	\end{center}
	\caption{Basic user study: performance of the examined methods for individual gestures and significance from ANOVA ($\alpha=0.05$)}
	\label{fig:userstudy1_gestures}\end{figure}

The relation between individual gestures and success rate is presented in \cref{fig:userstudy1_gestures}, basing on first averaging the per-level-performances for each action-subject-combination. The subject- and action-based variances and medians are depicted for each of the methods, again. It is noticeable that the performance trends were similar for kNN and reduced kNN. For them, the best success rates could be achieved for wrist extension (median of 100\% for both, mean of 90\% for kNN and 99\%for DSM-kNN with small standard deviation).

Wrist flexion was the second best detected gesture for the kNN-based methods (about 76\% mean for both), followed by power grasp (67\% median), and concluded by the pointing gesture with the worst performance (about 55\% mean).

For both kNN and DSM-kNN, the performance difference between wrist extension and pointing was significant. For DSM-kNN, also the comparison of wrist extension and power grasp yielded significance.

While RR exposed the same tendency of gesture performances as the kNN-based methods (on a lower baseline), for RR-RFF the pointing gesture yielded the best success rate on average (median 58\%, mean 51\%). Interestingly, wrist extension exposed the worst performance of gestures for RR-RFF (33\% median, 31\% mean). Wrist flexion and power grasp revealed the same tendency as described for the other methods. For the RR-based methods, no significance could be shown between different gestures.

With regard to the individual gestures, the group of kNN-based methods performed significantly better than the RR-group for power grasp ($p<0.05$) as well as wrist extension ($p<10^{-6}$). For wrist flexion, the same holds ($p<0.05$) with the exception of the difference between standard kNN and RR not being significant ($\alpha=0.05$). Concerning the pointing gesture, the kNN-based schemes as well as RR-RFF performed better than standard RR ($p<0.05$).

\begin{figure}[h]
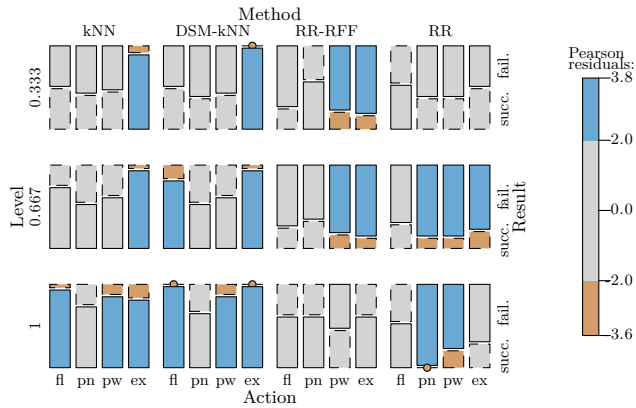

\hspace{-0.35cm}

 	\caption{Basic user study: mosaic plot illustrating significantly more successes for several gestures of different exertion levels for kNN-based methods with significantly more failures in certain cases for RR-based methods}
	\label{fig:study:userstudy1_mos}
\end{figure}

The overall relations are summarized in \cref{fig:study:userstudy1_mos}, where the contribution of factor combinations to significance are illustrated.

In \cref{tab:classTime_userstudy1}, the online classification times are given for the participants of the user study (ARM Cortex-A72), averaged for all classifications executed at sample rate during prediction, confirming the real-time control properties.

\begin{table}[]
	\centering
	\caption{Basic user study (5 classes, 12 subjects), online classification time per subject and method, averaged for all prediction samples, confirming real-time performance}
	\begin{tabular}{@{}ccccc@{}}
		\toprule
		$[ ms ]$ & kNN    & DSM-kNN & RR     & RR-RFF \\ \midrule
		& 3.692 & 0.004  & 0.001 & 0.028 \\
		& 3.775 & 0.004  & 0.001 & 0.035 \\
		& 3.663 & 0.004  & 0.002 & 0.036 \\
		& 3.673 & 0.004  & 0.001 & 0.034 \\
		& 4.343 & 0.005  & 0.001 & 0.035 \\
		& 3.717 & 0.004  & 0.001 & 0.034 \\
		& 3.731 & 0.004  & 0.001 & 0.034 \\
		& 3.733 & 0.004  & 0.001 & 0.035 \\
		& 4.337 & 0.005  & 0.001 & 0.034 \\
		& 3.647 & 0.004  & 0.001 & 0.035 \\
		& 3.678 & 0.004  & 0.001 & 0.033 \\
		& 3.674 & 0.004  & 0.001 & 0.036 \\ \midrule
		Mean & 3.805 & 0.004  & 0.001 & 0.034 \\
		\bottomrule
	\end{tabular}
	\label{tab:classTime_userstudy1}
\end{table}

\subsubsection{Extended User Study (Seven Classes)}
For the purpose of investigating the suitability of the developed methods when including even more gestures in the training, further experiments were conducted as an extension of the described user study. Four subjects who had no EMG experience before but participated in the basic user study were selected again (subjects S5, S7, S9 and S10).
On the one hand, the previous participation in the main part of the study might have influenced the impartiality. On the other hand, this might give interesting insights in the algorithms' performances in the case of low experience with EMG-based control.

Since standard RR showed to not perform well in the main part of the study, this was excluded in the extended evaluation in order to avoid participants' demotivation. Instead, the wrist rotation gestures pronation and supination were added.

For training the system, data were again gathered for two seconds per gesture at maximum sampling rate of 200\,Hz with three repetitions each. This was done for all considered classes (rs, pw, pn, fl, ex, pr, su) at full-intensity exertion. This results in 8400 training sample vectors (8 channels) per person ($= 7\cdot 3\cdot 2s\cdot 200\frac{1}{s}$).

By again repeating each task two times, the number of tasks performed per subject was 108 in total (6 gestures, 3 intensity levels, 3 learning methods, 2 repetitions). Besides these aspects, this part of the study was identical to the previous part. Again, the rest class was not explicitly tested.

\begin{figure}[h]
	\centering
	\begin{tikzpicture}[x=1pt,y=1pt,scale=0.8]
\definecolor{fillColor}{RGB}{255,255,255}
\path[use as bounding box,fill=fillColor,fill opacity=0.00] (0,0) rectangle (180.67,216.81);
\begin{scope}
\path[clip] (  0.00,  0.00) rectangle (180.67,216.81);

\path[] (  0.00,  0.00) rectangle (180.67,216.81);
\end{scope}
\begin{scope}
\path[clip] ( 31.44, 28.09) rectangle (175.17,211.31);

\path[] ( 31.44, 28.09) rectangle (175.17,211.31);
\definecolor{drawColor}{RGB}{255,255,255}

\path[draw=drawColor,line width= 0.3pt,line join=round] ( 31.44, 51.56) --
	(175.17, 51.56);

\path[draw=drawColor,line width= 0.3pt,line join=round] ( 31.44, 81.84) --
	(175.17, 81.84);

\path[draw=drawColor,line width= 0.3pt,line join=round] ( 31.44,112.13) --
	(175.17,112.13);

\path[draw=drawColor,line width= 0.3pt,line join=round] ( 31.44,142.41) --
	(175.17,142.41);

\path[draw=drawColor,line width= 0.3pt,line join=round] ( 31.44,172.70) --
	(175.17,172.70);

\path[draw=drawColor,line width= 0.3pt,line join=round] ( 31.44,202.98) --
	(175.17,202.98);
\definecolor{drawColor}{RGB}{190,190,190}

\path[draw=drawColor,line width= 0.6pt,line join=round] ( 31.44, 36.42) --
	(175.17, 36.42);

\path[draw=drawColor,line width= 0.6pt,line join=round] ( 31.44, 66.70) --
	(175.17, 66.70);

\path[draw=drawColor,line width= 0.6pt,line join=round] ( 31.44, 96.99) --
	(175.17, 96.99);

\path[draw=drawColor,line width= 0.6pt,line join=round] ( 31.44,127.27) --
	(175.17,127.27);

\path[draw=drawColor,line width= 0.6pt,line join=round] ( 31.44,157.56) --
	(175.17,157.56);

\path[draw=drawColor,line width= 0.6pt,line join=round] ( 31.44,187.84) --
	(175.17,187.84);
\definecolor{drawColor}{RGB}{0,125,197}

\path[draw=drawColor,line width= 0.6pt,line join=round] ( 53.90,179.43) --
	( 62.88,179.43);

\path[draw=drawColor,line width= 0.6pt,line join=round] ( 58.39,179.43) --
	( 58.39,133.16);

\path[draw=drawColor,line width= 0.6pt,line join=round] ( 53.90,133.16) --
	( 62.88,133.16);

\path[draw=drawColor,line width= 0.6pt,line join=round] ( 98.81,166.81) --
	(107.80,166.81);

\path[draw=drawColor,line width= 0.6pt,line join=round] (103.31,166.81) --
	(103.31,112.13);

\path[draw=drawColor,line width= 0.6pt,line join=round] ( 98.81,112.13) --
	(107.80,112.13);

\path[draw=drawColor,line width= 0.6pt,line join=round] (143.73,103.72) --
	(152.72,103.72);

\path[draw=drawColor,line width= 0.6pt,line join=round] (148.22,103.72) --
	(148.22, 36.42);

\path[draw=drawColor,line width= 0.6pt,line join=round] (143.73, 36.42) --
	(152.72, 36.42);

\path[draw=drawColor,line width= 0.6pt,line join=round] ( 58.39,166.81) -- ( 58.39,179.43);

\path[draw=drawColor,line width= 0.6pt,line join=round] ( 58.39,142.62) -- ( 58.39,133.16);
\definecolor{fillColor}{RGB}{148,188,228}

\path[draw=drawColor,line width= 0.6pt,line join=round,line cap=round,fill=fillColor] ( 41.54,166.81) --
	( 41.54,142.62) --
	( 75.23,142.62) --
	( 75.23,166.81) --
	( 41.54,166.81) --
	cycle;

\path[draw=drawColor,line width= 1.1pt,line join=round] ( 41.54,154.19) -- ( 75.23,154.19);

\path[draw=drawColor,line width= 0.6pt,line join=round] (103.31,163.65) -- (103.31,166.81);

\path[draw=drawColor,line width= 0.6pt,line join=round] (103.31,137.37) -- (103.31,112.13);

\path[draw=drawColor,line width= 0.6pt,line join=round,line cap=round,fill=fillColor] ( 86.46,163.65) --
	( 86.46,137.37) --
	(120.15,137.37) --
	(120.15,163.65) --
	( 86.46,163.65) --
	cycle;

\path[draw=drawColor,line width= 1.1pt,line join=round] ( 86.46,154.19) -- (120.15,154.19);

\path[draw=drawColor,line width= 0.6pt,line join=round] (148.22, 87.94) -- (148.22,103.72);

\path[draw=drawColor,line width= 0.6pt,line join=round] (148.22, 45.88) -- (148.22, 36.42);

\path[draw=drawColor,line width= 0.6pt,line join=round,line cap=round,fill=fillColor] (131.38, 87.94) --
	(131.38, 45.88) --
	(165.07, 45.88) --
	(165.07, 87.94) --
	(131.38, 87.94) --
	cycle;

\path[draw=drawColor,line width= 1.1pt,line join=round] (131.38, 65.86) -- (165.07, 65.86);

\path[draw=drawColor,line width= 0.4pt,line join=round,line cap=round] ( 55.61,155.24) -- ( 61.16,155.24);

\path[draw=drawColor,line width= 0.4pt,line join=round,line cap=round] ( 58.39,152.47) -- ( 58.39,158.02);

\path[draw=drawColor,line width= 0.4pt,line join=round,line cap=round] (100.53,146.83) -- (106.08,146.83);

\path[draw=drawColor,line width= 0.4pt,line join=round,line cap=round] (103.31,144.05) -- (103.31,149.60);

\path[draw=drawColor,line width= 0.4pt,line join=round,line cap=round] (145.45, 67.96) -- (151.00, 67.96);

\path[draw=drawColor,line width= 0.4pt,line join=round,line cap=round] (148.22, 65.19) -- (148.22, 70.74);
\definecolor{drawColor}{RGB}{0,0,0}

\path[draw=drawColor,line width= 0.6pt,line join=round,line cap=round] ( 58.39,192.55) -- ( 58.39,195.41);

\path[draw=drawColor,line width= 0.6pt,line join=round,line cap=round] ( 58.39,195.41) -- (103.31,195.41);

\path[draw=drawColor,line width= 0.6pt,line join=round,line cap=round] (103.31,195.41) -- (103.31,192.55);

\node[text=drawColor,anchor=base,inner sep=0pt, outer sep=0pt, scale=  0.85] at (114.54,204.81) {$p<.005$};

\path[draw=drawColor,line width= 0.6pt,line join=round,line cap=round] ( 80.85,200.12) -- ( 80.85,202.98);

\path[draw=drawColor,line width= 0.6pt,line join=round,line cap=round] ( 80.85,202.98) -- (148.22,202.98);

\path[draw=drawColor,line width= 0.6pt,line join=round,line cap=round] (148.22,202.98) -- (148.22,200.12);
\end{scope}
\begin{scope}
\path[clip] (  0.00,  0.00) rectangle (180.67,216.81);
\definecolor{drawColor}{RGB}{0,0,0}

\node[text=drawColor,anchor=base east,inner sep=0pt, outer sep=0pt, scale=  0.80] at ( 26.49, 33.66) {0.0};

\node[text=drawColor,anchor=base east,inner sep=0pt, outer sep=0pt, scale=  0.80] at ( 26.49, 63.95) {0.2};

\node[text=drawColor,anchor=base east,inner sep=0pt, outer sep=0pt, scale=  0.80] at ( 26.49, 94.23) {0.4};

\node[text=drawColor,anchor=base east,inner sep=0pt, outer sep=0pt, scale=  0.80] at ( 26.49,124.52) {0.6};

\node[text=drawColor,anchor=base east,inner sep=0pt, outer sep=0pt, scale=  0.80] at ( 26.49,154.80) {0.8};

\node[text=drawColor,anchor=base east,inner sep=0pt, outer sep=0pt, scale=  0.80] at ( 26.49,185.08) {1.0};
\end{scope}
\begin{scope}
\path[clip] (  0.00,  0.00) rectangle (180.67,216.81);
\definecolor{drawColor}{gray}{0.20}

\path[draw=drawColor,line width= 0.6pt,line join=round] ( 28.69, 36.42) --
	( 31.44, 36.42);

\path[draw=drawColor,line width= 0.6pt,line join=round] ( 28.69, 66.70) --
	( 31.44, 66.70);

\path[draw=drawColor,line width= 0.6pt,line join=round] ( 28.69, 96.99) --
	( 31.44, 96.99);

\path[draw=drawColor,line width= 0.6pt,line join=round] ( 28.69,127.27) --
	( 31.44,127.27);

\path[draw=drawColor,line width= 0.6pt,line join=round] ( 28.69,157.56) --
	( 31.44,157.56);

\path[draw=drawColor,line width= 0.6pt,line join=round] ( 28.69,187.84) --
	( 31.44,187.84);
\end{scope}
\begin{scope}
\path[clip] (  0.00,  0.00) rectangle (180.67,216.81);
\definecolor{drawColor}{gray}{0.20}

\path[draw=drawColor,line width= 0.6pt,line join=round] ( 58.39, 25.34) --
	( 58.39, 28.09);

\path[draw=drawColor,line width= 0.6pt,line join=round] (103.31, 25.34) --
	(103.31, 28.09);

\path[draw=drawColor,line width= 0.6pt,line join=round] (148.22, 25.34) --
	(148.22, 28.09);
\end{scope}
\begin{scope}
\path[clip] (  0.00,  0.00) rectangle (180.67,216.81);
\definecolor{drawColor}{RGB}{0,0,0}

\node[text=drawColor,anchor=base,inner sep=0pt, outer sep=0pt, scale=  0.80] at ( 58.39, 17.63) {kNN};

\node[text=drawColor,anchor=base,inner sep=0pt, outer sep=0pt, scale=  0.80] at (103.31, 17.63) {DSM-kNN};

\node[text=drawColor,anchor=base,inner sep=0pt, outer sep=0pt, scale=  0.80] at (148.22, 17.63) {RR-RFF};
\end{scope}
\begin{scope}
\path[clip] (  0.00,  0.00) rectangle (180.67,216.81);
\definecolor{drawColor}{RGB}{0,0,0}

\node[text=drawColor,anchor=base,inner sep=0pt, outer sep=0pt, scale=  0.90] at (103.31,  7.32) {Method};
\end{scope}
\begin{scope}
\path[clip] (  0.00,  0.00) rectangle (180.67,216.81);
\definecolor{drawColor}{RGB}{0,0,0}

\node[text=drawColor,rotate= 90.00,anchor=base,inner sep=0pt, outer sep=0pt, scale=  0.90] at ( 11.70,119.70) {Success Rate};
\end{scope}
\end{tikzpicture}
 	\caption{Extended user study: success rates (over subjects) and significance from ANOVA ($\alpha=0.05$), showing significant difference in success rates for kNN-based methods compared to RR-RFF}
	\label{fig:study:userstudy2_overall}
\end{figure}
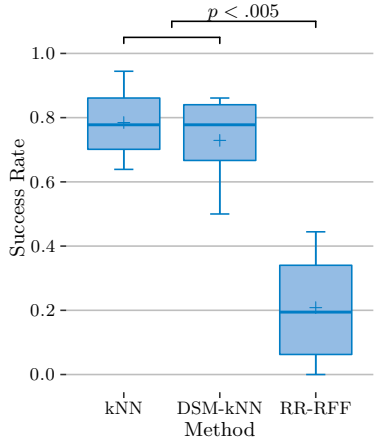

The results of the extended user study's evaluation are summarized in \cref{fig:study:userstudy2_overall}, after the per-level- and -gesture-performances were averaged for each subject-method combination to obtain the variance and median of the success rates based on the subjects. The kNN-based methods achieved success rate means and medians of over 70\%, while RR-RFF performed significantly worse (median 19\%, mean 21\%) with $p<0.005$. This time, the DSM-reduced kNN yielded slightly lower values than standard kNN (both medians and kNN mean at 78\%, but kNN-DSM mean at 73\%). 

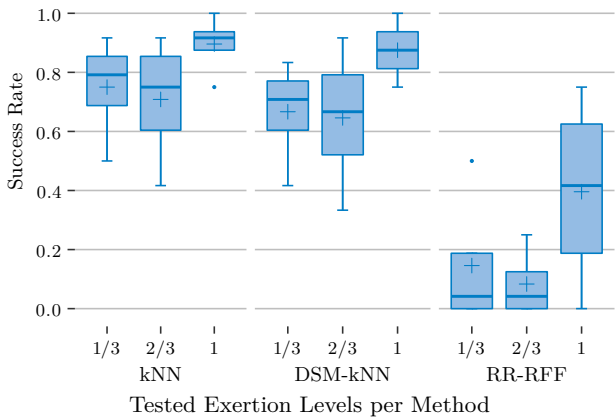
\begin{figure}[h]
	\begin{tikzpicture}[x=1pt,y=1pt]
\definecolor{fillColor}{RGB}{255,255,255}
\path[use as bounding box,fill=fillColor,fill opacity=0.00] (0,0) rectangle (101.18,180.67);
\begin{scope}
\path[clip] (  0.00,  0.00) rectangle (101.18,180.67);

\path[] (  0.00,  0.00) rectangle (101.18,180.68);
\end{scope}
\begin{scope}
\path[clip] ( 31.44, 28.09) rectangle ( 95.68,175.17);

\path[] ( 31.44, 28.09) rectangle ( 95.68,175.18);
\definecolor{drawColor}{RGB}{255,255,255}

\path[draw=drawColor,line width= 0.3pt,line join=round] ( 31.44, 45.92) --
	( 95.68, 45.92);

\path[draw=drawColor,line width= 0.3pt,line join=round] ( 31.44, 68.20) --
	( 95.68, 68.20);

\path[draw=drawColor,line width= 0.3pt,line join=round] ( 31.44, 90.49) --
	( 95.68, 90.49);

\path[draw=drawColor,line width= 0.3pt,line join=round] ( 31.44,112.78) --
	( 95.68,112.78);

\path[draw=drawColor,line width= 0.3pt,line join=round] ( 31.44,135.06) --
	( 95.68,135.06);

\path[draw=drawColor,line width= 0.3pt,line join=round] ( 31.44,157.35) --
	( 95.68,157.35);

\path[draw=drawColor,line width= 0.3pt,line join=round] ( 31.44,168.49) --
	( 95.68,168.49);
\definecolor{drawColor}{RGB}{190,190,190}

\path[draw=drawColor,line width= 0.6pt,line join=round] ( 31.44, 34.78) --
	( 95.68, 34.78);

\path[draw=drawColor,line width= 0.6pt,line join=round] ( 31.44, 57.06) --
	( 95.68, 57.06);

\path[draw=drawColor,line width= 0.6pt,line join=round] ( 31.44, 79.35) --
	( 95.68, 79.35);

\path[draw=drawColor,line width= 0.6pt,line join=round] ( 31.44,101.63) --
	( 95.68,101.63);

\path[draw=drawColor,line width= 0.6pt,line join=round] ( 31.44,123.92) --
	( 95.68,123.92);

\path[draw=drawColor,line width= 0.6pt,line join=round] ( 31.44,146.20) --
	( 95.68,146.20);
\definecolor{drawColor}{RGB}{0,125,197}

\path[draw=drawColor,line width= 0.6pt,line join=round] ( 41.47,136.92) --
	( 45.49,136.92);

\path[draw=drawColor,line width= 0.6pt,line join=round] ( 43.48,136.92) --
	( 43.48, 90.49);

\path[draw=drawColor,line width= 0.6pt,line join=round] ( 41.47, 90.49) --
	( 45.49, 90.49);

\path[draw=drawColor,line width= 0.6pt,line join=round] ( 61.55,136.92) --
	( 65.56,136.92);

\path[draw=drawColor,line width= 0.6pt,line join=round] ( 63.56,136.92) --
	( 63.56, 81.20);

\path[draw=drawColor,line width= 0.6pt,line join=round] ( 61.55, 81.20) --
	( 65.56, 81.20);

\path[draw=drawColor,line width= 0.6pt,line join=round] ( 81.63,146.20) --
	( 85.64,146.20);

\path[draw=drawColor,line width= 0.6pt,line join=round] ( 83.63,146.20) --
	( 83.63,132.28);

\path[draw=drawColor,line width= 0.6pt,line join=round] ( 81.63,132.28) --
	( 85.64,132.28);

\path[draw=drawColor,line width= 0.6pt,line join=round] ( 43.48,129.95) -- ( 43.48,136.92);

\path[draw=drawColor,line width= 0.6pt,line join=round] ( 43.48,111.38) -- ( 43.48, 90.49);
\definecolor{fillColor}{RGB}{148,188,228}

\path[draw=drawColor,line width= 0.6pt,line join=round,line cap=round,fill=fillColor] ( 35.95,129.95) --
	( 35.95,111.38) --
	( 51.01,111.38) --
	( 51.01,129.95) --
	( 35.95,129.95) --
	cycle;

\path[draw=drawColor,line width= 1.1pt,line join=round] ( 35.95,122.99) -- ( 51.01,122.99);

\path[draw=drawColor,line width= 0.6pt,line join=round] ( 63.56,129.95) -- ( 63.56,136.92);

\path[draw=drawColor,line width= 0.6pt,line join=round] ( 63.56,102.10) -- ( 63.56, 81.20);

\path[draw=drawColor,line width= 0.6pt,line join=round,line cap=round,fill=fillColor] ( 56.03,129.95) --
	( 56.03,102.10) --
	( 71.09,102.10) --
	( 71.09,129.95) --
	( 56.03,129.95) --
	cycle;

\path[draw=drawColor,line width= 1.1pt,line join=round] ( 56.03,118.35) -- ( 71.09,118.35);
\definecolor{fillColor}{RGB}{0,125,197}

\path[draw=drawColor,line width= 0.4pt,line join=round,line cap=round,fill=fillColor] ( 83.63,118.35) circle (  0.57);

\path[draw=drawColor,line width= 0.6pt,line join=round] ( 83.63,139.24) -- ( 83.63,146.20);

\path[draw=drawColor,line width= 0.6pt,line join=round] ( 83.63,132.28) -- ( 83.63,132.28);
\definecolor{fillColor}{RGB}{148,188,228}

\path[draw=drawColor,line width= 0.6pt,line join=round,line cap=round,fill=fillColor] ( 76.10,139.24) --
	( 76.10,132.28) --
	( 91.16,132.28) --
	( 91.16,139.24) --
	( 76.10,139.24) --
	cycle;

\path[draw=drawColor,line width= 1.1pt,line join=round] ( 76.10,136.92) -- ( 91.16,136.92);

\path[draw=drawColor,line width= 0.4pt,line join=round,line cap=round] ( 40.71,118.35) -- ( 46.26,118.35);

\path[draw=drawColor,line width= 0.4pt,line join=round,line cap=round] ( 43.48,115.57) -- ( 43.48,121.12);

\path[draw=drawColor,line width= 0.4pt,line join=round,line cap=round] ( 60.78,113.70) -- ( 66.33,113.70);

\path[draw=drawColor,line width= 0.4pt,line join=round,line cap=round] ( 63.56,110.93) -- ( 63.56,116.48);

\path[draw=drawColor,line width= 0.4pt,line join=round,line cap=round] ( 80.86,134.60) -- ( 86.41,134.60);

\path[draw=drawColor,line width= 0.4pt,line join=round,line cap=round] ( 83.63,131.82) -- ( 83.63,137.37);
\end{scope}
\begin{scope}
\path[clip] (  0.00,  0.00) rectangle (101.18,180.67);
\definecolor{drawColor}{RGB}{0,0,0}

\node[text=drawColor,anchor=base east,inner sep=0pt, outer sep=0pt, scale=  0.80] at ( 26.49, 32.02) {0.0};

\node[text=drawColor,anchor=base east,inner sep=0pt, outer sep=0pt, scale=  0.80] at ( 26.49, 54.31) {0.2};

\node[text=drawColor,anchor=base east,inner sep=0pt, outer sep=0pt, scale=  0.80] at ( 26.49, 76.59) {0.4};

\node[text=drawColor,anchor=base east,inner sep=0pt, outer sep=0pt, scale=  0.80] at ( 26.49, 98.88) {0.6};

\node[text=drawColor,anchor=base east,inner sep=0pt, outer sep=0pt, scale=  0.80] at ( 26.49,121.16) {0.8};

\node[text=drawColor,anchor=base east,inner sep=0pt, outer sep=0pt, scale=  0.80] at ( 26.49,143.45) {1.0};
\end{scope}
\begin{scope}
\path[clip] (  0.00,  0.00) rectangle (101.18,180.67);
\definecolor{drawColor}{gray}{0.20}

\path[draw=drawColor,line width= 0.6pt,line join=round] ( 28.69, 34.78) --
	( 31.44, 34.78);

\path[draw=drawColor,line width= 0.6pt,line join=round] ( 28.69, 57.06) --
	( 31.44, 57.06);

\path[draw=drawColor,line width= 0.6pt,line join=round] ( 28.69, 79.35) --
	( 31.44, 79.35);

\path[draw=drawColor,line width= 0.6pt,line join=round] ( 28.69,101.63) --
	( 31.44,101.63);

\path[draw=drawColor,line width= 0.6pt,line join=round] ( 28.69,123.92) --
	( 31.44,123.92);

\path[draw=drawColor,line width= 0.6pt,line join=round] ( 28.69,146.20) --
	( 31.44,146.20);
\end{scope}
\begin{scope}
\path[clip] (  0.00,  0.00) rectangle (101.18,180.67);
\definecolor{drawColor}{gray}{0.20}

\path[draw=drawColor,line width= 0.6pt,line join=round] ( 43.48, 25.34) --
	( 43.48, 28.09);

\path[draw=drawColor,line width= 0.6pt,line join=round] ( 63.56, 25.34) --
	( 63.56, 28.09);

\path[draw=drawColor,line width= 0.6pt,line join=round] ( 83.63, 25.34) --
	( 83.63, 28.09);
\end{scope}
\begin{scope}
\path[clip] (  0.00,  0.00) rectangle (101.18,180.67);
\definecolor{drawColor}{RGB}{0,0,0}

\node[text=drawColor,anchor=base,inner sep=0pt, outer sep=0pt, scale=  0.80] at ( 43.48, 17.63) {1/3};

\node[text=drawColor,anchor=base,inner sep=0pt, outer sep=0pt, scale=  0.80] at ( 63.56, 17.63) {2/3};

\node[text=drawColor,anchor=base,inner sep=0pt, outer sep=0pt, scale=  0.80] at ( 83.63, 17.63) {1};
\end{scope}
\begin{scope}
\path[clip] (  0.00,  0.00) rectangle (101.18,180.67);
\definecolor{drawColor}{RGB}{0,0,0}

\node[text=drawColor,anchor=base,inner sep=0pt, outer sep=0pt, scale=  0.90] at ( 63.56,  7.32) {kNN};
\end{scope}
\begin{scope}
\path[clip] (  0.00,  0.00) rectangle (101.18,180.67);
\definecolor{drawColor}{RGB}{0,0,0}

\node[text=drawColor,rotate= 90.00,anchor=base,inner sep=0pt, outer sep=0pt, scale=  0.90] at ( 11.70,101.63) {Success Rate};
\end{scope}
\end{tikzpicture}
 \hspace{-0.58cm}
	\begin{tikzpicture}[x=1pt,y=1pt]
\definecolor{fillColor}{RGB}{255,255,255}
\path[use as bounding box,fill=fillColor,fill opacity=0.00] (0,0) rectangle ( 79.50,180.67);
\begin{scope}
\path[clip] (  0.00,  0.00) rectangle ( 79.50,180.67);

\path[] (  0.00,  0.00) rectangle ( 79.50,180.68);
\end{scope}
\begin{scope}
\path[clip] (  8.25, 28.09) rectangle ( 74.00,175.17);

\path[] (  8.25, 28.09) rectangle ( 74.00,175.18);
\definecolor{drawColor}{RGB}{255,255,255}

\path[draw=drawColor,line width= 0.3pt,line join=round] (  8.25, 45.92) --
	( 74.00, 45.92);

\path[draw=drawColor,line width= 0.3pt,line join=round] (  8.25, 68.20) --
	( 74.00, 68.20);

\path[draw=drawColor,line width= 0.3pt,line join=round] (  8.25, 90.49) --
	( 74.00, 90.49);

\path[draw=drawColor,line width= 0.3pt,line join=round] (  8.25,112.78) --
	( 74.00,112.78);

\path[draw=drawColor,line width= 0.3pt,line join=round] (  8.25,135.06) --
	( 74.00,135.06);

\path[draw=drawColor,line width= 0.3pt,line join=round] (  8.25,157.35) --
	( 74.00,157.35);

\path[draw=drawColor,line width= 0.3pt,line join=round] (  8.25,168.49) --
	( 74.00,168.49);
\definecolor{drawColor}{RGB}{190,190,190}

\path[draw=drawColor,line width= 0.6pt,line join=round] (  8.25, 34.78) --
	( 74.00, 34.78);

\path[draw=drawColor,line width= 0.6pt,line join=round] (  8.25, 57.06) --
	( 74.00, 57.06);

\path[draw=drawColor,line width= 0.6pt,line join=round] (  8.25, 79.35) --
	( 74.00, 79.35);

\path[draw=drawColor,line width= 0.6pt,line join=round] (  8.25,101.63) --
	( 74.00,101.63);

\path[draw=drawColor,line width= 0.6pt,line join=round] (  8.25,123.92) --
	( 74.00,123.92);

\path[draw=drawColor,line width= 0.6pt,line join=round] (  8.25,146.20) --
	( 74.00,146.20);
\definecolor{drawColor}{RGB}{0,125,197}

\path[draw=drawColor,line width= 0.6pt,line join=round] ( 18.52,127.63) --
	( 22.63,127.63);

\path[draw=drawColor,line width= 0.6pt,line join=round] ( 20.58,127.63) --
	( 20.58, 81.20);

\path[draw=drawColor,line width= 0.6pt,line join=round] ( 18.52, 81.20) --
	( 22.63, 81.20);

\path[draw=drawColor,line width= 0.6pt,line join=round] ( 39.07,136.92) --
	( 43.18,136.92);

\path[draw=drawColor,line width= 0.6pt,line join=round] ( 41.12,136.92) --
	( 41.12, 71.92);

\path[draw=drawColor,line width= 0.6pt,line join=round] ( 39.07, 71.92) --
	( 43.18, 71.92);

\path[draw=drawColor,line width= 0.6pt,line join=round] ( 59.61,146.20) --
	( 63.72,146.20);

\path[draw=drawColor,line width= 0.6pt,line join=round] ( 61.67,146.20) --
	( 61.67,118.35);

\path[draw=drawColor,line width= 0.6pt,line join=round] ( 59.61,118.35) --
	( 63.72,118.35);

\path[draw=drawColor,line width= 0.6pt,line join=round] ( 20.58,120.67) -- ( 20.58,127.63);

\path[draw=drawColor,line width= 0.6pt,line join=round] ( 20.58,102.10) -- ( 20.58, 81.20);
\definecolor{fillColor}{RGB}{148,188,228}

\path[draw=drawColor,line width= 0.6pt,line join=round,line cap=round,fill=fillColor] ( 12.87,120.67) --
	( 12.87,102.10) --
	( 28.28,102.10) --
	( 28.28,120.67) --
	( 12.87,120.67) --
	cycle;

\path[draw=drawColor,line width= 1.1pt,line join=round] ( 12.87,113.70) -- ( 28.28,113.70);

\path[draw=drawColor,line width= 0.6pt,line join=round] ( 41.12,122.99) -- ( 41.12,136.92);

\path[draw=drawColor,line width= 0.6pt,line join=round] ( 41.12, 92.81) -- ( 41.12, 71.92);

\path[draw=drawColor,line width= 0.6pt,line join=round,line cap=round,fill=fillColor] ( 33.42,122.99) --
	( 33.42, 92.81) --
	( 48.83, 92.81) --
	( 48.83,122.99) --
	( 33.42,122.99) --
	cycle;

\path[draw=drawColor,line width= 1.1pt,line join=round] ( 33.42,109.06) -- ( 48.83,109.06);

\path[draw=drawColor,line width= 0.6pt,line join=round] ( 61.67,139.24) -- ( 61.67,146.20);

\path[draw=drawColor,line width= 0.6pt,line join=round] ( 61.67,125.31) -- ( 61.67,118.35);

\path[draw=drawColor,line width= 0.6pt,line join=round,line cap=round,fill=fillColor] ( 53.96,139.24) --
	( 53.96,125.31) --
	( 69.37,125.31) --
	( 69.37,139.24) --
	( 53.96,139.24) --
	cycle;

\path[draw=drawColor,line width= 1.1pt,line join=round] ( 53.96,132.28) -- ( 69.37,132.28);

\path[draw=drawColor,line width= 0.4pt,line join=round,line cap=round] ( 17.80,109.06) -- ( 23.35,109.06);

\path[draw=drawColor,line width= 0.4pt,line join=round,line cap=round] ( 20.58,106.29) -- ( 20.58,111.84);

\path[draw=drawColor,line width= 0.4pt,line join=round,line cap=round] ( 38.35,106.74) -- ( 43.90,106.74);

\path[draw=drawColor,line width= 0.4pt,line join=round,line cap=round] ( 41.12,103.96) -- ( 41.12,109.51);

\path[draw=drawColor,line width= 0.4pt,line join=round,line cap=round] ( 58.89,132.28) -- ( 64.44,132.28);

\path[draw=drawColor,line width= 0.4pt,line join=round,line cap=round] ( 61.67,129.50) -- ( 61.67,135.05);
\end{scope}
\begin{scope}
\path[clip] (  0.00,  0.00) rectangle ( 79.50,180.67);
\definecolor{drawColor}{gray}{0.20}

\path[draw=drawColor,line width= 0.6pt,line join=round] ( 20.58, 25.34) --
	( 20.58, 28.09);

\path[draw=drawColor,line width= 0.6pt,line join=round] ( 41.12, 25.34) --
	( 41.12, 28.09);

\path[draw=drawColor,line width= 0.6pt,line join=round] ( 61.67, 25.34) --
	( 61.67, 28.09);
\end{scope}
\begin{scope}
\path[clip] (  0.00,  0.00) rectangle ( 79.50,180.67);
\definecolor{drawColor}{RGB}{0,0,0}

\node[text=drawColor,anchor=base,inner sep=0pt, outer sep=0pt, scale=  0.80] at ( 20.58, 17.63) {1/3};

\node[text=drawColor,anchor=base,inner sep=0pt, outer sep=0pt, scale=  0.80] at ( 41.12, 17.63) {2/3};

\node[text=drawColor,anchor=base,inner sep=0pt, outer sep=0pt, scale=  0.80] at ( 61.67, 17.63) {1};
\end{scope}
\begin{scope}
\path[clip] (  0.00,  0.00) rectangle ( 79.50,180.67);
\definecolor{drawColor}{RGB}{0,0,0}

\node[text=drawColor,anchor=base,inner sep=0pt, outer sep=0pt, scale=  0.90] at ( 41.12,  7.32) {DSM-kNN};
\end{scope}
\end{tikzpicture}
 \hspace{-0.58cm}
	\begin{tikzpicture}[x=1pt,y=1pt]
\definecolor{fillColor}{RGB}{255,255,255}
\path[use as bounding box,fill=fillColor,fill opacity=0.00] (0,0) rectangle ( 79.50,180.67);
\begin{scope}
\path[clip] (  0.00,  0.00) rectangle ( 79.50,180.67);

\path[] (  0.00,  0.00) rectangle ( 79.50,180.68);
\end{scope}
\begin{scope}
\path[clip] (  8.25, 28.09) rectangle ( 74.00,175.17);

\path[] (  8.25, 28.09) rectangle ( 74.00,175.18);
\definecolor{drawColor}{RGB}{255,255,255}

\path[draw=drawColor,line width= 0.3pt,line join=round] (  8.25, 45.92) --
	( 74.00, 45.92);

\path[draw=drawColor,line width= 0.3pt,line join=round] (  8.25, 68.20) --
	( 74.00, 68.20);

\path[draw=drawColor,line width= 0.3pt,line join=round] (  8.25, 90.49) --
	( 74.00, 90.49);

\path[draw=drawColor,line width= 0.3pt,line join=round] (  8.25,112.78) --
	( 74.00,112.78);

\path[draw=drawColor,line width= 0.3pt,line join=round] (  8.25,135.06) --
	( 74.00,135.06);

\path[draw=drawColor,line width= 0.3pt,line join=round] (  8.25,157.35) --
	( 74.00,157.35);

\path[draw=drawColor,line width= 0.3pt,line join=round] (  8.25,168.49) --
	( 74.00,168.49);
\definecolor{drawColor}{RGB}{190,190,190}

\path[draw=drawColor,line width= 0.6pt,line join=round] (  8.25, 34.78) --
	( 74.00, 34.78);

\path[draw=drawColor,line width= 0.6pt,line join=round] (  8.25, 57.06) --
	( 74.00, 57.06);

\path[draw=drawColor,line width= 0.6pt,line join=round] (  8.25, 79.35) --
	( 74.00, 79.35);

\path[draw=drawColor,line width= 0.6pt,line join=round] (  8.25,101.63) --
	( 74.00,101.63);

\path[draw=drawColor,line width= 0.6pt,line join=round] (  8.25,123.92) --
	( 74.00,123.92);

\path[draw=drawColor,line width= 0.6pt,line join=round] (  8.25,146.20) --
	( 74.00,146.20);
\definecolor{drawColor}{RGB}{0,125,197}

\path[draw=drawColor,line width= 0.6pt,line join=round] ( 18.52, 55.67) --
	( 22.63, 55.67);

\path[draw=drawColor,line width= 0.6pt,line join=round] ( 20.58, 55.67) --
	( 20.58, 34.78);

\path[draw=drawColor,line width= 0.6pt,line join=round] ( 18.52, 34.78) --
	( 22.63, 34.78);

\path[draw=drawColor,line width= 0.6pt,line join=round] ( 39.07, 62.63) --
	( 43.18, 62.63);

\path[draw=drawColor,line width= 0.6pt,line join=round] ( 41.12, 62.63) --
	( 41.12, 34.78);

\path[draw=drawColor,line width= 0.6pt,line join=round] ( 39.07, 34.78) --
	( 43.18, 34.78);

\path[draw=drawColor,line width= 0.6pt,line join=round] ( 59.61,118.35) --
	( 63.72,118.35);

\path[draw=drawColor,line width= 0.6pt,line join=round] ( 61.67,118.35) --
	( 61.67, 34.78);

\path[draw=drawColor,line width= 0.6pt,line join=round] ( 59.61, 34.78) --
	( 63.72, 34.78);
\definecolor{fillColor}{RGB}{0,125,197}

\path[draw=drawColor,line width= 0.4pt,line join=round,line cap=round,fill=fillColor] ( 20.58, 90.49) circle (  0.57);

\path[draw=drawColor,line width= 0.6pt,line join=round] ( 20.58, 55.67) -- ( 20.58, 55.67);

\path[draw=drawColor,line width= 0.6pt,line join=round] ( 20.58, 34.78) -- ( 20.58, 34.78);
\definecolor{fillColor}{RGB}{148,188,228}

\path[draw=drawColor,line width= 0.6pt,line join=round,line cap=round,fill=fillColor] ( 12.87, 55.67) --
	( 12.87, 34.78) --
	( 28.28, 34.78) --
	( 28.28, 55.67) --
	( 12.87, 55.67) --
	cycle;

\path[draw=drawColor,line width= 1.1pt,line join=round] ( 12.87, 39.42) -- ( 28.28, 39.42);

\path[draw=drawColor,line width= 0.6pt,line join=round] ( 41.12, 48.70) -- ( 41.12, 62.63);

\path[draw=drawColor,line width= 0.6pt,line join=round] ( 41.12, 34.78) -- ( 41.12, 34.78);

\path[draw=drawColor,line width= 0.6pt,line join=round,line cap=round,fill=fillColor] ( 33.42, 48.70) --
	( 33.42, 34.78) --
	( 48.83, 34.78) --
	( 48.83, 48.70) --
	( 33.42, 48.70) --
	cycle;

\path[draw=drawColor,line width= 1.1pt,line join=round] ( 33.42, 39.42) -- ( 48.83, 39.42);

\path[draw=drawColor,line width= 0.6pt,line join=round] ( 61.67,104.42) -- ( 61.67,118.35);

\path[draw=drawColor,line width= 0.6pt,line join=round] ( 61.67, 55.67) -- ( 61.67, 34.78);

\path[draw=drawColor,line width= 0.6pt,line join=round,line cap=round,fill=fillColor] ( 53.96,104.42) --
	( 53.96, 55.67) --
	( 69.37, 55.67) --
	( 69.37,104.42) --
	( 53.96,104.42) --
	cycle;

\path[draw=drawColor,line width= 1.1pt,line join=round] ( 53.96, 81.20) -- ( 69.37, 81.20);

\path[draw=drawColor,line width= 0.4pt,line join=round,line cap=round] ( 17.80, 51.03) -- ( 23.35, 51.03);

\path[draw=drawColor,line width= 0.4pt,line join=round,line cap=round] ( 20.58, 48.25) -- ( 20.58, 53.80);

\path[draw=drawColor,line width= 0.4pt,line join=round,line cap=round] ( 38.35, 44.06) -- ( 43.90, 44.06);

\path[draw=drawColor,line width= 0.4pt,line join=round,line cap=round] ( 41.12, 41.29) -- ( 41.12, 46.84);

\path[draw=drawColor,line width= 0.4pt,line join=round,line cap=round] ( 58.89, 78.88) -- ( 64.44, 78.88);

\path[draw=drawColor,line width= 0.4pt,line join=round,line cap=round] ( 61.67, 76.11) -- ( 61.67, 81.66);
\end{scope}
\begin{scope}
\path[clip] (  0.00,  0.00) rectangle ( 79.50,180.67);
\definecolor{drawColor}{gray}{0.20}

\path[draw=drawColor,line width= 0.6pt,line join=round] ( 20.58, 25.34) --
	( 20.58, 28.09);

\path[draw=drawColor,line width= 0.6pt,line join=round] ( 41.12, 25.34) --
	( 41.12, 28.09);

\path[draw=drawColor,line width= 0.6pt,line join=round] ( 61.67, 25.34) --
	( 61.67, 28.09);
\end{scope}
\begin{scope}
\path[clip] (  0.00,  0.00) rectangle ( 79.50,180.67);
\definecolor{drawColor}{RGB}{0,0,0}

\node[text=drawColor,anchor=base,inner sep=0pt, outer sep=0pt, scale=  0.80] at ( 20.58, 17.63) {1/3};

\node[text=drawColor,anchor=base,inner sep=0pt, outer sep=0pt, scale=  0.80] at ( 41.12, 17.63) {2/3};

\node[text=drawColor,anchor=base,inner sep=0pt, outer sep=0pt, scale=  0.80] at ( 61.67, 17.63) {1};
\end{scope}
\begin{scope}
\path[clip] (  0.00,  0.00) rectangle ( 79.50,180.67);
\definecolor{drawColor}{RGB}{0,0,0}

\node[text=drawColor,anchor=base,inner sep=0pt, outer sep=0pt, scale=  0.90] at ( 41.12,  7.32) {RR-RFF};
\end{scope}
\end{tikzpicture}
 	\vspace{-0.7cm}
	\begin{center}
		\footnotesize Tested Exertion Levels per Method
	\end{center}
	\caption{Extended user study: performance of the examined methods for individual gesture exertion levels and significance from ANOVA ($\alpha=0.05$)}
	\label{fig:userstudy2_level}\end{figure}

In \cref{fig:userstudy2_level} it is observable that the lowest exertion levels did not show the worst performance for any of the methods. Instead, the sucess rates at $\nicefrac{1}{3}$ exertion level were similar to the $\nicefrac{2}{3}$ level but had slighly higher means and medians. The best behaviour could be reached at full intensity (kNN: 92\% median, 90\% mean; DSM-kNN: 88\% median and mean). All three tested methods showed the same tendency in terms of performance for individual levels -- with RR-RFF's success rates shifted towards a lower baseline (e.~g. for full intensity median 42\%, mean 40\%). RR-RFF could not outperform kNN or DSM-kNN at any level. Between the different levels of a single method there is no significance.

For each individual level, the success rates of RR-RFF and the group of kNN-based methods differ significantly ($p<0.05$), while there is no significance between kNN and DSM-kNN ($\alpha=0.05$).

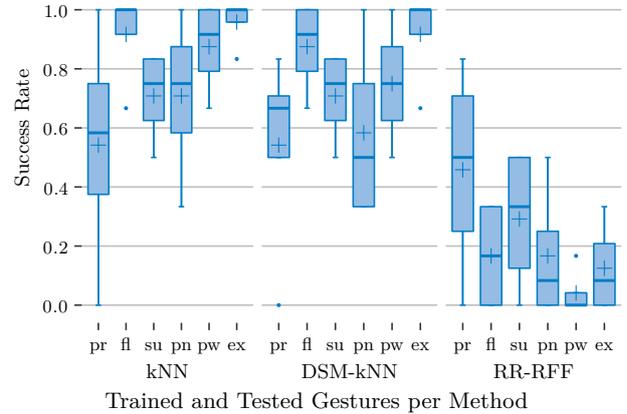
\begin{figure}[h]
	\begin{tikzpicture}[x=1pt,y=1pt]
\definecolor{fillColor}{RGB}{255,255,255}
\path[use as bounding box,fill=fillColor,fill opacity=0.00] (0,0) rectangle (101.18,180.67);
\begin{scope}
\path[clip] (  0.00,  0.00) rectangle (101.18,180.67);

\path[] (  0.00,  0.00) rectangle (101.18,180.68);
\end{scope}
\begin{scope}
\path[clip] ( 31.44, 28.09) rectangle ( 95.68,175.17);

\path[] ( 31.44, 28.09) rectangle ( 95.68,175.18);
\definecolor{drawColor}{RGB}{255,255,255}

\path[draw=drawColor,line width= 0.3pt,line join=round] ( 31.44, 45.92) --
	( 95.68, 45.92);

\path[draw=drawColor,line width= 0.3pt,line join=round] ( 31.44, 68.20) --
	( 95.68, 68.20);

\path[draw=drawColor,line width= 0.3pt,line join=round] ( 31.44, 90.49) --
	( 95.68, 90.49);

\path[draw=drawColor,line width= 0.3pt,line join=round] ( 31.44,112.78) --
	( 95.68,112.78);

\path[draw=drawColor,line width= 0.3pt,line join=round] ( 31.44,135.06) --
	( 95.68,135.06);

\path[draw=drawColor,line width= 0.3pt,line join=round] ( 31.44,157.35) --
	( 95.68,157.35);

\path[draw=drawColor,line width= 0.3pt,line join=round] ( 31.44,168.49) --
	( 95.68,168.49);
\definecolor{drawColor}{RGB}{190,190,190}

\path[draw=drawColor,line width= 0.6pt,line join=round] ( 31.44, 34.78) --
	( 95.68, 34.78);

\path[draw=drawColor,line width= 0.6pt,line join=round] ( 31.44, 57.06) --
	( 95.68, 57.06);

\path[draw=drawColor,line width= 0.6pt,line join=round] ( 31.44, 79.35) --
	( 95.68, 79.35);

\path[draw=drawColor,line width= 0.6pt,line join=round] ( 31.44,101.63) --
	( 95.68,101.63);

\path[draw=drawColor,line width= 0.6pt,line join=round] ( 31.44,123.92) --
	( 95.68,123.92);

\path[draw=drawColor,line width= 0.6pt,line join=round] ( 31.44,146.20) --
	( 95.68,146.20);
\definecolor{drawColor}{RGB}{0,125,197}

\path[draw=drawColor,line width= 0.6pt,line join=round] ( 36.62,146.20) --
	( 38.69,146.20);

\path[draw=drawColor,line width= 0.6pt,line join=round] ( 37.65,146.20) --
	( 37.65, 34.78);

\path[draw=drawColor,line width= 0.6pt,line join=round] ( 36.62, 34.78) --
	( 38.69, 34.78);

\path[draw=drawColor,line width= 0.6pt,line join=round] ( 46.98,146.20) --
	( 49.05,146.20);

\path[draw=drawColor,line width= 0.6pt,line join=round] ( 48.01,146.20) --
	( 48.01,136.92);

\path[draw=drawColor,line width= 0.6pt,line join=round] ( 46.98,136.92) --
	( 49.05,136.92);

\path[draw=drawColor,line width= 0.6pt,line join=round] ( 57.34,127.63) --
	( 59.41,127.63);

\path[draw=drawColor,line width= 0.6pt,line join=round] ( 58.38,127.63) --
	( 58.38, 90.49);

\path[draw=drawColor,line width= 0.6pt,line join=round] ( 57.34, 90.49) --
	( 59.41, 90.49);

\path[draw=drawColor,line width= 0.6pt,line join=round] ( 67.70,146.20) --
	( 69.77,146.20);

\path[draw=drawColor,line width= 0.6pt,line join=round] ( 68.74,146.20) --
	( 68.74, 71.92);

\path[draw=drawColor,line width= 0.6pt,line join=round] ( 67.70, 71.92) --
	( 69.77, 71.92);

\path[draw=drawColor,line width= 0.6pt,line join=round] ( 78.06,146.20) --
	( 80.14,146.20);

\path[draw=drawColor,line width= 0.6pt,line join=round] ( 79.10,146.20) --
	( 79.10,109.06);

\path[draw=drawColor,line width= 0.6pt,line join=round] ( 78.06,109.06) --
	( 80.14,109.06);

\path[draw=drawColor,line width= 0.6pt,line join=round] ( 88.42,146.20) --
	( 90.50,146.20);

\path[draw=drawColor,line width= 0.6pt,line join=round] ( 89.46,146.20) --
	( 89.46,141.56);

\path[draw=drawColor,line width= 0.6pt,line join=round] ( 88.42,141.56) --
	( 90.50,141.56);

\path[draw=drawColor,line width= 0.6pt,line join=round] ( 37.65,118.35) -- ( 37.65,146.20);

\path[draw=drawColor,line width= 0.6pt,line join=round] ( 37.65, 76.56) -- ( 37.65, 34.78);
\definecolor{fillColor}{RGB}{148,188,228}

\path[draw=drawColor,line width= 0.6pt,line join=round,line cap=round,fill=fillColor] ( 33.77,118.35) --
	( 33.77, 76.56) --
	( 41.54, 76.56) --
	( 41.54,118.35) --
	( 33.77,118.35) --
	cycle;

\path[draw=drawColor,line width= 1.1pt,line join=round] ( 33.77, 99.78) -- ( 41.54, 99.78);
\definecolor{fillColor}{RGB}{0,125,197}

\path[draw=drawColor,line width= 0.4pt,line join=round,line cap=round,fill=fillColor] ( 48.01,109.06) circle (  0.57);

\path[draw=drawColor,line width= 0.6pt,line join=round] ( 48.01,146.20) -- ( 48.01,146.20);

\path[draw=drawColor,line width= 0.6pt,line join=round] ( 48.01,136.92) -- ( 48.01,136.92);
\definecolor{fillColor}{RGB}{148,188,228}

\path[draw=drawColor,line width= 0.6pt,line join=round,line cap=round,fill=fillColor] ( 44.13,146.20) --
	( 44.13,136.92) --
	( 51.90,136.92) --
	( 51.90,146.20) --
	( 44.13,146.20) --
	cycle;

\path[draw=drawColor,line width= 1.1pt,line join=round] ( 44.13,146.20) -- ( 51.90,146.20);

\path[draw=drawColor,line width= 0.6pt,line join=round] ( 58.38,127.63) -- ( 58.38,127.63);

\path[draw=drawColor,line width= 0.6pt,line join=round] ( 58.38,104.42) -- ( 58.38, 90.49);

\path[draw=drawColor,line width= 0.6pt,line join=round,line cap=round,fill=fillColor] ( 54.49,127.63) --
	( 54.49,104.42) --
	( 62.26,104.42) --
	( 62.26,127.63) --
	( 54.49,127.63) --
	cycle;

\path[draw=drawColor,line width= 1.1pt,line join=round] ( 54.49,118.35) -- ( 62.26,118.35);

\path[draw=drawColor,line width= 0.6pt,line join=round] ( 68.74,132.28) -- ( 68.74,146.20);

\path[draw=drawColor,line width= 0.6pt,line join=round] ( 68.74, 99.78) -- ( 68.74, 71.92);

\path[draw=drawColor,line width= 0.6pt,line join=round,line cap=round,fill=fillColor] ( 64.85,132.28) --
	( 64.85, 99.78) --
	( 72.62, 99.78) --
	( 72.62,132.28) --
	( 64.85,132.28) --
	cycle;

\path[draw=drawColor,line width= 1.1pt,line join=round] ( 64.85,118.35) -- ( 72.62,118.35);

\path[draw=drawColor,line width= 0.6pt,line join=round] ( 79.10,146.20) -- ( 79.10,146.20);

\path[draw=drawColor,line width= 0.6pt,line join=round] ( 79.10,122.99) -- ( 79.10,109.06);

\path[draw=drawColor,line width= 0.6pt,line join=round,line cap=round,fill=fillColor] ( 75.21,146.20) --
	( 75.21,122.99) --
	( 82.99,122.99) --
	( 82.99,146.20) --
	( 75.21,146.20) --
	cycle;

\path[draw=drawColor,line width= 1.1pt,line join=round] ( 75.21,136.92) -- ( 82.99,136.92);
\definecolor{fillColor}{RGB}{0,125,197}

\path[draw=drawColor,line width= 0.4pt,line join=round,line cap=round,fill=fillColor] ( 89.46,127.63) circle (  0.57);

\path[draw=drawColor,line width= 0.6pt,line join=round] ( 89.46,146.20) -- ( 89.46,146.20);

\path[draw=drawColor,line width= 0.6pt,line join=round] ( 89.46,141.56) -- ( 89.46,141.56);
\definecolor{fillColor}{RGB}{148,188,228}

\path[draw=drawColor,line width= 0.6pt,line join=round,line cap=round,fill=fillColor] ( 85.58,146.20) --
	( 85.58,141.56) --
	( 93.35,141.56) --
	( 93.35,146.20) --
	( 85.58,146.20) --
	cycle;

\path[draw=drawColor,line width= 1.1pt,line join=round] ( 85.58,146.20) -- ( 93.35,146.20);

\path[draw=drawColor,line width= 0.4pt,line join=round,line cap=round] ( 34.88, 95.13) -- ( 40.43, 95.13);

\path[draw=drawColor,line width= 0.4pt,line join=round,line cap=round] ( 37.65, 92.36) -- ( 37.65, 97.91);

\path[draw=drawColor,line width= 0.4pt,line join=round,line cap=round] ( 45.24,136.92) -- ( 50.79,136.92);

\path[draw=drawColor,line width= 0.4pt,line join=round,line cap=round] ( 48.01,134.14) -- ( 48.01,139.69);

\path[draw=drawColor,line width= 0.4pt,line join=round,line cap=round] ( 55.60,113.70) -- ( 61.15,113.70);

\path[draw=drawColor,line width= 0.4pt,line join=round,line cap=round] ( 58.38,110.93) -- ( 58.38,116.48);

\path[draw=drawColor,line width= 0.4pt,line join=round,line cap=round] ( 65.96,113.70) -- ( 71.51,113.70);

\path[draw=drawColor,line width= 0.4pt,line join=round,line cap=round] ( 68.74,110.93) -- ( 68.74,116.48);

\path[draw=drawColor,line width= 0.4pt,line join=round,line cap=round] ( 76.32,132.28) -- ( 81.87,132.28);

\path[draw=drawColor,line width= 0.4pt,line join=round,line cap=round] ( 79.10,129.50) -- ( 79.10,135.05);

\path[draw=drawColor,line width= 0.4pt,line join=round,line cap=round] ( 86.69,141.56) -- ( 92.24,141.56);

\path[draw=drawColor,line width= 0.4pt,line join=round,line cap=round] ( 89.46,138.79) -- ( 89.46,144.34);
\end{scope}
\begin{scope}
\path[clip] (  0.00,  0.00) rectangle (101.18,180.67);
\definecolor{drawColor}{RGB}{0,0,0}

\node[text=drawColor,anchor=base east,inner sep=0pt, outer sep=0pt, scale=  0.80] at ( 26.49, 32.02) {0.0};

\node[text=drawColor,anchor=base east,inner sep=0pt, outer sep=0pt, scale=  0.80] at ( 26.49, 54.31) {0.2};

\node[text=drawColor,anchor=base east,inner sep=0pt, outer sep=0pt, scale=  0.80] at ( 26.49, 76.59) {0.4};

\node[text=drawColor,anchor=base east,inner sep=0pt, outer sep=0pt, scale=  0.80] at ( 26.49, 98.88) {0.6};

\node[text=drawColor,anchor=base east,inner sep=0pt, outer sep=0pt, scale=  0.80] at ( 26.49,121.16) {0.8};

\node[text=drawColor,anchor=base east,inner sep=0pt, outer sep=0pt, scale=  0.80] at ( 26.49,143.45) {1.0};
\end{scope}
\begin{scope}
\path[clip] (  0.00,  0.00) rectangle (101.18,180.67);
\definecolor{drawColor}{gray}{0.20}

\path[draw=drawColor,line width= 0.6pt,line join=round] ( 28.69, 34.78) --
	( 31.44, 34.78);

\path[draw=drawColor,line width= 0.6pt,line join=round] ( 28.69, 57.06) --
	( 31.44, 57.06);

\path[draw=drawColor,line width= 0.6pt,line join=round] ( 28.69, 79.35) --
	( 31.44, 79.35);

\path[draw=drawColor,line width= 0.6pt,line join=round] ( 28.69,101.63) --
	( 31.44,101.63);

\path[draw=drawColor,line width= 0.6pt,line join=round] ( 28.69,123.92) --
	( 31.44,123.92);

\path[draw=drawColor,line width= 0.6pt,line join=round] ( 28.69,146.20) --
	( 31.44,146.20);
\end{scope}
\begin{scope}
\path[clip] (  0.00,  0.00) rectangle (101.18,180.67);
\definecolor{drawColor}{gray}{0.20}

\path[draw=drawColor,line width= 0.6pt,line join=round] ( 37.65, 25.34) --
	( 37.65, 28.09);

\path[draw=drawColor,line width= 0.6pt,line join=round] ( 48.01, 25.34) --
	( 48.01, 28.09);

\path[draw=drawColor,line width= 0.6pt,line join=round] ( 58.38, 25.34) --
	( 58.38, 28.09);

\path[draw=drawColor,line width= 0.6pt,line join=round] ( 68.74, 25.34) --
	( 68.74, 28.09);

\path[draw=drawColor,line width= 0.6pt,line join=round] ( 79.10, 25.34) --
	( 79.10, 28.09);

\path[draw=drawColor,line width= 0.6pt,line join=round] ( 89.46, 25.34) --
	( 89.46, 28.09);
\end{scope}
\begin{scope}
\path[clip] (  0.00,  0.00) rectangle (101.18,180.67);
\definecolor{drawColor}{RGB}{0,0,0}

\node[text=drawColor,anchor=base,inner sep=0pt, outer sep=0pt, scale=  0.80] at ( 37.65, 17.63) {pr};

\node[text=drawColor,anchor=base,inner sep=0pt, outer sep=0pt, scale=  0.80] at ( 48.01, 17.63) {fl};

\node[text=drawColor,anchor=base,inner sep=0pt, outer sep=0pt, scale=  0.80] at ( 58.38, 17.63) {su};

\node[text=drawColor,anchor=base,inner sep=0pt, outer sep=0pt, scale=  0.80] at ( 68.74, 17.63) {pn};

\node[text=drawColor,anchor=base,inner sep=0pt, outer sep=0pt, scale=  0.80] at ( 79.10, 17.63) {pw};

\node[text=drawColor,anchor=base,inner sep=0pt, outer sep=0pt, scale=  0.80] at ( 89.46, 17.63) {ex};
\end{scope}
\begin{scope}
\path[clip] (  0.00,  0.00) rectangle (101.18,180.67);
\definecolor{drawColor}{RGB}{0,0,0}

\node[text=drawColor,anchor=base,inner sep=0pt, outer sep=0pt, scale=  0.90] at ( 63.56,  7.32) {kNN};
\end{scope}
\begin{scope}
\path[clip] (  0.00,  0.00) rectangle (101.18,180.67);
\definecolor{drawColor}{RGB}{0,0,0}

\node[text=drawColor,rotate= 90.00,anchor=base,inner sep=0pt, outer sep=0pt, scale=  0.90] at ( 11.70,101.63) {Success Rate};
\end{scope}
\end{tikzpicture}
 \hspace{-0.58cm}
	\begin{tikzpicture}[x=1pt,y=1pt]
\definecolor{fillColor}{RGB}{255,255,255}
\path[use as bounding box,fill=fillColor,fill opacity=0.00] (0,0) rectangle ( 79.50,180.67);
\begin{scope}
\path[clip] (  0.00,  0.00) rectangle ( 79.50,180.67);

\path[] (  0.00,  0.00) rectangle ( 79.50,180.68);
\end{scope}
\begin{scope}
\path[clip] (  8.25, 28.09) rectangle ( 74.00,175.17);

\path[] (  8.25, 28.09) rectangle ( 74.00,175.18);
\definecolor{drawColor}{RGB}{255,255,255}

\path[draw=drawColor,line width= 0.3pt,line join=round] (  8.25, 45.92) --
	( 74.00, 45.92);

\path[draw=drawColor,line width= 0.3pt,line join=round] (  8.25, 68.20) --
	( 74.00, 68.20);

\path[draw=drawColor,line width= 0.3pt,line join=round] (  8.25, 90.49) --
	( 74.00, 90.49);

\path[draw=drawColor,line width= 0.3pt,line join=round] (  8.25,112.78) --
	( 74.00,112.78);

\path[draw=drawColor,line width= 0.3pt,line join=round] (  8.25,135.06) --
	( 74.00,135.06);

\path[draw=drawColor,line width= 0.3pt,line join=round] (  8.25,157.35) --
	( 74.00,157.35);

\path[draw=drawColor,line width= 0.3pt,line join=round] (  8.25,168.49) --
	( 74.00,168.49);
\definecolor{drawColor}{RGB}{190,190,190}

\path[draw=drawColor,line width= 0.6pt,line join=round] (  8.25, 34.78) --
	( 74.00, 34.78);

\path[draw=drawColor,line width= 0.6pt,line join=round] (  8.25, 57.06) --
	( 74.00, 57.06);

\path[draw=drawColor,line width= 0.6pt,line join=round] (  8.25, 79.35) --
	( 74.00, 79.35);

\path[draw=drawColor,line width= 0.6pt,line join=round] (  8.25,101.63) --
	( 74.00,101.63);

\path[draw=drawColor,line width= 0.6pt,line join=round] (  8.25,123.92) --
	( 74.00,123.92);

\path[draw=drawColor,line width= 0.6pt,line join=round] (  8.25,146.20) --
	( 74.00,146.20);
\definecolor{drawColor}{RGB}{0,125,197}

\path[draw=drawColor,line width= 0.6pt,line join=round] ( 13.55,127.63) --
	( 15.67,127.63);

\path[draw=drawColor,line width= 0.6pt,line join=round] ( 14.61,127.63) --
	( 14.61, 90.49);

\path[draw=drawColor,line width= 0.6pt,line join=round] ( 13.55, 90.49) --
	( 15.67, 90.49);

\path[draw=drawColor,line width= 0.6pt,line join=round] ( 24.16,146.20) --
	( 26.28,146.20);

\path[draw=drawColor,line width= 0.6pt,line join=round] ( 25.22,146.20) --
	( 25.22,109.06);

\path[draw=drawColor,line width= 0.6pt,line join=round] ( 24.16,109.06) --
	( 26.28,109.06);

\path[draw=drawColor,line width= 0.6pt,line join=round] ( 34.76,127.63) --
	( 36.88,127.63);

\path[draw=drawColor,line width= 0.6pt,line join=round] ( 35.82,127.63) --
	( 35.82, 90.49);

\path[draw=drawColor,line width= 0.6pt,line join=round] ( 34.76, 90.49) --
	( 36.88, 90.49);

\path[draw=drawColor,line width= 0.6pt,line join=round] ( 45.37,146.20) --
	( 47.49,146.20);

\path[draw=drawColor,line width= 0.6pt,line join=round] ( 46.43,146.20) --
	( 46.43, 71.92);

\path[draw=drawColor,line width= 0.6pt,line join=round] ( 45.37, 71.92) --
	( 47.49, 71.92);

\path[draw=drawColor,line width= 0.6pt,line join=round] ( 55.97,146.20) --
	( 58.09,146.20);

\path[draw=drawColor,line width= 0.6pt,line join=round] ( 57.03,146.20) --
	( 57.03, 90.49);

\path[draw=drawColor,line width= 0.6pt,line join=round] ( 55.97, 90.49) --
	( 58.09, 90.49);

\path[draw=drawColor,line width= 0.6pt,line join=round] ( 66.57,146.20) --
	( 68.69,146.20);

\path[draw=drawColor,line width= 0.6pt,line join=round] ( 67.63,146.20) --
	( 67.63,136.92);

\path[draw=drawColor,line width= 0.6pt,line join=round] ( 66.57,136.92) --
	( 68.69,136.92);
\definecolor{fillColor}{RGB}{0,125,197}

\path[draw=drawColor,line width= 0.4pt,line join=round,line cap=round,fill=fillColor] ( 14.61, 34.78) circle (  0.57);

\path[draw=drawColor,line width= 0.6pt,line join=round] ( 14.61,113.70) -- ( 14.61,127.63);

\path[draw=drawColor,line width= 0.6pt,line join=round] ( 14.61, 90.49) -- ( 14.61, 90.49);
\definecolor{fillColor}{RGB}{148,188,228}

\path[draw=drawColor,line width= 0.6pt,line join=round,line cap=round,fill=fillColor] ( 10.64,113.70) --
	( 10.64, 90.49) --
	( 18.59, 90.49) --
	( 18.59,113.70) --
	( 10.64,113.70) --
	cycle;

\path[draw=drawColor,line width= 1.1pt,line join=round] ( 10.64,109.06) -- ( 18.59,109.06);

\path[draw=drawColor,line width= 0.6pt,line join=round] ( 25.22,146.20) -- ( 25.22,146.20);

\path[draw=drawColor,line width= 0.6pt,line join=round] ( 25.22,122.99) -- ( 25.22,109.06);

\path[draw=drawColor,line width= 0.6pt,line join=round,line cap=round,fill=fillColor] ( 21.24,146.20) --
	( 21.24,122.99) --
	( 29.19,122.99) --
	( 29.19,146.20) --
	( 21.24,146.20) --
	cycle;

\path[draw=drawColor,line width= 1.1pt,line join=round] ( 21.24,136.92) -- ( 29.19,136.92);

\path[draw=drawColor,line width= 0.6pt,line join=round] ( 35.82,127.63) -- ( 35.82,127.63);

\path[draw=drawColor,line width= 0.6pt,line join=round] ( 35.82,104.42) -- ( 35.82, 90.49);

\path[draw=drawColor,line width= 0.6pt,line join=round,line cap=round,fill=fillColor] ( 31.84,127.63) --
	( 31.84,104.42) --
	( 39.80,104.42) --
	( 39.80,127.63) --
	( 31.84,127.63) --
	cycle;

\path[draw=drawColor,line width= 1.1pt,line join=round] ( 31.84,118.35) -- ( 39.80,118.35);

\path[draw=drawColor,line width= 0.6pt,line join=round] ( 46.43,118.35) -- ( 46.43,146.20);

\path[draw=drawColor,line width= 0.6pt,line join=round] ( 46.43, 71.92) -- ( 46.43, 71.92);

\path[draw=drawColor,line width= 0.6pt,line join=round,line cap=round,fill=fillColor] ( 42.45,118.35) --
	( 42.45, 71.92) --
	( 50.40, 71.92) --
	( 50.40,118.35) --
	( 42.45,118.35) --
	cycle;

\path[draw=drawColor,line width= 1.1pt,line join=round] ( 42.45, 90.49) -- ( 50.40, 90.49);

\path[draw=drawColor,line width= 0.6pt,line join=round] ( 57.03,132.28) -- ( 57.03,146.20);

\path[draw=drawColor,line width= 0.6pt,line join=round] ( 57.03,104.42) -- ( 57.03, 90.49);

\path[draw=drawColor,line width= 0.6pt,line join=round,line cap=round,fill=fillColor] ( 53.05,132.28) --
	( 53.05,104.42) --
	( 61.01,104.42) --
	( 61.01,132.28) --
	( 53.05,132.28) --
	cycle;

\path[draw=drawColor,line width= 1.1pt,line join=round] ( 53.05,118.35) -- ( 61.01,118.35);
\definecolor{fillColor}{RGB}{0,125,197}

\path[draw=drawColor,line width= 0.4pt,line join=round,line cap=round,fill=fillColor] ( 67.63,109.06) circle (  0.57);

\path[draw=drawColor,line width= 0.6pt,line join=round] ( 67.63,146.20) -- ( 67.63,146.20);

\path[draw=drawColor,line width= 0.6pt,line join=round] ( 67.63,136.92) -- ( 67.63,136.92);
\definecolor{fillColor}{RGB}{148,188,228}

\path[draw=drawColor,line width= 0.6pt,line join=round,line cap=round,fill=fillColor] ( 63.66,146.20) --
	( 63.66,136.92) --
	( 71.61,136.92) --
	( 71.61,146.20) --
	( 63.66,146.20) --
	cycle;

\path[draw=drawColor,line width= 1.1pt,line join=round] ( 63.66,146.20) -- ( 71.61,146.20);

\path[draw=drawColor,line width= 0.4pt,line join=round,line cap=round] ( 11.84, 95.13) -- ( 17.39, 95.13);

\path[draw=drawColor,line width= 0.4pt,line join=round,line cap=round] ( 14.61, 92.36) -- ( 14.61, 97.91);

\path[draw=drawColor,line width= 0.4pt,line join=round,line cap=round] ( 22.44,132.28) -- ( 27.99,132.28);

\path[draw=drawColor,line width= 0.4pt,line join=round,line cap=round] ( 25.22,129.50) -- ( 25.22,135.05);

\path[draw=drawColor,line width= 0.4pt,line join=round,line cap=round] ( 33.05,113.70) -- ( 38.60,113.70);

\path[draw=drawColor,line width= 0.4pt,line join=round,line cap=round] ( 35.82,110.93) -- ( 35.82,116.48);

\path[draw=drawColor,line width= 0.4pt,line join=round,line cap=round] ( 43.65, 99.78) -- ( 49.20, 99.78);

\path[draw=drawColor,line width= 0.4pt,line join=round,line cap=round] ( 46.43, 97.00) -- ( 46.43,102.55);

\path[draw=drawColor,line width= 0.4pt,line join=round,line cap=round] ( 54.26,118.35) -- ( 59.80,118.35);

\path[draw=drawColor,line width= 0.4pt,line join=round,line cap=round] ( 57.03,115.57) -- ( 57.03,121.12);

\path[draw=drawColor,line width= 0.4pt,line join=round,line cap=round] ( 64.86,136.92) -- ( 70.41,136.92);

\path[draw=drawColor,line width= 0.4pt,line join=round,line cap=round] ( 67.63,134.14) -- ( 67.63,139.69);
\end{scope}
\begin{scope}
\path[clip] (  0.00,  0.00) rectangle ( 79.50,180.67);
\definecolor{drawColor}{gray}{0.20}

\path[draw=drawColor,line width= 0.6pt,line join=round] ( 14.61, 25.34) --
	( 14.61, 28.09);

\path[draw=drawColor,line width= 0.6pt,line join=round] ( 25.22, 25.34) --
	( 25.22, 28.09);

\path[draw=drawColor,line width= 0.6pt,line join=round] ( 35.82, 25.34) --
	( 35.82, 28.09);

\path[draw=drawColor,line width= 0.6pt,line join=round] ( 46.43, 25.34) --
	( 46.43, 28.09);

\path[draw=drawColor,line width= 0.6pt,line join=round] ( 57.03, 25.34) --
	( 57.03, 28.09);

\path[draw=drawColor,line width= 0.6pt,line join=round] ( 67.63, 25.34) --
	( 67.63, 28.09);
\end{scope}
\begin{scope}
\path[clip] (  0.00,  0.00) rectangle ( 79.50,180.67);
\definecolor{drawColor}{RGB}{0,0,0}

\node[text=drawColor,anchor=base,inner sep=0pt, outer sep=0pt, scale=  0.80] at ( 14.61, 17.63) {pr};

\node[text=drawColor,anchor=base,inner sep=0pt, outer sep=0pt, scale=  0.80] at ( 25.22, 17.63) {fl};

\node[text=drawColor,anchor=base,inner sep=0pt, outer sep=0pt, scale=  0.80] at ( 35.82, 17.63) {su};

\node[text=drawColor,anchor=base,inner sep=0pt, outer sep=0pt, scale=  0.80] at ( 46.43, 17.63) {pn};

\node[text=drawColor,anchor=base,inner sep=0pt, outer sep=0pt, scale=  0.80] at ( 57.03, 17.63) {pw};

\node[text=drawColor,anchor=base,inner sep=0pt, outer sep=0pt, scale=  0.80] at ( 67.63, 17.63) {ex};
\end{scope}
\begin{scope}
\path[clip] (  0.00,  0.00) rectangle ( 79.50,180.67);
\definecolor{drawColor}{RGB}{0,0,0}

\node[text=drawColor,anchor=base,inner sep=0pt, outer sep=0pt, scale=  0.90] at ( 41.12,  7.32) {DSM-kNN};
\end{scope}
\end{tikzpicture}
 \hspace{-0.58cm}
	\begin{tikzpicture}[x=1pt,y=1pt]
\definecolor{fillColor}{RGB}{255,255,255}
\path[use as bounding box,fill=fillColor,fill opacity=0.00] (0,0) rectangle ( 79.50,180.67);
\begin{scope}
\path[clip] (  0.00,  0.00) rectangle ( 79.50,180.67);

\path[] (  0.00,  0.00) rectangle ( 79.50,180.68);
\end{scope}
\begin{scope}
\path[clip] (  8.25, 28.09) rectangle ( 74.00,175.17);

\path[] (  8.25, 28.09) rectangle ( 74.00,175.18);
\definecolor{drawColor}{RGB}{255,255,255}

\path[draw=drawColor,line width= 0.3pt,line join=round] (  8.25, 45.92) --
	( 74.00, 45.92);

\path[draw=drawColor,line width= 0.3pt,line join=round] (  8.25, 68.20) --
	( 74.00, 68.20);

\path[draw=drawColor,line width= 0.3pt,line join=round] (  8.25, 90.49) --
	( 74.00, 90.49);

\path[draw=drawColor,line width= 0.3pt,line join=round] (  8.25,112.78) --
	( 74.00,112.78);

\path[draw=drawColor,line width= 0.3pt,line join=round] (  8.25,135.06) --
	( 74.00,135.06);

\path[draw=drawColor,line width= 0.3pt,line join=round] (  8.25,157.35) --
	( 74.00,157.35);

\path[draw=drawColor,line width= 0.3pt,line join=round] (  8.25,168.49) --
	( 74.00,168.49);
\definecolor{drawColor}{RGB}{190,190,190}

\path[draw=drawColor,line width= 0.6pt,line join=round] (  8.25, 34.78) --
	( 74.00, 34.78);

\path[draw=drawColor,line width= 0.6pt,line join=round] (  8.25, 57.06) --
	( 74.00, 57.06);

\path[draw=drawColor,line width= 0.6pt,line join=round] (  8.25, 79.35) --
	( 74.00, 79.35);

\path[draw=drawColor,line width= 0.6pt,line join=round] (  8.25,101.63) --
	( 74.00,101.63);

\path[draw=drawColor,line width= 0.6pt,line join=round] (  8.25,123.92) --
	( 74.00,123.92);

\path[draw=drawColor,line width= 0.6pt,line join=round] (  8.25,146.20) --
	( 74.00,146.20);
\definecolor{drawColor}{RGB}{0,125,197}

\path[draw=drawColor,line width= 0.6pt,line join=round] ( 13.55,127.63) --
	( 15.67,127.63);

\path[draw=drawColor,line width= 0.6pt,line join=round] ( 14.61,127.63) --
	( 14.61, 34.78);

\path[draw=drawColor,line width= 0.6pt,line join=round] ( 13.55, 34.78) --
	( 15.67, 34.78);

\path[draw=drawColor,line width= 0.6pt,line join=round] ( 24.16, 71.92) --
	( 26.28, 71.92);

\path[draw=drawColor,line width= 0.6pt,line join=round] ( 25.22, 71.92) --
	( 25.22, 34.78);

\path[draw=drawColor,line width= 0.6pt,line join=round] ( 24.16, 34.78) --
	( 26.28, 34.78);

\path[draw=drawColor,line width= 0.6pt,line join=round] ( 34.76, 90.49) --
	( 36.88, 90.49);

\path[draw=drawColor,line width= 0.6pt,line join=round] ( 35.82, 90.49) --
	( 35.82, 34.78);

\path[draw=drawColor,line width= 0.6pt,line join=round] ( 34.76, 34.78) --
	( 36.88, 34.78);

\path[draw=drawColor,line width= 0.6pt,line join=round] ( 45.37, 90.49) --
	( 47.49, 90.49);

\path[draw=drawColor,line width= 0.6pt,line join=round] ( 46.43, 90.49) --
	( 46.43, 34.78);

\path[draw=drawColor,line width= 0.6pt,line join=round] ( 45.37, 34.78) --
	( 47.49, 34.78);

\path[draw=drawColor,line width= 0.6pt,line join=round] ( 55.97, 39.42) --
	( 58.09, 39.42);

\path[draw=drawColor,line width= 0.6pt,line join=round] ( 57.03, 39.42) --
	( 57.03, 34.78);

\path[draw=drawColor,line width= 0.6pt,line join=round] ( 55.97, 34.78) --
	( 58.09, 34.78);

\path[draw=drawColor,line width= 0.6pt,line join=round] ( 66.57, 71.92) --
	( 68.69, 71.92);

\path[draw=drawColor,line width= 0.6pt,line join=round] ( 67.63, 71.92) --
	( 67.63, 34.78);

\path[draw=drawColor,line width= 0.6pt,line join=round] ( 66.57, 34.78) --
	( 68.69, 34.78);

\path[draw=drawColor,line width= 0.6pt,line join=round] ( 14.61,113.70) -- ( 14.61,127.63);

\path[draw=drawColor,line width= 0.6pt,line join=round] ( 14.61, 62.63) -- ( 14.61, 34.78);
\definecolor{fillColor}{RGB}{148,188,228}

\path[draw=drawColor,line width= 0.6pt,line join=round,line cap=round,fill=fillColor] ( 10.64,113.70) --
	( 10.64, 62.63) --
	( 18.59, 62.63) --
	( 18.59,113.70) --
	( 10.64,113.70) --
	cycle;

\path[draw=drawColor,line width= 1.1pt,line join=round] ( 10.64, 90.49) -- ( 18.59, 90.49);

\path[draw=drawColor,line width= 0.6pt,line join=round] ( 25.22, 71.92) -- ( 25.22, 71.92);

\path[draw=drawColor,line width= 0.6pt,line join=round] ( 25.22, 34.78) -- ( 25.22, 34.78);

\path[draw=drawColor,line width= 0.6pt,line join=round,line cap=round,fill=fillColor] ( 21.24, 71.92) --
	( 21.24, 34.78) --
	( 29.19, 34.78) --
	( 29.19, 71.92) --
	( 21.24, 71.92) --
	cycle;

\path[draw=drawColor,line width= 1.1pt,line join=round] ( 21.24, 53.35) -- ( 29.19, 53.35);

\path[draw=drawColor,line width= 0.6pt,line join=round] ( 35.82, 90.49) -- ( 35.82, 90.49);

\path[draw=drawColor,line width= 0.6pt,line join=round] ( 35.82, 48.70) -- ( 35.82, 34.78);

\path[draw=drawColor,line width= 0.6pt,line join=round,line cap=round,fill=fillColor] ( 31.84, 90.49) --
	( 31.84, 48.70) --
	( 39.80, 48.70) --
	( 39.80, 90.49) --
	( 31.84, 90.49) --
	cycle;

\path[draw=drawColor,line width= 1.1pt,line join=round] ( 31.84, 71.92) -- ( 39.80, 71.92);

\path[draw=drawColor,line width= 0.6pt,line join=round] ( 46.43, 62.63) -- ( 46.43, 90.49);

\path[draw=drawColor,line width= 0.6pt,line join=round] ( 46.43, 34.78) -- ( 46.43, 34.78);

\path[draw=drawColor,line width= 0.6pt,line join=round,line cap=round,fill=fillColor] ( 42.45, 62.63) --
	( 42.45, 34.78) --
	( 50.40, 34.78) --
	( 50.40, 62.63) --
	( 42.45, 62.63) --
	cycle;

\path[draw=drawColor,line width= 1.1pt,line join=round] ( 42.45, 44.06) -- ( 50.40, 44.06);
\definecolor{fillColor}{RGB}{0,125,197}

\path[draw=drawColor,line width= 0.4pt,line join=round,line cap=round,fill=fillColor] ( 57.03, 53.35) circle (  0.57);

\path[draw=drawColor,line width= 0.6pt,line join=round] ( 57.03, 39.42) -- ( 57.03, 39.42);

\path[draw=drawColor,line width= 0.6pt,line join=round] ( 57.03, 34.78) -- ( 57.03, 34.78);
\definecolor{fillColor}{RGB}{148,188,228}

\path[draw=drawColor,line width= 0.6pt,line join=round,line cap=round,fill=fillColor] ( 53.05, 39.42) --
	( 53.05, 34.78) --
	( 61.01, 34.78) --
	( 61.01, 39.42) --
	( 53.05, 39.42) --
	cycle;

\path[draw=drawColor,line width= 1.1pt,line join=round] ( 53.05, 34.78) -- ( 61.01, 34.78);

\path[draw=drawColor,line width= 0.6pt,line join=round] ( 67.63, 57.99) -- ( 67.63, 71.92);

\path[draw=drawColor,line width= 0.6pt,line join=round] ( 67.63, 34.78) -- ( 67.63, 34.78);

\path[draw=drawColor,line width= 0.6pt,line join=round,line cap=round,fill=fillColor] ( 63.66, 57.99) --
	( 63.66, 34.78) --
	( 71.61, 34.78) --
	( 71.61, 57.99) --
	( 63.66, 57.99) --
	cycle;

\path[draw=drawColor,line width= 1.1pt,line join=round] ( 63.66, 44.06) -- ( 71.61, 44.06);

\path[draw=drawColor,line width= 0.4pt,line join=round,line cap=round] ( 11.84, 85.85) -- ( 17.39, 85.85);

\path[draw=drawColor,line width= 0.4pt,line join=round,line cap=round] ( 14.61, 83.07) -- ( 14.61, 88.62);

\path[draw=drawColor,line width= 0.4pt,line join=round,line cap=round] ( 22.44, 53.35) -- ( 27.99, 53.35);

\path[draw=drawColor,line width= 0.4pt,line join=round,line cap=round] ( 25.22, 50.57) -- ( 25.22, 56.12);

\path[draw=drawColor,line width= 0.4pt,line join=round,line cap=round] ( 33.05, 67.28) -- ( 38.60, 67.28);

\path[draw=drawColor,line width= 0.4pt,line join=round,line cap=round] ( 35.82, 64.50) -- ( 35.82, 70.05);

\path[draw=drawColor,line width= 0.4pt,line join=round,line cap=round] ( 43.65, 53.35) -- ( 49.20, 53.35);

\path[draw=drawColor,line width= 0.4pt,line join=round,line cap=round] ( 46.43, 50.57) -- ( 46.43, 56.12);

\path[draw=drawColor,line width= 0.4pt,line join=round,line cap=round] ( 54.26, 39.42) -- ( 59.80, 39.42);

\path[draw=drawColor,line width= 0.4pt,line join=round,line cap=round] ( 57.03, 36.64) -- ( 57.03, 42.19);

\path[draw=drawColor,line width= 0.4pt,line join=round,line cap=round] ( 64.86, 48.70) -- ( 70.41, 48.70);

\path[draw=drawColor,line width= 0.4pt,line join=round,line cap=round] ( 67.63, 45.93) -- ( 67.63, 51.48);
\end{scope}
\begin{scope}
\path[clip] (  0.00,  0.00) rectangle ( 79.50,180.67);
\definecolor{drawColor}{gray}{0.20}

\path[draw=drawColor,line width= 0.6pt,line join=round] ( 14.61, 25.34) --
	( 14.61, 28.09);

\path[draw=drawColor,line width= 0.6pt,line join=round] ( 25.22, 25.34) --
	( 25.22, 28.09);

\path[draw=drawColor,line width= 0.6pt,line join=round] ( 35.82, 25.34) --
	( 35.82, 28.09);

\path[draw=drawColor,line width= 0.6pt,line join=round] ( 46.43, 25.34) --
	( 46.43, 28.09);

\path[draw=drawColor,line width= 0.6pt,line join=round] ( 57.03, 25.34) --
	( 57.03, 28.09);

\path[draw=drawColor,line width= 0.6pt,line join=round] ( 67.63, 25.34) --
	( 67.63, 28.09);
\end{scope}
\begin{scope}
\path[clip] (  0.00,  0.00) rectangle ( 79.50,180.67);
\definecolor{drawColor}{RGB}{0,0,0}

\node[text=drawColor,anchor=base,inner sep=0pt, outer sep=0pt, scale=  0.80] at ( 14.61, 17.63) {pr};

\node[text=drawColor,anchor=base,inner sep=0pt, outer sep=0pt, scale=  0.80] at ( 25.22, 17.63) {fl};

\node[text=drawColor,anchor=base,inner sep=0pt, outer sep=0pt, scale=  0.80] at ( 35.82, 17.63) {su};

\node[text=drawColor,anchor=base,inner sep=0pt, outer sep=0pt, scale=  0.80] at ( 46.43, 17.63) {pn};

\node[text=drawColor,anchor=base,inner sep=0pt, outer sep=0pt, scale=  0.80] at ( 57.03, 17.63) {pw};

\node[text=drawColor,anchor=base,inner sep=0pt, outer sep=0pt, scale=  0.80] at ( 67.63, 17.63) {ex};
\end{scope}
\begin{scope}
\path[clip] (  0.00,  0.00) rectangle ( 79.50,180.67);
\definecolor{drawColor}{RGB}{0,0,0}

\node[text=drawColor,anchor=base,inner sep=0pt, outer sep=0pt, scale=  0.90] at ( 41.12,  7.32) {RR-RFF};
\end{scope}
\end{tikzpicture}
 	\vspace{-0.7cm}
	\begin{center}
		\footnotesize Trained and Tested Gestures per Method
	\end{center}
	\caption{Extended user study: performance of the examined methods for individual gestures and significance from ANOVA ($\alpha=0.05$)}
	\label{fig:userstudy2_gestures}\end{figure}

The examination of the individual gestures, see \cref{fig:userstudy2_gestures}, exposes a behaviour that was comparable between kNN and DSM-kNN. Wrist flexion and extension achieved the highest success rates (100\% median for both methods). Pointing and pronation performed worst here (medians of 75\% as well as 58\% for kNN and 50\% as well as 67\% for DSM-kNN). In contrast, for RR-RFF pronation performed the best with similar success rates (median 50\%) as kNN and DSM-kNN, while power exposed severe issues (median 0\%, mean 4\%). The analysis of variances between the success rates of gesture performances for a single method could not show any significance within the method.

However, significant differences could be found between the methods for individual gestures: For power grasp ($p<0.0005$), extension ($p<0.0001$), flexion ($p<0.001$) as well as supination ($p<0.05$), RR-RFF was significantly worse than both kNN and DSM-kNN. There was no significant success rate difference for neither pronation nor pointing ($\alpha=0.05$).

\begin{figure}
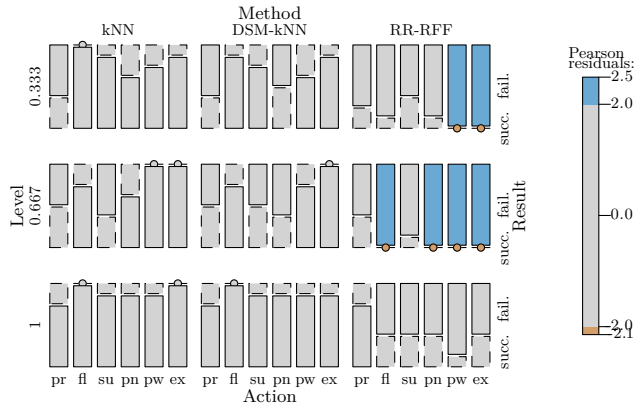

\hspace{-0.35cm}

 	\caption{Extended user study: mosaic plot illustrating significantly more failures for several gestures of different exertion levels for RR-RFF}
	\label{fig:study:userstudy2_mos}
\end{figure}

In \cref{fig:study:userstudy2_mos} the effects of the combinations of factors on significance are summarized. When examining the success rates for individual gestures at specific exertion levels, it can be seen that RR-RFF contributed to significantly more failures than the group of kNN-based methods at levels of $\nicefrac{1}{3}$ and $\nicefrac{2}{3}$, specifically for power grasp and wrist extension.

As for the basic user study, the computation times needed for each classification were measured and averaged per user and method. These are presented in \cref{tab:classTime_userstudy2}, again providing a confirmation for the real-time capability of the proposed methods.

\begin{table}[]
	\centering
	\caption{Extended user study (7 classes, 4 subjects), online classification time performance per subject, averaged for all prediction samples}
	\begin{tabular}{@{}cccc@{}}
		\toprule
		$[ ms ]$ & kNN    & DSM-kNN & RR-RFF \\ \midrule
		& 6.010 & 0.006  & 0.033 \\
		& 6.039 & 0.007  & 0.040 \\
		& 6.051 & 0.006  & 0.040 \\
		& 6.048 & 0.007  & 0.037 \\ \midrule
		Mean & 6.037 & 0.006 & 0.037 \\ \bottomrule
	\end{tabular}
	\label{tab:classTime_userstudy2}
\end{table}

\section{User Study Discussion}
\label{discussion}
Overall, it could be shown in the user studies that both the standard kNN scheme as well as the DSM-reduced technique yielded significantly higher success rates than RR-RFF and RR in most of the scenarios. The behaviour that kNN-based methods performed significantly better at higher exertion levels in the basic user study could be due to the fact that gestures of low intensity are more often subject to misclassification. This might result from a too high rest magnitude threshold which causes movements with low signal amplitudes being classified as rest state. In the extended user study, there was no significance for this difference. However, the effect could probably be curtailed in general by a learning process where the subjects would get used to the specific behaviour of the algorithm and adapt to it. Furthermore, the limb position effect might influence the results with respect to the average rest signal magnitude, although the subjects sat in a standardized pose.

Since the principle idea behind the use of Random Fourier Features is to fit cosine functions in the regression space, this might lead to unwanted behaviour at intermediate levels while showing better performance for especially the full gesture exertion (and slightly also for low levels). For RR, a similar principle holds, with the exception of using linear instead of cosine functions. Probably, the assumption of linear dependency is valid for small intensities, hence showing better success rates for $\nicefrac{1}{3}$ exertion level when comparing RR to RR-RFF. For increasing intensities, the proportionality behaviour might change to other functional dependencies. Nevertheless, the regression approach should fit the level of full intensity since it was trained on that. This means a higher number of successes for full intensity gestures. Regarding the basic user study, it has to be noted that RR-RFF performed significantly better than RR at full gesture exertion. There, also kNN and DSM-kNN had significantly higher success rates than the RR-based methods at intermediate and full gesture level. In the extended user study, kNN and DSM-kNN were significantly better than RR-RFF at all gesture levels.

For the kNN-based methods, the highest success rates were achieved for wrist extension (and in the second user study also for wrist flexion). The reason for this could be that power grasp and pointing gesture are most probably mainly exerted by the same group of muscles but the wrist gestures are not -- thus leading to better separability of those classes. In the basic user study, precisely the success rate difference between pointing gesture and wrist extension was significant; for DSM-kNN, also power grasp differed significantly from wrist extension. This might point towards the described explanation. In the extended user study, there was no significance for that.

Since the muscle groups activated in pointing gesture and power grasp are spatially close to each other from a biomechanical point of view, it could be explained that these two yielded the lowest success rates in the first user study, probably due to misclassifications between the two classes (they differed not significantly). The gestures are only distinguished in one degree of freedom (index finger), while the other degrees of freedom are the same. In the extended user study, also pronation and supination performed at the same lower level, however, there was no significance provable.
Standard RR showed the same behaviour in terms of individual gesture performances as the kNN-based methods (at a generally reduced success rate baseline), with the exception of wrist extension performing worse than flexion, although this difference was not significant in any case. Extension and flexion address the same degree of freedom which might therefore cause smaller deviations. With that, also the differences between the wrist movements in RR-RFF could be explained. One reason for pointing yielding the most successes in RR-RFF (although not significantly) could be that it was exerted by the subjects in a different manner than when the other methods were tested. Basically, the exertion can take place by also using the muscle group used for extension (stressing the index finger movement), instead of the muscle group used for flexion and power (where the activity patterns of the flexed fingers are stressed). If this was the case for RR-RFF, this might also explain the reduced performance of wrist extension. Nevertheless, the question would be why this would have been only the case for RR-RFF. A reason therefore might be traced back to the specific properties of Random Fourier Features when subjects try to reduce overshooting or similar.
However, for the success rate differences between individual gestures in RR and RR-RFF no significance could be substantiated.

In the extended study, pointing and pronation performed with the lowest success rates for the kNN-based methods, probably due to potentially addressing the same groups of muscles by these gestures, although there was no significance. The observation that RR-RFF performed worst in the extended user study (with the main exception of pronation where there was no significance between RR-RFF and the other schemes) could potentially be related to its capability of predicting multiple degrees of freedom in parallel. This might be leading to unstable predictions when it comes to predicting only a single degree of freedom. In order to realize a higher extent of comparability, only single degrees of freedom might be checked in the target achievement tests instead of all of them in future experiments. All in all, the extended user study could confirm the general observations made in the study's previous main part so that further in-depth experiments including more than the originally tested four gestures are recommended. Except for RR-RFF, there was no drop of success rate in comparison to the basic user study.

The fact that the DSM-reduced kNN did not perform worse (and in some cases even better) than the non-reduced kNN could be attributed to a possibly better subject's adaptability to the algorithm since there are less samples available whose decision borders are more clearly defined than when making use of the whole non-reduced dataset with possibly more abrupt changes in the decision borders the user can hardly learn. Another reason might be that noisy instances are discarded in the reduction process so that samples leading to misclassifications and worsening the performance are not considered anymore. Nevertheless, the slightly better performance of DSM-kNN might also be just resulting from stochastic factors. There is no statistical significance for that observation.

The generally better performance of the chosen kNN-based techniques can be also addressed to the fact that they are classification-based methods -- extended by proportionality schemes. This is why in the presented manner they are not suited for simultaneous control, i.~e. predicting mixed states of different gestures. In contrast, the RR-based methods also consider these states as an inherent property of regression. This means, as soon as multiple degrees of freedom are trained, RR and RR-RFF can get influenced by multiple degrees of freedom in parallel, although in the prediction tasks only a certain degree is tested at once. Therefore, the advantage of simultaneous control is at the expense of stability and robustness, and vice versa for kNN.

In order to find a good measure of comparability with regard to the development of the success rate over time for the individual methods, the available time slots during one experiment for a subject have been split into eight time subgroups. The differences over time are rather minor. The main perceivable difference originated from the choice of method. However, some interpretation of minor tendencies is given in the following.

It could be observed for the kNN-based methods that the performance in the first time slot was below the ones in the later timeslots. This could potentially be explained by a learning effect, i.~e. the subject adapting to the specific behaviour of the algorithm. The same effect could explain that in DSM-kNN the highest success rates were achieved towards the end of the experiment. At the very end, the performance decreased again, potentially due to muscle fatigue setting in. It could be seen that RR-RFF and RR exposed the same monotonicity when it comes to the mean performance over time, namely a sequence of possible learning effect, muscle fatigue and learning effect again. The learning effect might have set in again after each break. Apart from this, the RR-based methods showed a constant median of 33\% success rate over time (with the exception of the first time slot in RR). However, to gain representative insights, it might be useful to look into the time performance of individual gestures.

\section{Conclusions and Outlook}
In this work, a detailed examination of kNN-based learning techniques in the context of electromyographically controlled prostheses was conducted.

Summing up, with the proposed and implemented algorithms, all requirements stated in \cref{tab:req1} could be fulfilled.

First, the influence of several parameters on the block-wise cross-validation was examined for kNN classification. This showed that setting $k=1$ yielded excellent results, sometimes causing a ceiling effect. Accuracies often close to 100\%, always higher than 95\% for gesture subsets from (rs, pw, pn, fl, ex, pr, su) could be achieved, thus satisfying requirement R1.

The analysis of numbers $k$ on a higher scale was based on the inspiration that for increasing $k$s the upper probability bound of classification error decreases from about twice to once the Bayes probability of error \cite{topten}. Furthermore, previous work observed that ``the standard deviations tend to decrease as k-values increase'' \cite{KIM2011740}. Independent of the mentioned bounds, the experiments showed that the overall performance did not enhance for increasing $k$s -- as opposed to the expectations. All in all, relative $k$s until 5\% can be used without explicit drops in accuracy.

With the choice of $k=1$ the runtime complexity of the algorithm is reduced to linear time since instead of sorting distances (with logarithmic-linear time in the best case) a minimum search is sufficient, favouring the applicability on embedded systems.

In contrast to the Mahalanobis distance, the distance metrics based on the Minkowski norm proved well. In some cases, a higher order of norm yielded better results. This was the case when not considering the wrist rotation gestures where the Chebyshev distance performed the best. With pronation and supination included, there was the reverse effect, i. e. the Manhattan norm performing best. The Euclidean distance seems to be a good compromise to equalize both effects, although its calculation (multiplications) is slightly more computationally expensive than those of the Manhattan and Chebyshev norms.

The chosen factor of distance weighting seemed to not heavily influence the classification accuracy if $k$ was low. Nonetheless, higher exponents in the factor's divisor showed drastic improvements for a high $k$ so that this might be considered when choosing $k_{rel}$ (i. e. the proportion of $k$ and the total sample size) over 5\%. A weighting factor of $\nicefrac{1}{d^2}$ seemed to be sufficient in any case. When specifically referring to computational requirements (R5.1), a weighting of $\nicefrac{1}{d}$ might be of preference due to less multiplication operations. For $k_{rel}<5\%$, applying no weighting would be even more advantageous due to the reduced computational demand without loss in accuracy.

Regarding the pilot experiments, the offset scaling showed the best effect in optimizing the trade-off between a high value range of exertion levels and low-intensity gestures still being reachable. A scale offset divisor set to 5 could increase the success rate up to over 90\% both for the gesture set (rs, pw, pn, fl, ex) as well as the specifically problematic set of additionally including another degree of freedom in the form of pronation and supination. Nevertheless, possibilities for a usage of more sophisticated proportionality schemes have to be evaluated. Instead of using linear dependencies, other functions to realize interpolation might be tested.

The approach of rest magnitude thresholding could overcome the problem related to gestures exerted with less than half the full intensity being detected as rest state. A value of about 2.5 times the average magnitude across all rest training samples showed the best results.

The adaptions made to kNN have no influence on kNN's original incrementality so that requirement R4 is inherently guaranteed.

The motivation behind investigating prototype reduction mechanisms was to cope with the inherent issue of instance-based learning manifesting in very high computational demands during prediction. The concept of prototype reduction promised to reduce these demands by preponing calculations to the training phase where the amount of data to be processed in prediction is reduced, thus accomplishing both requirement R5.1 and requirement R5.2.

From the multitude of algorithms proposed in literature, the DSM algorithm singled out as highly appropriate. It is deterministic with regard to the size of the final reduced prototype set to be generated (memory determinism, requirement R5.2), yielded high cross-validation accuracies using EMG datasets captured with the Myo armband (requirement R1) at a low amount of time needed for reduction (requirement R5.3), and is considered to be incremental (requirement R4).

So far, the best results for DSM were shown when using the centres of the classes as initial prototypes. This was implemented by means of calculating the class-wise means. Nevertheless, this can lead to misclassification in the case of overlapping classes, specifically if they are concentric \cite[p.~335]{Trends}, why it is proposed to use the median instead. This might be examined in further research.

Furthermore, requirement R5.4 is also met due to DSM's elementary composition of an initialization of prototypes in the class centres and a correction phase shifting them by either penalizing or rewarding them depending on different basic criteria. As it is meant to be used with the proposed kNN implementation which guarantees this requirement, R2 is also fulfilled.

In the final user study, requirement R3 was evaluated for both the standard kNN approach (extended by the introduced adaptions) and kNN applied on the dataset reduced by DSM. It could be shown that the kNN methods performed significantly better than the ridge regression methods. Within the groups themselves, there was no statistical significance determinable. 

Interestingly, DSM-kNN and kNN performed equally well; DSM-kNN sometimes even better, even though a reduction of over 99\% was achieved by relying on only seven prototypes in total. By this, requirement R3 as to user satisfaction in real scenarios is fulfilled. The extended user study on additionally including the wrist rotation gestures, achieving very good success rates, might give further motivation to deeper analyze this influence in the context of representative studies.

Regarding the measured online timing behaviour in the proposed set-up, it could be shown that DSM reduces kNN's classification times by three orders of magnitude. With that, it achieves the same order as standard RR and is one order of magnitude faster than RR-RFF. This means, DSM-kNN is excellently suited for real-time control \cite{Farrell}.

It could be shown that DSM-kNN is an appropriate method to be integrated into wearable  prosthetic devices. Its properties lead to fulfilling non-functional requirements with respect to dependability and energy consumption, among other properties, favouring battery-powered portable myocontrol implementations.

A limitation of the conducted user study is the combined comparison of simultaneous and non-simultaneous control. The advantage of higher stability and robustness in the kNN-based methods comes at the disadvantage of not allowing to predict multiple degrees of freedom in parallel. On the contrary, the RR-based methods are subject to instabilities because of their tendency towards simultaneous predicting multiple degrees of freedoms. 
Approaches of how to handle mixed states in the case of kNN could comprise explicit learning on mixed gestures, or implicit learning by automatically creating mixtures of gestures. Another approach is suggested in \cite{Amsu}: if a non-simultaneous control method yields a low precision for the current gesture, it is switched to a simultaneous control scheme.

For a further evaluation of the kNN-based methods, a user study with handicapped subjects is of high importance. Additionally, besides using the visual feedback of the hand model, experiments with prosthetic devices have to conducted to identify the potential of the methods in terms of helpfulness for amputees. Longer-term studies may provide information about the influence of potential electrode shift as well as how to counteract this effect (as in \cite{prahm}).

Such experiments could also reveal further insights with regard to preprocessing, choice of the features (potentially combined with feature selection for horizontal data reduction), additional modalities, or coping with the limb position effect (where the armband's integrated IMU might be useful). Since we solely rely on the linear envelope, a combination of our work with an embedded feature analysis \cite{Raurale} seems promising to be investigated.

All in all, this paper confirmed the suitability of nearest neighbour learning techniques in the context of proportional myocontrol. Specifically, the results of using Decision Surface Mapping at very high reduction rates (>99\%) motivate to further look into this promising method.

\section*{Ethics Declarations}
All procedures performed in the studies that involved human participants were approved by the internal committee for personal data protection of the German Aerospace Center (DLR) and followed the World Medical Association's Declaration of Helsinki. Each participant was informed about the experimental process beforehand and signed an informed consent form.
\subsection*{Conflict of Interests}
The authors declare no competing interests.

\bibliographystyle{spmpsci2}
\bibliography{bibliography}

\begin{thebibliography}{100}
\providecommand{\url}[1]{{#1}}
\providecommand{\urlprefix}{URL }
\expandafter\ifx\csname urlstyle\endcsname\relax
  \providecommand{\doi}[1]{DOI~\discretionary{}{}{}#1}\else
  \providecommand{\doi}{DOI~\discretionary{}{}{}\begingroup
  \urlstyle{rm}\Url}\fi

\bibitem{lvqnn}
Aggarwal, C.C.: Data classification: Algorithms and Applications.
\newblock New York: Chapman and Hall/CRC (2014).
\newblock \doi{10.1201/b17320}

\bibitem{Ahmad}
{Ahmad}, S.A., {Chappell}, P.H.: Surface emg classification using moving
  approximate entropy.
\newblock In: 2007 International Conference on Intelligent and Advanced
  Systems, pp. 1163--1167 (2007).
\newblock \doi{10.1109/ICIAS.2007.4658567}

\bibitem{Ajiboye}
Ajiboye, A., Weir, R.: A heuristic fuzzy logic approach to emg pattern
  recognition for multifunctional prosthesis control.
\newblock IEEE transactions on neural systems and rehabilitation engineering :
  a publication of the IEEE Engineering in Medicine and Biology Society
  \textbf{13}, 280--91 (2005).
\newblock \doi{10.1109/TNSRE.2005.847357}

\bibitem{5767304}
{Al-Faiz}, M.Z., {Ali}, A.A., {Miry}, A.H.: A k-nearest neighbor based
  algorithm for human arm movements recognition using emg signals.
\newblock In: 2010 1st International Conference on Energy, Power and Control
  (EPC-IQ), pp. 159--167 (2010)

\bibitem{Amsu}
{Amsuess}, S., {Vujaklija}, I., {Goebel}, P., {Roche}, A.D., {Graimann}, B.,
  {Aszmann}, O.C., {Farina}, D.: Context-dependent upper limb prosthesis
  control for natural and robust use.
\newblock IEEE Transactions on Neural Systems and Rehabilitation Engineering
  \textbf{24}(7), 744--753 (2016).
\newblock \doi{10.1109/TNSRE.2015.2454240}

\bibitem{Antfolk2011ACB}
Antfolk, C., Sebelius, F.: A comparison between three pattern recognition
  algorithms for decoding finger movements using surface emg.
\newblock In: MyoElectric Controls/Powered Prosthetics Symposium (2011)

\bibitem{Arjunan}
{Arjunan}, S.P., {Kumar}, D.K.: Fractal based modelling and analysis of
  electromyography (emg) to identify subtle actions.
\newblock In: 2007 29th Annual International Conference of the IEEE Engineering
  in Medicine and Biology Society, pp. 1961--1964 (2007).
\newblock \doi{10.1109/IEMBS.2007.4352702}

\bibitem{Arvetti}
Arvetti, M., Gini, G., Folgheraiter, M.: Classification of emg signals through
  wavelet analysis and neural networks for controlling an active hand
  prosthesis.
\newblock In: 2007 IEEE 10th International Conference on Rehabilitation
  Robotics, ICORR'07, pp. 531--536 (2007).
\newblock \doi{10.1109/ICORR.2007.4428476}

\bibitem{Bajr}
Bajramovic, F., Mattern, F., Butko, N., Denzler, J.: A comparison of nearest
  neighbor search algorithms for generic object recognition.
\newblock In: Proceedings of the 8th International Conference on Advanced
  Concepts For Intelligent Vision Systems, ACIVS'06, pp. 1186--1197.
  Springer-Verlag, Berlin, Heidelberg (2006).
\newblock \doi{10.1007/11864349_108}

\bibitem{BARZILAY2011678}
Barzilay, O., Wolf, A.: A fast implementation for emg signal linear envelope
  computation.
\newblock Journal of Electromyography and Kinesiology \textbf{21}(4), 678 --
  682 (2011).
\newblock \doi{https://doi.org/10.1016/j.jelekin.2011.04.004}

\bibitem{6346923}
{Boschmann}, A., {Platzner}, M.: Reducing classification accuracy degradation
  of pattern recognition based myoelectric control caused by electrode shift
  using a high density electrode array.
\newblock In: 2012 Annual International Conference of the IEEE Engineering in
  Medicine and Biology Society, pp. 4324--4327 (2012).
\newblock \doi{10.1109/EMBC.2012.6346923}

\bibitem{AMPSO}
Cervantes, A., Galv{\'a}n, I., Isasi, P.: An adaptive michigan approach pso for
  nearest prototype classification.
\newblock In: J.~Mira, J.R. {\'A}lvarez (eds.) Nature Inspired Problem-Solving
  Methods in Knowledge Engineering, pp. 287--296. Springer Berlin Heidelberg,
  Berlin, Heidelberg (2007)

\bibitem{Chan}
{Chan}, A.D.C., {Englehart}, K.B.: Continuous myoelectric control for powered
  prostheses using hidden markov models.
\newblock IEEE Transactions on Biomedical Engineering \textbf{52}(1), 121--124
  (2005).
\newblock \doi{10.1109/TBME.2004.836492}

\bibitem{Chang}
Chang, G.C., Kang, W.J., Luh, J.J., Cheng, C.K., Lai, J.S., Chen, J.J.J., Kuo,
  T.S.: Real-time implementation of electromyogram pattern recognition as a
  control command of man-machine interface.
\newblock Medical Engineering \& Physics \textbf{18}(7), 529 -- 537 (1996).
\newblock \doi{https://doi.org/10.1016/1350-4533(96)00006-9}

\bibitem{Chen}
Chen, C., Jóźwik, A.: A sample set condensation algorithm for the class
  sensitive artificial neural network.
\newblock Pattern Recognition Letters \textbf{17}(8), 819 -- 823 (1996).
\newblock \doi{https://doi.org/10.1016/0167-8655(96)00041-4}

\bibitem{8122765}
{Chen}, H., {Zhang}, Y., {Zhang}, Z., {Fang}, Y., {Liu}, H., {Yao}, C.:
  Exploring the relation between emg sampling frequency and hand motion
  recognition accuracy.
\newblock In: 2017 IEEE International Conference on Systems, Man, and
  Cybernetics (SMC), pp. 1139--1144 (2017).
\newblock \doi{10.1109/SMC.2017.8122765}

\bibitem{5704586}
{Cipriani}, C., {Antfolk}, C., {Controzzi}, M., {Lundborg}, G., {Rosen}, B.,
  {Carrozza}, M.C., {Sebelius}, F.: Online myoelectric control of a dexterous
  hand prosthesis by transradial amputees.
\newblock IEEE Transactions on Neural Systems and Rehabilitation Engineering
  \textbf{19}(3), 260--270 (2011).
\newblock \doi{10.1109/TNSRE.2011.2108667}

\bibitem{CoverHart}
Cover, T.M., Hart, P.E.: Nearest neighbor pattern classification.
\newblock IEEE Transactions on Information Theory \textbf{13}(1), 21--27
  (2006).
\newblock \doi{10.1109/TIT.1967.1053964}

\bibitem{MSE}
Decaestecker, C.: Finding prototypes for nearest neighbour classification by
  means of gradient descent and deterministic annealing.
\newblock Pattern Recognition \textbf{30}(2), 281 -- 288 (1997).
\newblock \doi{https://doi.org/10.1016/S0031-3203(96)00072-6}

\bibitem{Della}
Dellacasa~Bellingegni, A., Gruppioni, E., Colazzo, G., Davalli, A., Sacchetti,
  R., Guglielmelli, E., Zollo, L.: Nlr, mlp, svm, and lda: a comparative
  analysis on emg data from people with trans-radial amputation.
\newblock Journal of NeuroEngineering and Rehabilitation \textbf{14}(1), 82
  (2017).
\newblock \doi{10.1186/s12984-017-0290-6}

\bibitem{Dening}
Dening, D., Gray, F., Haralick, R.: Prosthesis control using a nearest neighbor
  electromyographic pattern classifier.
\newblock Biomedical Engineering, IEEE Transactions on \textbf{30}, 356 -- 360
  (1983).
\newblock \doi{10.1109/TBME.1983.325138}

\bibitem{Englehart2003}
{Englehart}, K., {Hudgins}, B.: A robust, real-time control scheme for
  multifunction myoelectric control.
\newblock IEEE Transactions on Biomedical Engineering \textbf{50}(7), 848--854
  (2003).
\newblock \doi{10.1109/TBME.2003.813539}

\bibitem{Englehart1999}
Englehart, K., Hudgins, B., Parker, P., Stevenson, M.: Classification of the
  myoelectric signal using time-frequency based representations.
\newblock Medical Engineering \& Physics \textbf{21}(6), 431 -- 438 (1999).
\newblock \doi{https://doi.org/10.1016/S1350-4533(99)00066-1}

\bibitem{s18082553}
Esposito, D., Andreozzi, E., Fratini, A., Gargiulo, G.D., Savino, S., Niola,
  V., Bifulco, P.: A piezoresistive sensor to measure muscle contraction and
  mechanomyography.
\newblock Sensors \textbf{18}(8) (2018).
\newblock \doi{10.3390/s18082553}

\bibitem{Farrell}
Farrell, T.R., Weir, R.F.: The optimal controller delay for myoelectric
  prostheses.
\newblock IEEE Transactions on Neural Systems and Rehabilitation Engineering
  \textbf{15}(1), 111--118 (2007).
\newblock \doi{10.1109/TNSRE.2007.891391}

\bibitem{Farry1993}
Farry, K.A., Walker, I.D., Baraniuk, R.G.: Myoelectric teleoperation of a
  complex robotic hand.
\newblock IEEE Trans. Robotics and Automation \textbf{12}, 775--788 (1993)

\bibitem{ENPC}
Fern{\'a}ndez, F., Isasi, P.: Evolutionary design of nearest prototype
  classifiers.
\newblock Journal of Heuristics \textbf{10}(4), 431--454 (2004).
\newblock \doi{10.1023/B:HEUR.0000034715.70386.5b}

\bibitem{PSCSA}
Garain, U.: Prototype reduction using an artificial immune model.
\newblock Pattern Anal. Appl. \textbf{11}, 353--363 (2008).
\newblock \doi{10.1007/s10044-008-0106-1}

\bibitem{PS}
{García}, S., {Derrac}, J., {Cano}, J., {Herrera}, F.: Prototype selection for
  nearest neighbor classification: Taxonomy and empirical study.
\newblock IEEE Transactions on Pattern Analysis and Machine Intelligence
  \textbf{34}(3), 417--435 (2012).
\newblock \doi{10.1109/TPAMI.2011.142}

\bibitem{Geethanjali2015}
Geethanjali, P.: Comparative study of pca in classification of multichannel emg
  signals.
\newblock Australasian Physical {\&} Engineering Sciences in Medicine
  \textbf{38}(2), 331--343 (2015).
\newblock \doi{10.1007/s13246-015-0343-8}

\bibitem{Geethanjali2011}
Geethanjali, P., Ray, K.K.: Identification of motion from multi-channel emg
  signals for control of prosthetic hand.
\newblock Australasian Physical {\&} Engineering Sciences in Medicine
  \textbf{34}(3), 419--427 (2011).
\newblock \doi{10.1007/s13246-011-0079-z}

\bibitem{5396091}
{Geethanjali}, P., {Ray}, K.K., {Shanmuganathan}, P.V.: Actuation of prosthetic
  drive using emg signal.
\newblock In: TENCON 2009 - 2009 IEEE Region 10 Conference, pp. 1--5 (2009).
\newblock \doi{10.1109/TENCON.2009.5396091}

\bibitem{DSM}
Geva, S., Sitte, J.: Adaptive nearest neighbor pattern classifier.
\newblock IEEE transactions on neural networks / a publication of the IEEE
  Neural Networks Council \textbf{2}, 318--22 (1991).
\newblock \doi{10.1109/72.80344}

\bibitem{Gijsberts}
Gijsberts, A., Bohra, R., Sierra~Gonzalez, D., Werner, A., Nowak, M., Caputo,
  B., Roa, M., Castellini, C.: Stable myoelectric control of a hand prosthesis
  using non-linear incremental learning.
\newblock Frontiers in Neurorobotics \textbf{8}, 8 (2014).
\newblock \doi{10.3389/fnbot.2014.00008}

\bibitem{Gijs}
Gijsberts, A., Metta, G.: Incremental learning of robot dynamics using random
  features.
\newblock pp. 951--956 (2011).
\newblock \doi{10.1109/ICRA.2011.5980191}

\bibitem{Gini}
Gini, G., Arvetti, M., Somlai, I., Folgheraiter, M.: Acquisition and analysis
  of emg signals to recognize multiple hand movements for prosthetic
  applications.
\newblock Appl. Bionics Biomechanics \textbf{9}(2), 145--155 (2012).
\newblock \doi{10.3233/ABB-2011-0024}.
\newblock \urlprefix\url{http://dx.doi.org/10.3233/ABB-2011-0024}

\bibitem{4584252}
{Glette}, K., {Gruber}, T., {Kaufmann}, P., {Torresen}, J., {Sick}, B.,
  {Platzner}, M.: Comparing evolvable hardware to conventional classifiers for
  electromyographic prosthetic hand control.
\newblock In: 2008 NASA/ESA Conference on Adaptive Hardware and Systems, pp.
  32--39 (2008).
\newblock \doi{10.1109/AHS.2008.12}

\bibitem{glvq}
{Gonzalez}, A.I., {Grana}, M., {D'Anjou}, A.: An analysis of the glvq
  algorithm.
\newblock IEEE Transactions on Neural Networks \textbf{6}(4), 1012--1016
  (1995).
\newblock \doi{10.1109/72.392266}

\bibitem{Guler}
G\"{u}ler, N.F., Ko\c{c}er, S.: Classification of emg signals using pca and
  fft.
\newblock J. Med. Syst. \textbf{29}(3), 241--250 (2005).
\newblock \doi{10.1007/s10916-005-5184-7}.
\newblock \urlprefix\url{http://dx.doi.org/10.1007/s10916-005-5184-7}

\bibitem{BTS}
{Hamamoto}, Y., {Uchimura}, S., {Tomita}, S.: A bootstrap technique for nearest
  neighbor classifier design.
\newblock IEEE Transactions on Pattern Analysis and Machine Intelligence
  \textbf{19}(1), 73--79 (1997).
\newblock \doi{10.1109/34.566814}

\bibitem{Hannaford}
{Hannaford}, B., {Lehman}, S.: Short time fourier analysis of the
  electromyogram: Fast movements and constant contraction.
\newblock IEEE Transactions on Biomedical Engineering \textbf{BME-33}(12),
  1173--1181 (1986).
\newblock \doi{10.1109/TBME.1986.325697}

\bibitem{7437907}
{Haris}, M., {Chakraborty}, P., {Rao}, B.V.: Emg signal based finger movement
  recognition for prosthetic hand control.
\newblock In: 2015 Communication, Control and Intelligent Systems (CCIS), pp.
  194--198 (2015).
\newblock \doi{10.1109/CCIntelS.2015.7437907}

\bibitem{Hu}
Hu, X., Wang, Z., Ren, X.: Classification of surface emg signal with fractal
  dimension.
\newblock Journal of Zhejiang University. Science. B \textbf{6}, 844--8 (2005).
\newblock \doi{10.1631/jzus.2005.B0844}

\bibitem{Huang2004OptimizedGM}
Huang, Y., Englehart, K., Hudgins, B., Chan, A.D.C.: Optimized gaussian mixture
  models for upper limb motion classification.
\newblock The 26th Annual International Conference of the IEEE Engineering in
  Medicine and Biology Society \textbf{1}, 72--75 (2004)

\bibitem{Hudgins}
{Hudgins}, B., {Parker}, P., {Scott}, R.N.: A new strategy for multifunction
  myoelectric control.
\newblock IEEE Transactions on Biomedical Engineering \textbf{40}(1), 82--94
  (1993).
\newblock \doi{10.1109/10.204774}

\bibitem{Jeong2013}
Jeong, E.c., Kim, S.j., Song, Y.r., Lee, S.m.: Comparison of wrist motion
  classification methods using surface electromyogram.
\newblock Journal of Central South University \textbf{20}(4), 960--968 (2013).
\newblock \doi{10.1007/s11771-013-1571-2}

\bibitem{Jiang}
{Jiang}, M.W., {Wang}, R.C., {Wang}, J.Z., {Jin}, D.W.: A method of recognizing
  finger motion using wavelet transform of surface emg signal.
\newblock In: 2005 IEEE Engineering in Medicine and Biology 27th Annual
  Conference, pp. 2672--2674 (2005).
\newblock \doi{10.1109/IEMBS.2005.1617020}

\bibitem{Kakoty}
Kakoty, N.M., Hazarika, S.M.: Classification of grasp types through wavelet
  decomposition of emg signals.
\newblock 2009 2nd International Conference on Biomedical Engineering and
  Informatics pp. 1--5 (2009)

\bibitem{Kartsch}
{Kartsch}, V., {Benatti}, S., {Mancini}, M., {Magno}, M., {Benini}, L.: Smart
  wearable wristband for emg based gesture recognition powered by solar energy
  harvester.
\newblock In: 2018 IEEE International Symposium on Circuits and Systems
  (ISCAS), pp. 1--5 (2018).
\newblock \doi{10.1109/ISCAS.2018.8351727}

\bibitem{7591042}
{Khushaba}, R.N., {Al-Timemy}, A., {Al-Ani}, A., {Al-Jumaily}, A.: Myoelectric
  feature extraction using temporal-spatial descriptors for multifunction
  prosthetic hand control.
\newblock In: 2016 38th Annual International Conference of the IEEE Engineering
  in Medicine and Biology Society (EMBC), pp. 1696--1699 (2016).
\newblock \doi{10.1109/EMBC.2016.7591042}

\bibitem{KHUSHABA201210731}
Khushaba, R.N., Kodagoda, S., Takruri, M., Dissanayake, G.: Toward improved
  control of prosthetic fingers using surface electromyogram (emg) signals.
\newblock Expert Systems with Applications \textbf{39}(12), 10731 -- 10738
  (2012).
\newblock \doi{https://doi.org/10.1016/j.eswa.2012.02.192}

\bibitem{Kim2008}
Kim, J., Mastnik, S., Andr{\'e}, E.: Emg-based hand gesture recognition for
  realtime biosignal interfacing.
\newblock In: Proceedings of the 13th International Conference on Intelligent
  User Interfaces, IUI '08, pp. 30--39. ACM, New York, NY, USA (2008).
\newblock \doi{10.1145/1378773.1378778}

\bibitem{KIM2011740}
Kim, K.S., Choi, H.H., Moon, C.S., Mun, C.W.: Comparison of k-nearest neighbor,
  quadratic discriminant and linear discriminant analysis in classification of
  electromyogram signals based on the wrist-motion directions.
\newblock Current Applied Physics \textbf{11}(3), 740 -- 745 (2011).
\newblock \doi{10.1016/j.cap.2010.11.051}

\bibitem{HYB}
Kim, S.W., Oommen, B.J.: A brief taxonomy and ranking of creative prototype
  reduction schemes.
\newblock Pattern Analysis {\&} Applications \textbf{6}(3), 232--244 (2003).
\newblock \doi{10.1007/s10044-003-0191-0}

\bibitem{Kirlangic}
{Kirlangic}, M.E., {Denizhan}, Y.: Fractal modelling for pattern recognition
  via artificial neural networks.
\newblock In: 2000 IEEE International Conference on Acoustics, Speech, and
  Signal Processing. Proceedings (Cat. No.00CH37100), vol.~6, pp. 3610--3613
  vol.6 (2000).
\newblock \doi{10.1109/ICASSP.2000.860183}

\bibitem{LVQ}
{Kohonen}, T.: The self-organizing map.
\newblock Proceedings of the IEEE \textbf{78}(9), 1464--1480 (1990).
\newblock \doi{10.1109/5.58325}

\bibitem{Kuiken2016}
Kuiken, T.A., Miller, L.A., Turner, K., Hargrove, L.J.: A comparison of pattern
  recognition control and direct control of a multiple degree-of-freedom
  transradial prosthesis.
\newblock {IEEE} Journal of Translational Engineering in Health and Medicine
  \textbf{4}, 1--8 (2016).
\newblock \doi{10.1109/jtehm.2016.2616123}

\bibitem{Kusner}
Kusner, M.J., Tyree, S., Weinberger, K., Agrawal, K.: Stochastic neighbor
  compression.
\newblock In: Proceedings of the 31st International Conference on International
  Conference on Machine Learning - Volume 32, ICML'14, pp. II--622--II--630.
  JMLR.org (2014).
\newblock \doi{10.5555/3044805.3044962}

\bibitem{Kuzborskij}
{Kuzborskij}, I., {Gijsberts}, A., {Caputo}, B.: On the challenge of
  classifying 52 hand movements from surface electromyography.
\newblock In: 2012 Annual International Conference of the IEEE Engineering in
  Medicine and Biology Society, pp. 4931--4937 (2012).
\newblock \doi{10.1109/EMBC.2012.6347099}

\bibitem{ICPL}
Lam, W., Keung, C.K., Liu, D.: Discovering useful concept prototypes for
  classification based on filtering and abstraction.
\newblock Pattern Analysis and Machine Intelligence, IEEE Transactions on
  \textbf{24}, 1075-- 1090 (2002).
\newblock \doi{10.1109/TPAMI.2002.1023804}

\bibitem{LVQPRU}
Li, J., Manry, M.T., Yu, C., Wilson, D.R.: Prototype classifier design with
  pruning.
\newblock International Journal on Artificial Intelligence Tools
  \textbf{14}(01n02), 261--280 (2005).
\newblock \doi{10.1142/S0218213005002090}

\bibitem{7860925}
{Li}, Q.X., {Chan}, P.P.K., {Zhou}, D., {Fang}, Y., {Liu}, H., {Yeung}, D.S.:
  Improving robustness against electrode shift of semg based hand gesture
  recognition using online semi-supervised learning.
\newblock In: 2016 International Conference on Machine Learning and Cybernetics
  (ICMLC), vol.~1, pp. 344--349 (2016).
\newblock \doi{10.1109/ICMLC.2016.7860925}

\bibitem{nih}
Library, U.S.N.: Electromyography mesh descriptor data 2019 (1999).
\newblock
  \urlprefix\url{https://meshb.nlm.nih.gov/record/ui?name=Electromyography}

\bibitem{discr}
Liu, H., Hussain, F., Tan, C.L., Dash, M.: Discretization: An enabling
  technique.
\newblock Data Min. Knowl. Discov. \textbf{6}, 393--423 (2002).
\newblock \doi{10.1023/A:1016304305535}

\bibitem{MGauss}
Lozano, M., Sotoca, J.M., S\'{a}nchez, J.S., Pla, F., Pkalska, E., Duin,
  R.P.W.: Experimental study on prototype optimisation algorithms for
  prototype-based classification in vector spaces.
\newblock Pattern Recogn. \textbf{39}(10), 1827--1838 (2006).
\newblock \doi{10.1016/j.patcog.2006.04.005}

\bibitem{Zhizeng}
{Luo Zhizeng}, {Gao Jian}: Using singular eigenvalues of wavelet coefficient as
  the input of svm to recognize motion patterns of the hand.
\newblock In: 2005 International Conference on Neural Networks and Brain,
  vol.~3, pp. 1477--1481 (2005).
\newblock \doi{10.1109/ICNNB.2005.1614910}

\bibitem{Maitrot}
Maitrot, A., Lucas, M.F., Doncarli, C., Farina, D.: Signal-dependent wavelets
  for electromyogram classification.
\newblock Medical \& Biological Engineering \& Computing \textbf{43}(4),
  487--492 (2005).
\newblock \doi{10.1007/BF02344730}.
\newblock Erratum in: Med Bio Eng Comput. 2007. 45(8):807

\bibitem{mw}
Merriam-Webster: Definition of electromyograph (1944).
\newblock
  \urlprefix\url{https://www.merriam-webster.com/dictionary/electromyography}

\bibitem{Micera2000OnAI}
Micera, S., Sabatini, A.M., Dario, P.: On automatic identification of
  upper-limb movements using small-sized training sets of emg signals.
\newblock Medical engineering \& physics \textbf{22 8}, 527--33 (2000)

\bibitem{Lei}
{Min Lei}, {Zhi-Zhong Wang}, {Li-Yu Cai}, {Hai-Hong Zhang}, {Hua Cai}: An emg
  classifying method based on bayes' criterion.
\newblock In: Proceedings of the 20th Annual International Conference of the
  IEEE Engineering in Medicine and Biology Society. Vol.20 Biomedical
  Engineering Towards the Year 2000 and Beyond (Cat. No.98CH36286), vol.~5, pp.
  2625--2626 vol.5 (1998).
\newblock \doi{10.1109/IEMBS.1998.744998}

\bibitem{Nagata}
{Nagata}, K., {Adno}, K., {Magatani}, K., {Yamada}, M.: A classification method
  of hand movements using multi channel electrode.
\newblock In: 2005 IEEE Engineering in Medicine and Biology 27th Annual
  Conference, pp. 2375--2378 (2005).
\newblock \doi{10.1109/IEMBS.2005.1616944}

\bibitem{PSO}
Nanni, L., Lumini, A.: Particle swarm optimization for prototype reduction.
\newblock Neurocomputing \textbf{72}, 1092--1097 (2009).
\newblock \doi{10.1016/j.neucom.2008.03.008}

\bibitem{7748960}
{Negi}, S., {Kumar}, Y., {Mishra}, V.M.: Feature extraction and classification
  for emg signals using linear discriminant analysis.
\newblock In: 2016 2nd International Conference on Advances in Computing,
  Communication, Automation (ICACCA) (Fall), pp. 1--6 (2016).
\newblock \doi{10.1109/ICACCAF.2016.7748960}

\bibitem{Nowak2023}
Nowak, M., Bongers, R.M., van~der Sluis, C.K., Albu-Sch\"{a}ffer, A.,
  Castellini, C.: Simultaneous assessment and training of an upper-limb amputee
  using incremental machine-learning-based myocontrol: a single-case
  experimental design.
\newblock Journal of {NeuroEngineering} and Rehabilitation \textbf{20}(1)
  (2023).
\newblock \doi{10.1186/s12984-023-01171-2}

\bibitem{LVQTC}
Odorico, R.: Learning vector quantization with training count (lvqtc).
\newblock Neural networks : the official journal of the International Neural
  Network Society \textbf{10}(6), 1083—1088 (1997).
\newblock \doi{10.1016/s0893-6080(97)00012-9}

\bibitem{Paek2013}
{Paek}, A.Y., {Brown}, J.D., {Gillespie}, R.B., {O'Malley}, M.K., {Shewokis},
  P.A., {Contreras-Vidal}, J.L.: Reconstructing surface emg from scalp eeg
  during myoelectric control of a closed looped prosthetic device.
\newblock In: 2013 35th Annual International Conference of the IEEE Engineering
  in Medicine and Biology Society (EMBC), pp. 5602--5605 (2013).
\newblock \doi{10.1109/EMBC.2013.6610820}

\bibitem{Peerdeman}
Peerdeman, B., Boere, D., Witteveen, H., Veld, R., Hermens, H., Stramigioli,
  S., Rietman, J., Veltink, P., Misra, S.: Myoelectric forearm prostheses:
  State of the art from a user-centered perspective.
\newblock Journal of rehabilitation research and development \textbf{48},
  719--37 (2011).
\newblock \doi{10.1682/JRRD.2010.08.0161}

\bibitem{Trends}
Perez, J.C., Vidal, E.: Constructive design of lvq and dsm classifiers.
\newblock In: J.~Mira, J.~Cabestany, A.~Prieto (eds.) New Trends in Neural
  Computation, pp. 334--339. Springer Berlin Heidelberg, Berlin, Heidelberg
  (1993)

\bibitem{PHINYOMARK20134832}
Phinyomark, A., Quaine, F., Charbonnier, S., Serviere, C., Tarpin-Bernard, F.,
  Laurillau, Y.: Emg feature evaluation for improving myoelectric pattern
  recognition robustness.
\newblock Expert Systems with Applications \textbf{40}(12), 4832 -- 4840
  (2013).
\newblock \doi{https://doi.org/10.1016/j.eswa.2013.02.023}

\bibitem{prahm}
{Prahm}, C., {Schulz}, A., {Paaßen}, B., {Schoisswohl}, J., {Kaniusas}, E.,
  {Dorffner}, G., {Hammer}, B., {Aszmann}, O.: Counteracting electrode shifts
  in upper-limb prosthesis control via transfer learning.
\newblock IEEE Transactions on Neural Systems and Rehabilitation Engineering
  \textbf{27}(5), 956--962 (2019)

\bibitem{Geethanjali}
Purushothaman, G.: Myoelectric control of prosthetic hands: State-of-the-art
  review.
\newblock Medical Devices: Evidence and Research \textbf{Volume 9}, 247--255
  (2016).
\newblock \doi{10.2147/MDER.S91102}

\bibitem{Purushothaman2016}
Purushothaman, G., Ray, K.K.: Motion control of drives for prosthetic hand
  using continuous myoelectric signals.
\newblock Journal of The Institution of Engineers (India): Series B
  \textbf{97}(1), 55--60 (2016).
\newblock \doi{10.1007/s40031-014-0172-2}

\bibitem{Raurale}
Raurale, S.A., McAllister, J., del Rincon, J.M.: Real-time embedded emg signal
  analysis for wrist-hand pose identification.
\newblock IEEE Transactions on Signal Processing \textbf{68}, 2713--2723
  (2020).
\newblock \doi{10.1109/TSP.2020.2985299}

\bibitem{Rekhi}
{Rekhi}, N.S., {Singh}, H., {Arora}, A.S., {Rekhi}, A.K.: Analysis of emg
  signal using wavelet coefficients for upper limb function.
\newblock In: 2009 2nd IEEE International Conference on Computer Science and
  Information Technology, pp. 357--361 (2009).
\newblock \doi{10.1109/ICCSIT.2009.5234929}

\bibitem{Ren}
{Ren}, X., {Huang}, H., {Deng}, L.: Muap classification based on wavelet packet
  and fuzzy clustering technique.
\newblock In: 2009 3rd International Conference on Bioinformatics and
  Biomedical Engineering, pp. 1--4 (2009).
\newblock \doi{10.1109/ICBBE.2009.5163091}

\bibitem{Robinson}
Robinson, C.P., Li, B., Meng, Q., Pain, M.T.: Pattern classification of hand
  movements using time domain features of electromyography.
\newblock In: Proceedings of the 4th International Conference on Movement
  Computing, MOCO '17, pp. 27:1--27:6. ACM, New York, NY, USA (2017).
\newblock \doi{10.1145/3077981.3078031}

\bibitem{Saridis1982}
{Saridis}, G.N., {Gootee}, T.P.: Emg pattern analysis and classification for a
  prosthetic arm.
\newblock IEEE Transactions on Biomedical Engineering \textbf{BME-29}(6),
  403--412 (1982).
\newblock \doi{10.1109/TBME.1982.324954}

\bibitem{Scheme2013}
Scheme, E., Englehart, K.: Training strategies for mitigating the effect of
  proportional control on classification in pattern
  recognition{\textendash}based myoelectric control.
\newblock {JPO} Journal of Prosthetics and Orthotics \textbf{25}(2), 76--83
  (2013).
\newblock \doi{10.1097/jpo.0b013e318289950b}

\bibitem{Scheme}
{Scheme}, E., {Lock}, B., {Hargrove}, L., {Hill}, W., {Kuruganti}, U.,
  {Englehart}, K.: Motion normalized proportional control for improved pattern
  recognition-based myoelectric control.
\newblock IEEE Transactions on Neural Systems and Rehabilitation Engineering
  \textbf{22}(1), 149--157 (2014)

\bibitem{Shin}
Shin, S., Langari, R., Tafrershi, R.: A performance comparison of emg
  classification methods for hand and finger motion (2014).
\newblock \doi{10.1115/DSCC2014-5993}

\bibitem{Du}
{Sijiang Du}, {Vuskovic}, M.: Temporal vs. spectral approach to feature
  extraction from prehensile emg signals.
\newblock In: Proceedings of the 2004 IEEE International Conference on
  Information Reuse and Integration, 2004. IRI 2004., pp. 344--350 (2004).
\newblock \doi{10.1109/IRI.2004.1431485}

\bibitem{RMHC}
Skalak, D.B.: Prototype and feature selection by sampling and random mutation
  hill climbing algorithms.
\newblock In: W.W. Cohen, H.~Hirsh (eds.) Machine Learning Proceedings 1994,
  pp. 293 -- 301. Morgan Kaufmann, San Francisco (CA) (1994).
\newblock \doi{https://doi.org/10.1016/B978-1-55860-335-6.50043-X}

\bibitem{Sueaseenak}
{Sueaseenak}, D., {Wibirama}, S., {Chanwimalueang}, T., {Pintavirooj}, C.,
  {Sangworasil}, M.: Comparison study of muscular-contraction classification
  between independent component analysis and artificial neural network.
\newblock In: 2008 International Symposium on Communications and Information
  Technologies, pp. 468--472 (2008).
\newblock \doi{10.1109/ISCIT.2008.4700236}

\bibitem{Saridis1984}
{Sukhan Lee}, {Saridis}, G.: The control of a prosthetic arm by emg pattern
  recognition.
\newblock IEEE Transactions on Automatic Control \textbf{29}(4), 290--302
  (1984).
\newblock \doi{10.1109/TAC.1984.1103521}

\bibitem{szib}
Sziburis, T.: Nearest-neighbour-based learning techniques for proportional
  myocontrol in prosthetics.
\newblock Master's thesis, University of Trento, Università degli Studi di
  Trento (2019).
\newblock \urlprefix\url{https://elib.dlr.de/133564}.
\newblock At German Aerospace Center (DLR)

\bibitem{szibBIOSIGNALS}
Sziburis, T., Nowak, M., Brunelli, D.: Prototype reduction on semg data for
  instance-based gesture learning towards real-time prosthetic control.
\newblock In: Proceedings of the 14th International Joint Conference on
  Biomedical Engineering Systems and Technologies - Volume 2: BIOSIGNALS,, pp.
  299--305. INSTICC, SciTePress (2021).
\newblock \doi{10.5220/0010327500002865}

\bibitem{szibICNR}
Sziburis, T., Nowak, M., Brunelli, D.: Knn learning techniques for proportional
  myocontrol in prosthetics.
\newblock In: D.~Torricelli, M.~Akay, J.L. Pons (eds.) Converging Clinical and
  Engineering Research on Neurorehabilitation IV, pp. 679--683. Springer
  International Publishing, Cham (2022)

\bibitem{6487520}
{Tello}, R.M.G., {Bastos-Filho}, T., {Costa}, R.M., {Frizera-Neto}, A.,
  {Arjunan}, S., {Kumar}, D.: Towards semg classification based on bayesian and
  k-nn to control a prosthetic hand.
\newblock In: 2013 ISSNIP Biosignals and Biorobotics Conference: Biosignals and
  Robotics for Better and Safer Living (BRC), pp. 1--6 (2013).
\newblock \doi{10.1109/BRC.2013.6487520}

\bibitem{PG}
{Triguero}, I., {Derrac}, J., {Garcia}, S., {Herrera}, F.: A taxonomy and
  experimental study on prototype generation for nearest neighbor
  classification.
\newblock IEEE Transactions on Systems, Man, and Cybernetics, Part C
  (Applications and Reviews) \textbf{42}(1), 86--100 (2012).
\newblock \doi{10.1109/TSMCC.2010.2103939}

\bibitem{Triguero}
Triguero, I., González, S., Moyano, J., García, S., Alcala-Fdez, J., Luengo,
  J., Fernández, A., Del~Jesus, M.J., Sanchez, L., Herrera, F.: Keel 3.0: An
  open source software for multi-stage analysis in data mining.
\newblock International Journal of Computational Intelligence Systems
  \textbf{10}(1), 1238--1249 (2017).
\newblock \doi{10.2991/ijcis.10.1.82}

\bibitem{Visconti}
Visconti, P., Gaetani, F., Zappatore, G., Primiceri, P.: Technical features and
  functionalities of myo armband: An overview on related literature and
  advanced applications of myoelectric armbands mainly focused on arm
  prostheses.
\newblock International Journal on Smart Sensing and Intelligent Systems
  \textbf{11}, 1--25 (2018).
\newblock \doi{10.21307/ijssis-2018-005}

\bibitem{Vujaklija}
Vujaklija, I., Farina, D., Aszmann, O.: New developments in prosthetic arm
  systems.
\newblock Orthopedic Research and Reviews \textbf{20168}, 31--39 (2016).
\newblock \doi{10.2147/ORR.S71468}

\bibitem{Kang}
{Wen-Juh Kang}, {Jiue-Rou Shiu}, {Cheng-Kung Cheng}, {Jin-Shin Lai}, {Hen-Wai
  Tsao}, {Te-Son Kuo}: The application of cepstral coefficients and maximum
  likelihood method in emg pattern recognition [movements classification].
\newblock IEEE Transactions on Biomedical Engineering \textbf{42}(8), 777--785
  (1995).
\newblock \doi{10.1109/10.398638}

\bibitem{Winter}
Winter, D.A.: Biomechanics and motor control of human movement, 4 edn.
\newblock John Wiley \& Sons, Hoboken, N.J. (2005)

\bibitem{topten}
Wu, X., Kumar, V., Ross~Quinlan, J., Ghosh, J., Yang, Q., Motoda, H.,
  McLachlan, G.J., Ng, A., Liu, B., Yu, P.S., Zhou, Z.H., Steinbach, M., Hand,
  D.J., Steinberg, D.: Top 10 algorithms in data mining.
\newblock Knowl. Inf. Syst. \textbf{14}(1), 1--37 (2007).
\newblock \doi{10.1007/s10115-007-0114-2}

\bibitem{1519588}
{Yonghong Huang}, {Englehart}, K.B., {Hudgins}, B., {Chan}, A.D.C.: A gaussian
  mixture model based classification scheme for myoelectric control of powered
  upper limb prostheses.
\newblock IEEE Transactions on Biomedical Engineering \textbf{52}(11),
  1801--1811 (2005).
\newblock \doi{10.1109/TBME.2005.856295}

\bibitem{Zhang1991ClusteringAA}
Zhang, L.Q., Shiavi, R., Hunt, M.A., Chen, J.J.: Clustering analysis and
  pattern discrimination of emg linear envelopes.
\newblock IEEE Transactions on Biomedical Engineering \textbf{38}, 777--784
  (1991)

\bibitem{Zhang2013}
Zhang, Z., Wong, C., Yang, G.Z.: Forearm functional movement recognition using
  spare channel surface electromyography.
\newblock pp. 1--6 (2013).
\newblock \doi{10.1109/BSN.2013.6575507}

\end{thebibliography}

\newpage
\appendix

\section{DSM Runtime Complexity} \label{a_DSM}
\begin{table}[H]
	\centering
	\caption{Time complexity of DSM, training phase consists of initialization and reduction}
	\begin{adjustbox}{angle=90}
\begin{tabular}{@{}llcl@{}}
			\toprule
			\multicolumn{4}{c}{\textbf{Initialization}}                                                                               \\ \midrule
			& \multicolumn{1}{c}{Step}                          & Complexity & \multicolumn{1}{c}{Comment}               \\ \midrule
			& create new prototype set (signal in X/label in Y) & $\mathcal{O}(1)$       & if static arrays used                     \\
			$C$ times:     & get samples of specific class                     & $\mathcal{O}(N)$       &                                           \\
			& average these samples                             & $\mathcal{O}(N)$       &                                           \\
			& set row in prototype set                          & $\mathcal{O}(1)$       & if implemented as array with indexing     \\
			$(M-C)$ times: & if selection from specific class: get its samples & $\mathcal{O}(N)$       &                                           \\
			& get random number                                 & $\mathcal{O}(1)$       &                                           \\
\end{tabular}
	\end{adjustbox}
	\label{tab:compl}
\end{table}
\newpage
\section{Further Offline Cross-Validation Results}
\begin{figure}[ht]
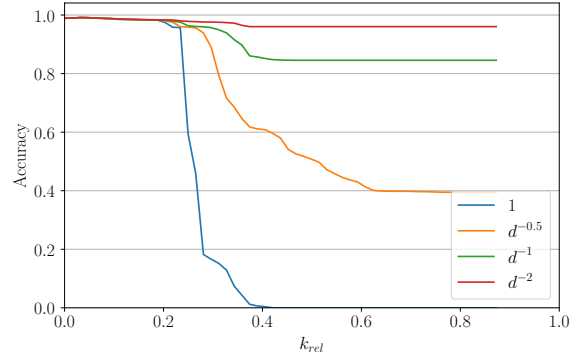
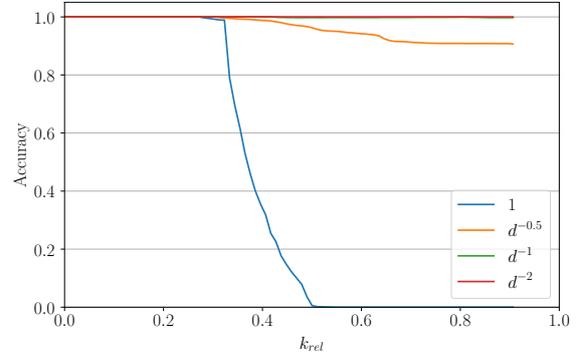

	\centering
	\begin{subfigure}{\linewidth}
		\begin{adjustbox}{max width=\columnwidth}
			\begingroup \makeatletter %
\makeatother \endgroup  		\end{adjustbox}
		\caption{Euclidean distance}\label{fig:cv:dw1}
	\end{subfigure}\\
\begin{subfigure}{\linewidth}
		\begin{adjustbox}{max width=\columnwidth}
			\begingroup \makeatletter %
\makeatother \endgroup  		\end{adjustbox}
		\caption{Chebyshev distance}\label{fig:cv:dw3}
	\end{subfigure}
	\caption{Distance weighting influence on cross-validation accuracy for sets of four gestures (rs, pw, fl, ex)}\label{fig:cv:dw_a}\end{figure}

\begin{figure}[ht]
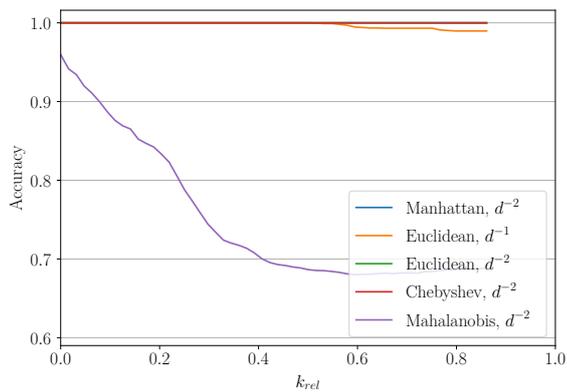
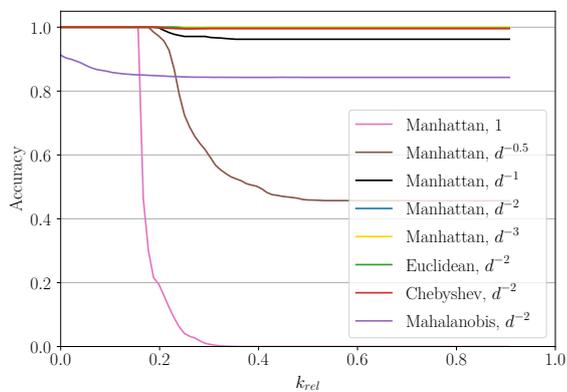
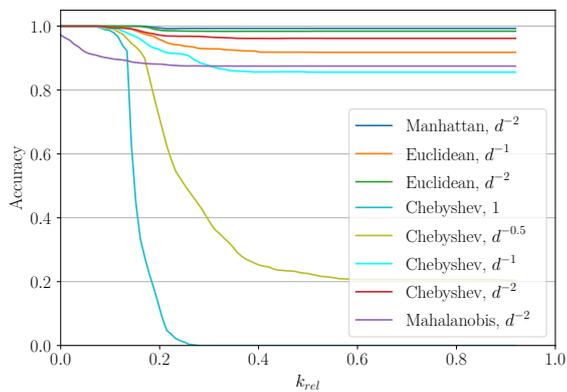

	\centering
	\begin{subfigure}{\linewidth}
		\begin{adjustbox}{max width=\columnwidth}
			\begingroup \makeatletter %
\makeatother \endgroup  \end{adjustbox}
		\caption{Gestures (rs, pw, fl, ex)}\label{fig:cv:dm1}
	\end{subfigure}\\
	\begin{subfigure}{\linewidth}
		\begin{adjustbox}{max width=\columnwidth}
			\begingroup \makeatletter %
\makeatother \endgroup  		\end{adjustbox}
		\caption{Gestures (rs, pw, fl, ex, pr, su)}\label{fig:cv:dm4}
	\end{subfigure}\\
	\begin{subfigure}{\linewidth}
		\begin{adjustbox}{max width=\columnwidth}
			\begingroup \makeatletter %
\makeatother \endgroup  		\end{adjustbox}
		\caption{Gestures (rs, pw, pn, fl, ex, pr, su)}\label{fig:cv:dm3}
	\end{subfigure}
	\caption{Distance metric and weighting influence on cross-validation accuracy}\label{fig:cv:dm_a}\end{figure}

\onecolumn
\begin{figure}
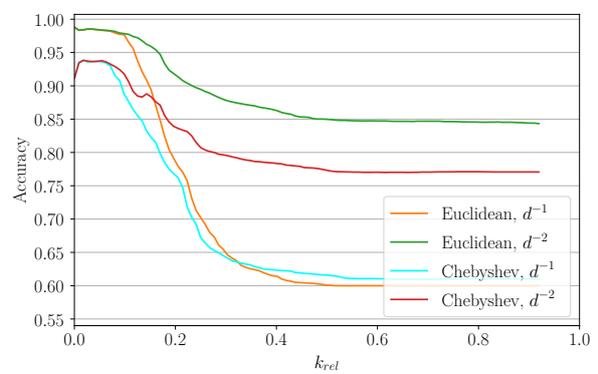
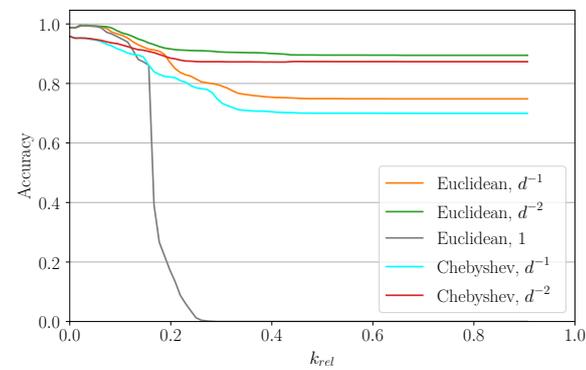
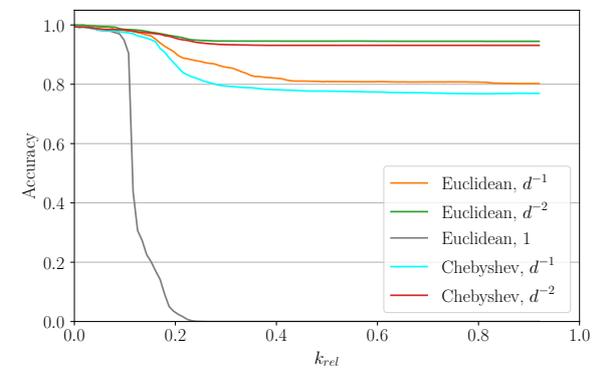
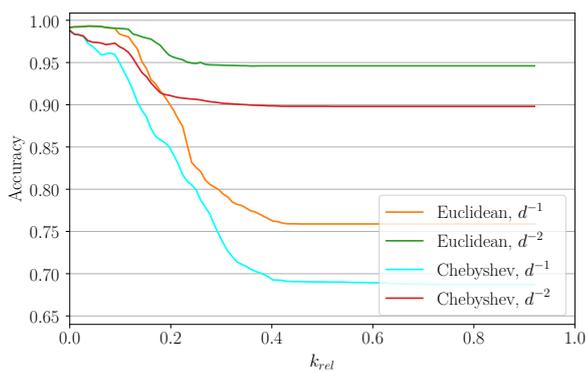
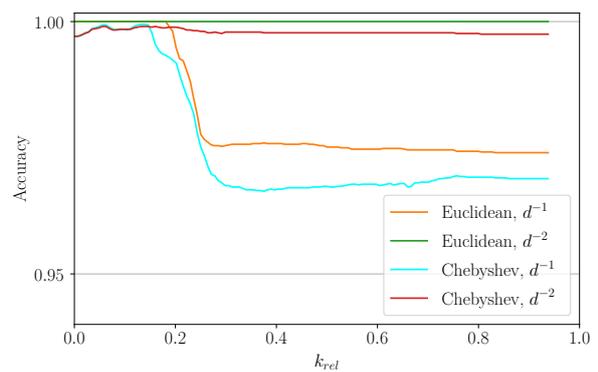

	\centering
	\begin{subfigure}{0.493\linewidth}
		\begin{adjustbox}{max width=\columnwidth}
			\begingroup \makeatletter %
\makeatother \endgroup  \end{adjustbox}
		\caption{Gestures (rs, pw, fl, ex)}\label{fig:cv:mw1}
	\end{subfigure}
	\begin{subfigure}{0.493\linewidth}
		\begin{adjustbox}{max width=\columnwidth}
			\begingroup \makeatletter %
\makeatother \endgroup  \end{adjustbox}
		\caption{Gestures (rs, pw, pn, fl, ex)}\label{fig:cv:mw2}
	\end{subfigure}\\
	\vspace{0.2cm}
	\begin{subfigure}{0.493\linewidth}
		\begin{adjustbox}{max width=\columnwidth}
			\begingroup \makeatletter %
\makeatother \endgroup  \end{adjustbox}
		\caption{Gestures (rs, pw, fl, ex, pr, su)}\label{fig:cv:mw3}
	\end{subfigure}
	\begin{subfigure}{0.493\linewidth}
		\begin{adjustbox}{max width=\columnwidth}
			\begingroup \makeatletter %
\makeatother \endgroup  \end{adjustbox}
		\caption{Gestures (rs, pw, pn, fl, ex, pr, su)}\label{fig:cv:mw4}
	\end{subfigure}\\
	\vspace{0.2cm}
	\begin{subfigure}{0.493\linewidth}
		\begin{adjustbox}{max width=\columnwidth}
			\begingroup \makeatletter %
\makeatother \endgroup  		\end{adjustbox}
		\caption{Gestures (rs, pw, fl, ex, pr, su)}\label{fig:cv:mw5}
	\end{subfigure}
	\begin{subfigure}{0.493\linewidth}
		\begin{adjustbox}{max width=\columnwidth}
			\begingroup \makeatletter %
\makeatother \endgroup  		\end{adjustbox}
		\caption{Gestures (rs, pw, pn, fl, ex, pr, su)}\label{fig:cv:mw6}
	\end{subfigure}\\
	\vspace{0.2cm}
	\begin{subfigure}{0.493\linewidth}
		\begin{adjustbox}{max width=\columnwidth}
			\begingroup \makeatletter %
\makeatother \endgroup  		\end{adjustbox}
		\caption{Gestures (rs, pw, pn, fl, ex, pr, su)}\label{fig:cv:mw7}
	\end{subfigure}
	\begin{subfigure}{0.493\linewidth}
		\begin{adjustbox}{max width=\columnwidth}
			\begingroup \makeatletter %
\makeatother \endgroup  \end{adjustbox}
		\caption{Gestures (rs, pw, fl, ex, pr, su)}\label{fig:cv:mw8}
	\end{subfigure}
	\caption{Influence of metrics/weightings on cross-validation accuracy, further datasets}
	\label{fig:cv:mw}\end{figure}

\end{document}